\documentclass{article}
\PassOptionsToPackage{numbers, compress}{natbib}
\usepackage{arxiv}

\usepackage[utf8]{inputenc} 
\usepackage[T1]{fontenc}    
\usepackage{hyperref}       
\usepackage{url}            
\usepackage{booktabs}       
\usepackage{amsfonts}       
\usepackage{nicefrac}       
\usepackage{microtype}      
\usepackage{lipsum}		
\usepackage{graphicx}
\usepackage{natbib}
\usepackage{doi}

\usepackage{multibib}
\usepackage{xcolor}   
\usepackage{siunitx}  
\usepackage{paralist}
\usepackage{caption}
\usepackage{subcaption}

\usepackage{amsmath}

\newcommand{\longtitle}{MetMamba: Regional Weather Forecasting with Spatial-Temporal Mamba Model}
\newcommand{\longtitlenostyle}{\emph{MetMamba}: Regional Weather Forecasting with Spatial-Temporal Mamba Model}
\renewcommand{\shorttitle}{\textit{MetMamba}: Regional Weather Forecasting}

\newcommand{\company}{Beijing PRESKY Technology Co., Ltd. \\}
\newcommand{\companyaddress}{Beijing \\}
\newcommand{\companyemail}[1]{\texttt{{#1}@cnpresky.com} \\} 

\newcommand{\rolloutlen}{N_\text{Rollout}}
\newcommand{\inittime}{t_{\text{Init}}}
\newcommand{\variableweight}{\omega}

\newcommand{\loss}{\mathcal{L}}
\newcommand{\weatherstatedim}{S}
\newcommand{\interiorgrid}{\mathbb{G}}
\newcommand{\setsize}[1]{\left| #1 \right|}
\newcommand{\weatherstate}{X}
\newcommand{\pred}{\hat{\weatherstate}}

\newcommand{\boundary}{\mathbb{B}}

\newcommand{\set}[1]{\{#1\}}

\newcommand{\al}[2][my_equation]{\begin{align}\label{eq:#1}#2\end{align}}
\newcommand{\ngridpoints}{N}

\newcommand{\reffig}[1]{Figure~\ref{fig:#1}}

\newcommand{\refapp}[1]{Appendix \ref{sec:\detokenize{#1}}}

\newcommand{\spatialsubfig}[2]{
\begin{subfigure}[b]{0.49\textwidth}
     \centering
     \includegraphics[width=\textwidth]{#1}
     \caption{\qty{#2}{\hour}}
 \end{subfigure}
}

\newcommand{\normalsubfig}[2]{
\begin{subfigure}[b]{0.49\textwidth}
     \centering
     \includegraphics[width=\textwidth]{#1}
     \caption{#2}
 \end{subfigure}
}

\newcommand{\normalspatialsubfig}[3]{
\begin{subfigure}[b]{0.49\textwidth}
     \centering
     \includegraphics[width=\textwidth]{#1}
     \vspace{-2em}
     \caption{#2 \qty{#3}{\hour}}
 \end{subfigure}
}

\usepackage{amsmath}
\usepackage{cleveref}

\DeclareSIUnit\year{yr}

\title{\longtitle}

\author{
  \href{https://orcid.org/0009-0008-0461-8701}{\includegraphics[scale=0.06]{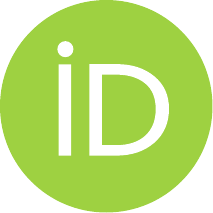}\hspace{1mm}Haoyu~Qin} \\
  \company
  \companyaddress
  \companyemail{qinhaoyu}	
	\And
  Yungang~Chen\thanks{Corresponding Author} \\
  \company
  \companyaddress
  \companyemail{chenyungang}	
	\And
  Qianchuan~Jiang\\
  \company
  \companyaddress
  \companyemail{jiangqianchuan}	
  \And
  Pengchao~Sun\\
  \company
  \companyaddress
  \companyemail{sunpengchao}
  \And
  Xiancai~Ye\\
  \company
  \companyaddress
  \companyemail{yexiancai}
  \And
  Chao~Lin\\
  \company
  \companyaddress
  \companyemail{linchao}
}

\hypersetup{
  pdftitle={\longtitlenostyle},
  pdfsubject={q-bio.NC, q-bio.QM},
  pdfauthor={Haoyu Qin},
}

\begin{document}
\maketitle

\begin{abstract}
Deep Learning based Weather Prediction (DLWP) models have been improving
rapidly over the last few years, surpassing state of the art numerical weather 
forecasts by significant margins. 
While much of the optimization effort is focused on training curriculum 
to extend forecast range in the global context, two aspects remains less explored:
limited area modeling and better backbones for weather forecasting.
We show in this paper that MetMamba, a DLWP model built on a state-of-the-art state-space model, Mamba, 
offers notable performance gains and unique advantages over other popular
backbones using traditional attention mechanisms and neural operators. 
We also demonstrate the feasibility of deep learning based limited area modeling
via coupled training with a global host model.

\end{abstract}


\section{Introduction}
\label{sec:introduction}

Weather forecasting has attracted increasing attention as climate change brought
more heat waves, tropical cyclones, heavy rainfalls and other extreme weather
events, affecting millions of people worldwide \cite{hashim2016climate}
\cite{de2013climate}. Researchers have
devised sophisticated numerical schemes and equations to capture complex weather dynamics
to improve forecast accuracy\cite{kalnay2003atmospheric}. However, its immense
computational complexity often necessitates the use of large scale compute clusters,
incurring notable energy cost and long processing time, making ensemble forecasts, critical
for predicting such events\cite{bellprat2019towards}, expensive or infeasible.

Trained on European Centre for Medium-range Weather Forecast (ECMWF)'s
reanalysis product ECMWF Reanalysis v5 (ERA5) dataset, deep learning based weather prediction (DLWP) 
models have shown promising performance, with FourCastNet\cite{pathak2022fourcastnet} being the
first data-driven model that directly competes with ECMWF's Integrate Forecasting
System (IFS), followed by a series of other
models\cite{lam2022graphcast}\cite{bi2022pangu}\cite{chen2023fuxi}\cite{chen2023fengwu} that
outperforms IFS in various metrics \cite{rasp2023weatherbench}, these models
demonstrates excellent ability in capturing global weather trends, with a
fraction of the computation requirement. 

Though a coarser grid global forecast is crucial for gaining meaningful insights into longer horizon
outlooks, extreme weather events often come in the form of convection,
which are sub-grid dynamics not reflected in coarser grid data,
even with advancements in fine grid global forecast, such
events also requires localized tunning\cite{baldwin2002properties}, making
limited area model (LAM) pivotal in many applications. We conduct this research within 
the LAM context, partly due to its greatly reduced compute, but also aim
to explore the feasibility of deep learning for Limited Area Modeling (LAM).

The Mamba model\cite{gu2023mamba} exhibits remarkable performance with linear 
complexity and unique advantages on contiguous data, its vision variants\cite{liu2024vmambavisualstatespace}
also features an effective receptive field similar to that of convolution networks,
making it a promising alternative to the transformer dominated DLWP models, 
enabling deeper models and finer resolutions to be used, capturing more complex
dynamics.

\section{Related Work}
\label{sec:related-work}

\subsection{Neural Operators}
\label{subsec:fourier-neural-operator-networks}
Neural Operators have shown great efficiency in solving partial differential equations,
effective in a range of synthetic and physics-based tasks\cite{li2020fourier}\cite{xiong2023koopman}. 
It has also shown similar efficiency in vision tasks as a token mixer when used in a Vision 
Transformer (ViT) like architecture\cite{guibas2021adaptive}.
FourCastNet builds on this model and delivers the first DLWP model capable of matching the performance of
leading numerical weather prediction (NWP) system, ECMWF's IFS. Further developments have been made by
adopting spherical harmonics to better address the characteristics of global weather\cite{bonev2023spherical}.

\subsection{Transformers}
Transformers, known for its ability to associate information in various parts of 
the input data via the attention mechanism\cite{vaswani2017attention}, are initially
designed for natural language processing tasks, have quickly gain wide adoption
in computer vision with Vision Transformers\cite{dosovitskiy2020image}. Its quadratic 
complexity \eqref{equ:attention} however, poses as a significant hurdle in its application, particularly in vision. 
Different approaches to alleviate this issue have been proposed, most popular among them is
Swin-Transformer\cite{liu2021swin}, which uses a switching window mechanism to reduce the 
number of required tokens to achieve a global receptive field. Sub-quadratic 
approximations of attention like Linformer\cite{wang2020linformerselfattentionlinearcomplexity}
are also a viable surrogate. In the DLWP space, where the compute efficiency is of a lesser concern,
the use of transformers is ubiquitous\cite{bi2022pangu}\cite{chen2023fuxi}\cite{nguyen2023climax}\cite{chen2023fengwu}\cite{nguyen2023scaling},
favoured by its modeling power. But as noted by \cite{nguyen2023scaling}, 
the patching step which are used to reduce the number of tokens
also leads to degradation in performance. 

\begin{equation}
\label{equ:attention}
\text{Attention}(Q, K, V) = \text{softmax}\left(\frac{QK^T}{\sqrt{d_k}}\right) V
\end{equation}

\subsection{Mamba Model}
\label{subsec:mamba-models}
Mamba\cite{gu2023mamba} model is a class of State Space Models (SSMs), which is originally 
designed to process continuous signals, bearing close similarity in its formulation to RNNs. (Eq.\eqref{eq:ssm-formulation}) 
Its primary contribution, aside from parallelization and hardware level optimization, 
is enabling SSMs to process inputs in a time dependent manner, 
making it a viable alternative to attention 
\cite{dao2024transformersssmsgeneralizedmodels}
Modifying it for vision use cases involves transferring its 1D-Scan pattern to 
a 2d space\cite{liu2024vmambavisualstatespace}\cite{zhu2024vision}, which achieves better performance
than other leading vision transformer models.
\begin{align}
  \label{eq:ssm-formulation}
  h_t &= \overline{A}h_{t-1} + \overline{B}x_t \\
  y_t &= C x_t
\end{align}

\subsection{Limited Area Modeling}
\label{subsec:limited-area-modeling}
The paradigm of using Lateral Boundary Conditions (LBCs) from a global model
in Limited Area Modeling (LAM) is a common practise in running more location
specific limited area models. 

The usage of LBCs, specifically from forecast fields of a global model, has yet
to be tried in the DLWP context, with neural-LAM\cite{oskarsson2023graph} integrating
ground truth field as LBCs. The main objectives of our paper are to  
evaluate a high performance LAM with LBCs from a global model, optimizing the 
coupling of the two models, and quantifying performance gains in a more realistic scenario.

\section{Methodology}
\label{sec:methodology}

\subsection{Dataset}
The dataset we used in this paper is the ECMWF Reanalysis v5
(ERA5)\cite{hersbach2023era5}\cite{hersbach2023era5_pressure} data. 
We collected ERA5 data from the year 2006 through 2022 in 6 hour intervals, 
using the area spanning: \ang{11.25}\textasciitilde\ang{60}N,
\ang{73}\textasciitilde\ang{135.75}E. We use year 2006 to 2020 for training, 
2021 for validation and 2022 for testing.

For the lateral boundary condition, we chose the latest iteration of FourCastNet (SFNO), 
distributed with Earth2Studio\cite{Geneva_NVIDIA_Earth2Studio_2024}.
Among models with wide adoption and benchmarks, it is the only one with a continuous 
6 hour rollout scheme, specific humidity as its vapor related variable and 
sufficient years outside of the training set for LBC generation.
The LBCs are produced by initializing the model with ERA5 global data from 2018 onward,
making a relatively smaller (2018\textasciitilde2021) for LBCs adaptation training,
avoiding the years where the global model was trained on. 
See Appendix~\ref{sec:data-details} for details regarding the dataset used.

\begin{figure}
    \centering
    \includegraphics[width=1.\linewidth]{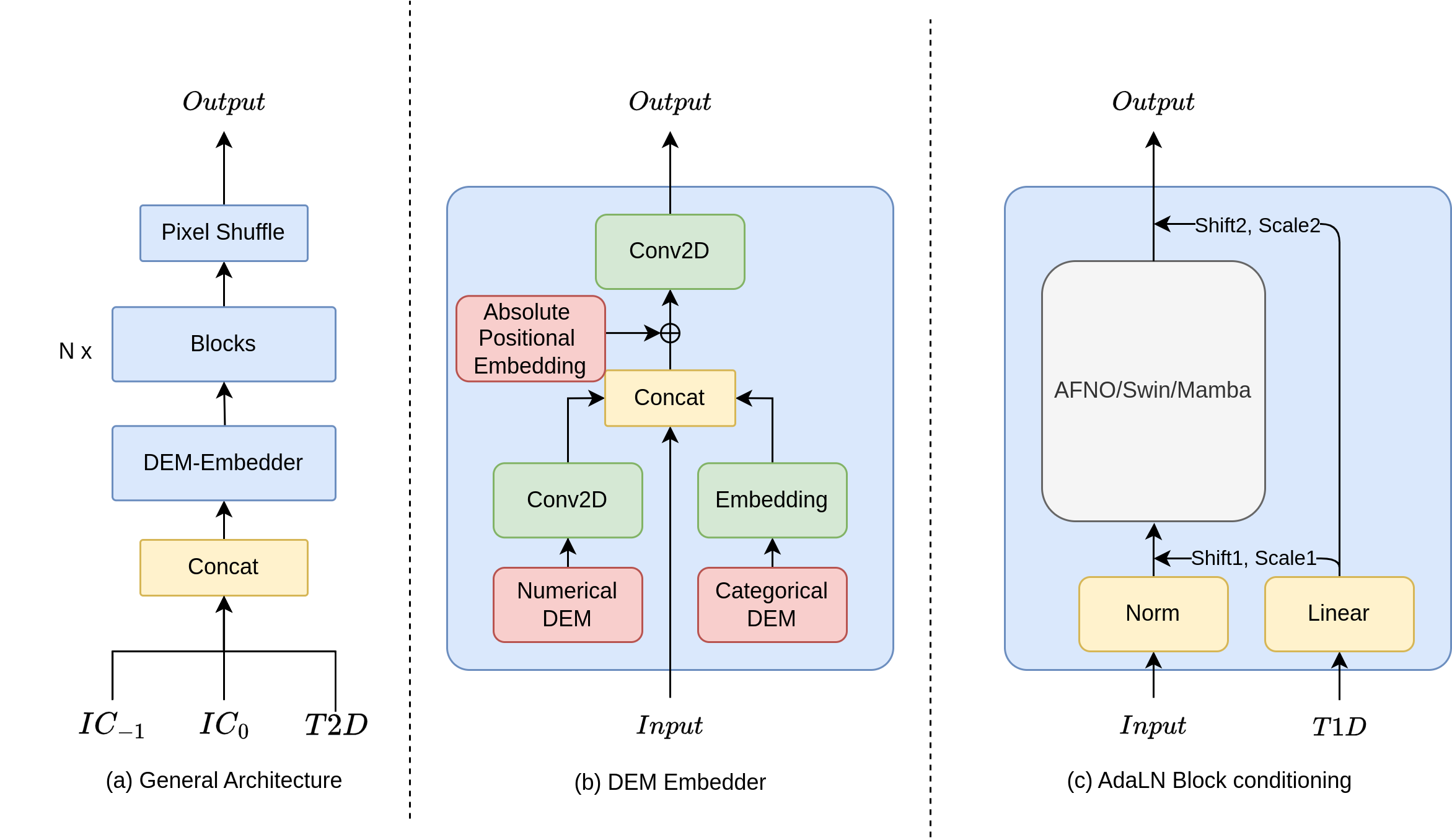}
    \caption{
      (a) general model architecture:
      the DLWP-LAM takes primary inputs: two initial conditions 
      \(IC_{0}\) at initialization time \(t_{0}\) and \(IC_{-1}\) at previous 
      (6 hours before)
      initialization time \(t_{0-6h}\), and auxiliary inputs: constants from
      Digital Elevation Model (DEM) and elapsed time, 
      they are processed by a DEM specific embedder, a conditioning 
      module or by simple concatenation. The input tensor is then processed by 
      \(N\) blocks. The output is decoded with a 
      simple linear layer and pixel shuffle operation.
      (b) DEM Embedder: an embedder that processes different information from the DEM accordingly.
      (c) AdaLN Block: a block that integrates elapsed year time via the use of adaLN, this will
      regulate all blocks in the process step.
}
\label{fig:lam-general-arch}
\vspace{-1em}
\end{figure}

\subsection{Models}

To illustrate the performance advantages of a Mamba based backbone for weather, 
we designed a general network architecture with three different types of block as
their core processing unit: Swin, Swin+AFNO and Mamba. 
The models share similar parameter count and depth. The models all take two 
initial conditions and relevant boundary conditions as input.
For model implementation details, ablation studies of relevant modules, 
see Appendix~\ref{model-details}

\paragraph{General Architecture} We applied the same design to all models, using
a 3D convolution module encoding the two input initial conditions, with Mamba-3D
being the exception as it can natively process spatial temporal data. Each model
features the same Digital Elevation Model(DEM) embedder and AdaLN conditioning
to process auxiliary topographical and temporal information. No down or up sampling 
are used, with model embedding size and latent space resolution staying the same
throughout the main processing blocks, which we have found to be beneficial to 
the general modeling capability. See Figure \ref{fig:lam-general-arch}

\paragraph{Swin Block} This is a Swin-Transformer-V2 block used in 
Swin-Transformer-V2\cite{liu2022swin} and also by FuXi\cite{chen2023fuxi}, 
with no modifications, we refer to their paper for implementation details.
LAM using this block is referred to as MetSwin.

\paragraph{Swin-AFNO Fusion Block} Adaptive Fourier Neural Operator, despite its 
efficiency, has rather limited skill ceiling in our earlier LAM based experiments, 
consistently performing worse than all other types of model.
This is also reflected in benchmarks by several global DLWP papers.
It is also prune to generate defects and even model collapse\cite{bonev2023spherical}.

We design a dual branch model that mixes latent 
information between Fourier Neural Operator and SwinV2 branches using a 
FFT cross attention mechanism \cite{chen2021crossvit}, 
achieving better results than using  either of the single branch models, while 
alleviating the limitation of AFNO in the limited area setting.
LAM using this block is referred to as MetSwinAFNOFusion.

\paragraph{Mamba-3D Block} We adopt the cross-scan and cross-merge methods used
by VMamba\cite{liu2024vmambavisualstatespace}, and extend these methods to the 
3 dimensional space, to better integrate information from spatial and temporal 
dimensions, similarly, a 3D depth-wise convolution module is also used. 
We delineate the design choices of Mamba-3D in Appendix \ref{mamba-3d}
LAM using this block is referred to as MetMamba or MetMamba-3D,
its 2 dimensional counterpart with basic 2D spatial scan pattern is referred to 
as MetMamba-2D.

\paragraph{DEM encoder} We encode different static information including DEM data 
used in the Weather Research and Forecast (WRF) Model, latitude
and longitude coordinates into the model using a specific DEM encoder.
We separate these data based on their datatype, categorical data 
(e.g. soil class, land sea mask) is handled by specific embedding layers, 
and numerical data is processed by simple linear layers.

\paragraph{AdaLN conditioning}Following FiLM\cite{perez2017filmvisualreasoninggeneral} and
Stormer \cite{nguyen2023scaling},  we encode 1 dimensional time data: elapsed year time 
to better regulate model against seasonal variations in weather dynamics.

\begin{figure}
  \centering
  \includegraphics[width=1.\linewidth]{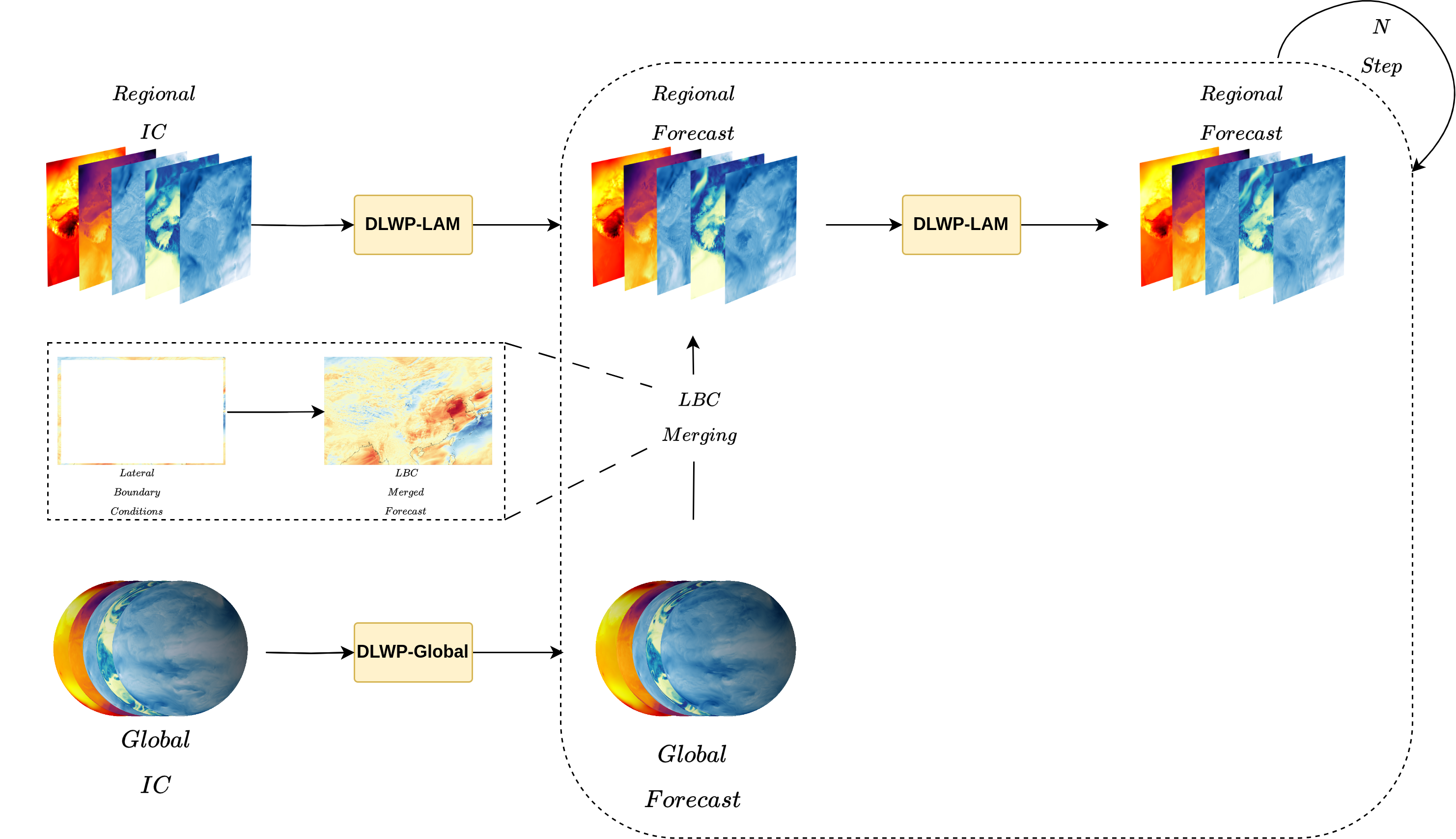}
  \caption{Lateral Boundary Condition merging scheme, used in LBC adaptation training and inference rollout}
  \vspace{-1em}
  \label{fig:lbc_merging_scheme}
\end{figure}

\subsection{Lateral Boundary Condition}
As shown in \reffig{lbc_merging_scheme}, for a forecast rollout of $L$ steps, we run the global model first for $L$ steps. 
The LAM is then executed, where each output result would have its data points on all 
four sides (LBCs) replaced by the data points sampled from the
global model in the same location, with each side having a fixed width of $W$. This is 
set to be 28 grid points (\textasciitilde \SI{780}{\kilo\meter}). 
The effect of the size of $W$ is discussed in Appendix \ref{subsec:effects-of-lateral-boundary-conditions}.

\subsection{Loss Function}
We use Mean Squared Error for the loss function. The loss is calculated by 
summing and averaging all loss values for each weather variable across all pressure levels, 
auto-regressive rollout time steps
on grid points within the area constrained by the Lateral Boundary Condition.

It is weighted generally following the scheme of GraphCast, 
with height-based weights focusing on surface and near ground variables,
and latitude-based weights assigning higher multipliers to grid points with lower latitude.
All weights are fixed in all experiments presented in this paper.
\al[loss_function]{
    \loss = 
    \frac{1}{\rolloutlen}\sum_{t=\inittime + 1}^{\inittime + \rolloutlen} 
    \frac{1}{\setsize{\interiorgrid}} \sum_{v \in \interiorgrid} 
    \sum_{i=1}^{\weatherstatedim}
    \variableweight_i
    \left(\pred^t_{v,i} - \weatherstate^t_{v,i}\right)^2
}
where 
\begin{itemize}
    \item $\inittime$ is the initial time step the forecast starts from,
    \item $\rolloutlen$ is the number of rollout steps used during training,
    \item $\interiorgrid = \set{1, \dots, \ngridpoints} \setminus \boundary$ is the set of grid nodes not on the boundary,
    \item $\variableweight_i$ is a weight associated with the variable $i$. This varies depending on variable category, pressure level and latitude.
\end{itemize}

\subsection{Training Curriculum}

\paragraph{Pre-Training} The model is trained first to predict the weather state of the next time step for 
30 epochs each consisting of 10957 gradient updates with the AdamW optimizer, which follows 
a linear warmup and cosine annealing schedule.

\paragraph{LBC Adaptive Auto-Regressive Training} 

As noted by \cite{davies2014lateral}, the convergence of the LBCs with the LAMs is the more
important factor when producing high fidelity results, than the well-posedness of the LBCs.
We thus introduced these LBCs into the training process.
LBCs sampled from the global model FourCastNet(SFNO) are merged with the predicted 
results of the LAM and used in turn as input to the LAM. This process replaces 
the auto-regressive training step usually found in the global DLWPs. 
We set this auto-regressive step count to be 4. 


\subsection{Evaluation}
We evaluate all LAM models, FourCastNet(SFNO) and relevant ablation studies with initial conditions
from test year 2021. In the total 1439 initial conditions, only 1418 of them are used, with the first 1 and last 20
initial conditions reserved for model initialization and forecast truncation. The models are run and evaluated for 20 
steps. Only Root Mean Squared Error (RMSE) is provided as a performance metric for simplicity. We note that this 
metric is calculated following the convention by WeatherBench\cite{rasp2024weatherbench2benchmarkgeneration}, 
with averaging as the last step of the calculation, latitude weighting scheme is not used.

\subsection{Environment}

The model and revelant training, evaluation and inference apparatus are developed using
PyTorch \cite{Ansel_PyTorch_2_Faster_2024}, xarray
\cite{Hoyer_xarray_N-D_labeled_2017}.

These models are trained on two Nvidia A100-40GB cards utilizing PyTorch's Fully
Sharded Distributed Parallel, gradient checkpointing and Automatic Mixed
Precision in \texttt{bf16} datatype.

\section{Results}
\label{sec:results}

\begin{figure}[htbp]
  \centering
  \includegraphics[width=\textwidth,height=\textheight,keepaspectratio]{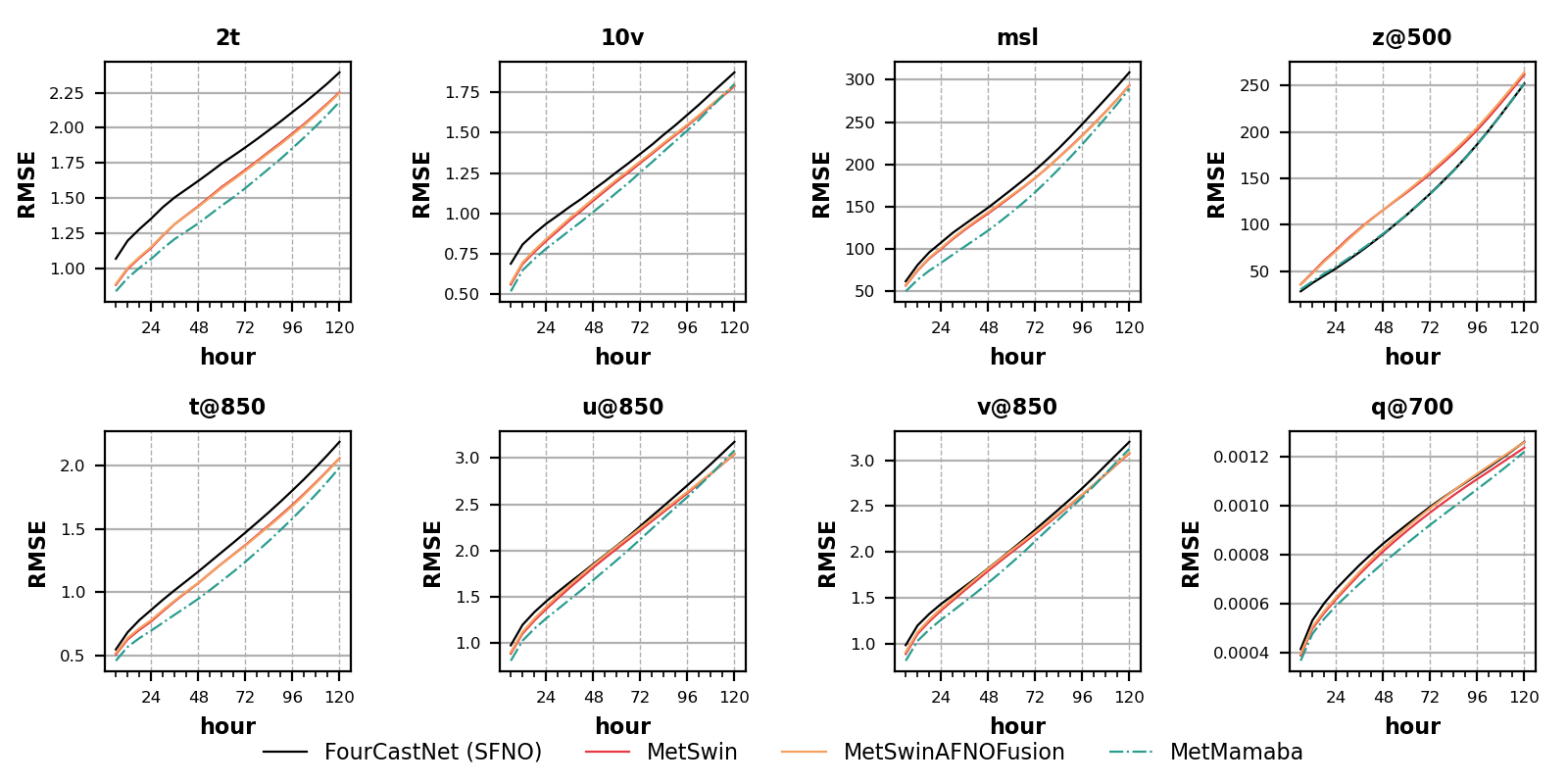}
  \caption{
    \textbf{Root Mean Square Error (RMSE) of headline variables},
    comparing against ground truth ERA5 in 1418 5-day forecasts, 
    evaluated on the test set of year 2022, for 3 types of DLWP-LAMs discussed 
    in this paper. 
    We omit variable $10u$ here for its similarity with the variable $10v$.
  }
  \label{fig:overview_comparison_vs_fcnv3}
\end{figure}

As illustrated in \reffig{overview_comparison_vs_fcnv3}, all of our DLWP-LAMs
achieved better performance using the LBCs from FourCastNet (SFNO),
when compared to FourCastNet (SFNO) itself, along the majority of the 
headline variables, using only around 40 percent (\SI{15}{\year}) of the data which the global model is trained on (\SI{37}{\year}).

The Mamba based model has a notable performance margin over the other two
LAMs, able to match or surpass on all 69 of the predicted variables from its global host model,
while simultaneously resolving more detail than its counterparts. 

We also note that the DLWP-LAMs are prune to generating artifacts at long lead times,
which is a known phenomenon in both constrained DLWP-LAMs and unconstrained global DLWP models,
a manifestation of the inherent flaws in the model architecture, i.e. checkerboard artifacts in
Swin-Transformer based models and complete collapse of AFNO based models.
This is alleviated by more robust auto-regressive training. We note that the Mamba
based model have no noticeable artifacts, hinting at its unique suitability in 
weather forecasting.

For detailed evaluations, forecast examples and spectra analysis, see \refapp{detailed-results}.

\section{Discussion}

\paragraph{Mamba for Weather} The Mamba block, when used as a drop-in replacement
in our existing LAMs offers a notable performance uplift. This can be attributed to its
ability to both generate global receptive field and associate nearby tokens more efficiently
than convolution networks. A reduced direct association between disjointed and distant tokens 
than vision transformers, which we suspect is less common in the field of meteorology, could also
be a contributing factor. Its mathematical formulation being more suited to process continuous 
signals is another valid hypothesis. We also note that the cross scan pattern for VMamba and MetMamba is natively 
transferable and more suitable to a global geometry.

\begin{figure}[htbp]
    \centering
    \includegraphics[width=\textwidth,height=\textheight,keepaspectratio]{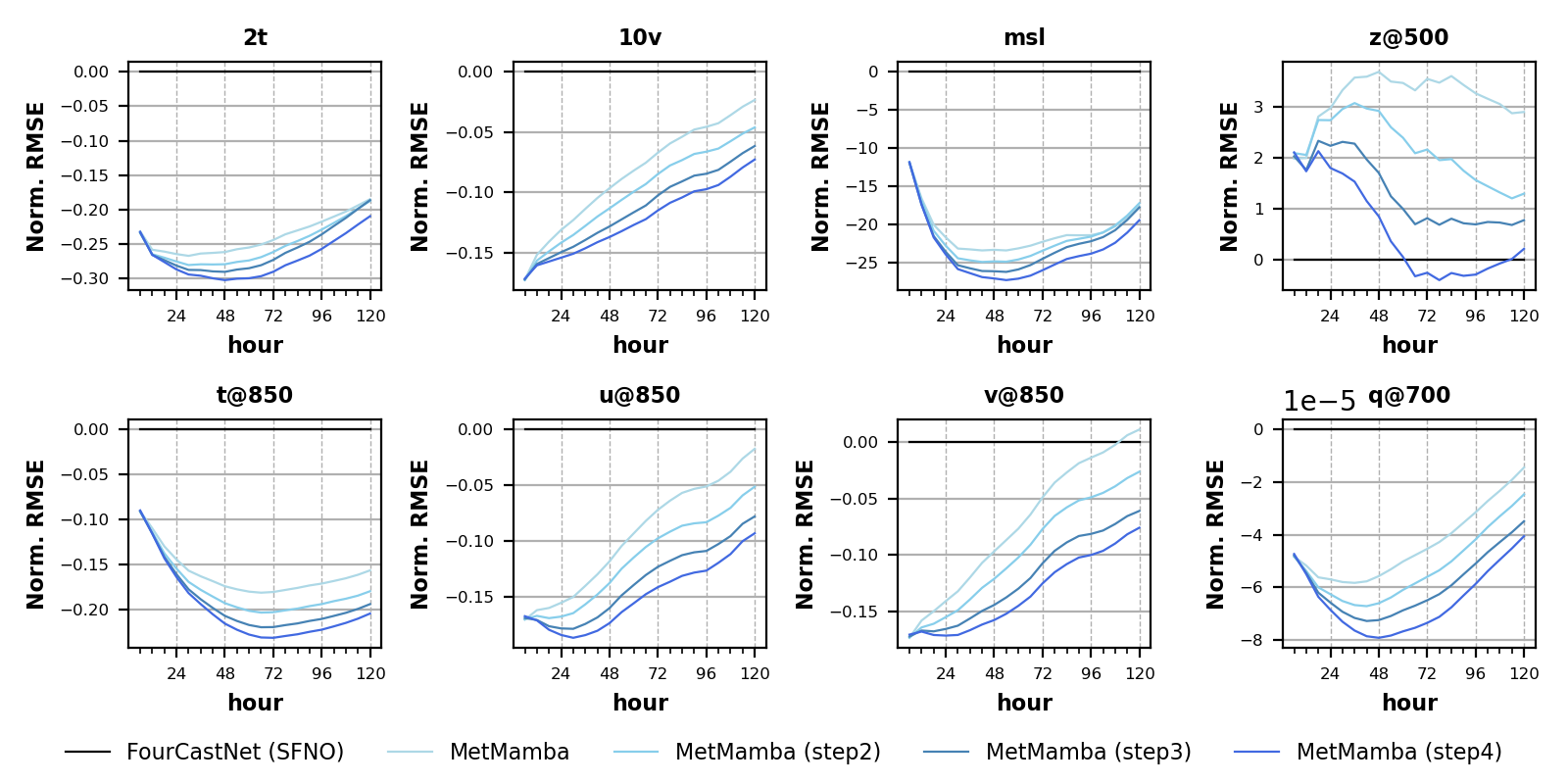}
    \caption{
      \textbf{
      4 steps of auto-regressive training:
      Normalized Root Mean Square Error (Norm. RMSE) of headline variables},
      comparing against FourCastNet (SFNO) in 1418 5-day forecasts, 
      evaluated on the test set of year 2022.
      We omit variable $10u$ here for its
      similarity with the variable $10v$.
      The LAM model shows improvements on long lead times as the 
      LBC-Merging auto-regressive training progresses.}
    \label{fig:effect_of_ar_training}
\end{figure}

\paragraph{Training for a Limited Area Model}
Auto-regressive training is a universally adopted practise in almost all of the 
global DLWP models. Mimicking the model coupling step in numerical LAM systems, 
we train the DLWP-LAMs with LBC data merged from the host global model, 
exposing them to the characteristics of the LBCs, significantly improving 
the compatibility between the two, enhancing the forecast capability of the LAM.
The effects of this adaptive training process is shown in \reffig{effect_of_ar_training}. 

\paragraph{Auxiliary Information Encoding}: 
Previous models have used simple linear layers to process elevation, soil, land-sea mask and other time related information.
We use specialized modules according to process each type of data, See Appendix~\ref{subsec:ablation} for effects of these modules.

\paragraph{Performance Headroom} We note the significant smaller dataset 
used in this paper (\SI{15}{\year}), compared to the regular 
(\textasciitilde \SI{40}{\year}) used to train other DLWP models, and the limited 
hyper-parameters tuning applied, the performance 
of these models could continue to scale with more data and better tuned hyper-parameters. 
Improvements in the host global model can also help extend the forecast horizon of the LAMs.

\begin{ack}
The ECMWF Reanalysis v5 (ERA5)~\cite{hersbach2023era5}~\cite{hersbach2023era5_pressure}
dataset is used in this paper to train various models, it is retrieved via Copernicus Climate Data Store.

We thank European Centre for Medium-Range Weather Forecasts (ECMWF) and its C3S service for their efforts
in collecting, archiving, and disseminating the data, without which studies like this
would not have been possible.

\end{ack}

\bibliographystyle{unsrtnat}
\bibliography{main}  
\newpage
\appendix
\section{Data Details}
\label{sec:data-details}

\begin{table}[b]
\centering
\small
\addtolength{\tabcolsep}{-5.pt}
\begin{tabular}{c|c|c|c|c}
\toprule 
\textbf{Type} & \textbf{Variable Name} & \textbf{Description} & \textbf{ECMWF Parameter ID} & \textbf{Role} \\
\midrule
Atmospheric & \texttt{z@---} & Geopotential (at pressure level \texttt{---}) & 129 & Input/Ouptut \\
Atmospheric & \texttt{t@---} & Temperature (at pressure level \texttt{---}) & 130 & Input/Ouptut \\
Atmospheric & \texttt{u@---} & \texttt{u} component wind (at pressure level \texttt{---}) & 131 & Input/Ouptut \\
Atmospheric & \texttt{v@---} & \texttt{v} component wind (at pressure level \texttt{---}) & 132 & Input/Ouptut \\
Atmospheric & \texttt{q@---} & Specific humidity (at pressure level \texttt{---}) & 133 & Input/Ouptut \\
\hline
Surface & \texttt{msl} & Mean sea level pressure & 151 & Input/Ouptut \\
Surface & \texttt{10u} & 10 metre \texttt{u} wind component & 165 & Input/Ouptut \\
Surface & \texttt{10v} & 10 metre \texttt{v} wind component & 166 & Input/Ouptut \\
Surface & \texttt{2t} & 2 metre temperature & 167 & Input/Ouptut \\
\hline
Static & \texttt{Longitude} & Sine normalized & n/a & Input \\
Static & \texttt{Longitude} & Cosine normalized & n/a & Input \\
Static & \texttt{Latitude} & Sine normalized & n/a & Input \\
Static & \texttt{LANDMASK} & Land sea mask & n/a & Input \\
Static & \texttt{LU\_INDEX} & Land use index & n/a & Input \\
Static & \texttt{SNOALB} & Maximum snow albedo (MODIS) & n/a & Input \\
Static & \texttt{SANDFRAC} & Sand content in soil & n/a & Input \\
Static & \texttt{HGT\_M} & Topography height (GMTED2010) & n/a & Input \\
Static & \texttt{SCB\_DOM} & Dominant soil class & n/a & Input \\
Static & \texttt{SCT\_DOM} & Dominant surface cover type & n/a & Input \\
\hline
Clock & \texttt{Local Time of Day} & Sine normalized & n/a & Input \\
Clock & \texttt{Local Time of Day} & Cosine normalized & n/a & Input \\
Clock & \texttt{Elapsed Year Progress} & Sine normalized (1D) & n/a & Input \\
Clock & \texttt{Elapsed Year Progress} & Cosine normalized (1D) & n/a & Input \\
Clock & \texttt{Forecast lead time} & Fixed value of 6 (z-score normalized) & n/a & Input \\
\bottomrule
\end{tabular}
\normalsize
\vspace{1em}
\caption{
  Pressure level variables, surface variables and auxiliary used and predicted by the model and their notation style, 
}
\label{tab:data-details}
\end{table}

In this appendix we give more details on the dataset and other auxiliary 
features used in our experiments. 

\paragraph{Data collection, notation and processing} We collected ERA5\cite{hersbach2023era5}\cite{hersbach2023era5_pressure} 
data with ECMWF's Copernicus Climate Change Service (cdsapi), targeting 
the area spanning: \ang{11.25}\textasciitilde\ang{60}N,
\ang{73}\textasciitilde\ang{135.75}E,
atmospheric variables: geopotential \texttt{z@---}, temperature \texttt{t@---}, 
$u$-component wind \texttt{u@---}, $v$-component wind \texttt{v@---} specific humidity \texttt{q@---},
10 metre $u$ wind component \texttt{10u}, 10 metre $v$ wind component \texttt{10v}, mean sea level pressure \texttt{msl},
2 metre temperature \texttt{2t}. We calculate the mean and standard deviation for all variables throughout the 
training subset, and perform z-score normalization for all of the variables prior to model input. 
For all variables and notations, see {Table~\ref{tab:data-details}}.

\paragraph{Topographical Data} Digital Elevation Model data is sampled from
the Weather Research and Forecast (WRF) model, containing land-sea mask, soil 
type, snow albedo etc. Positional information is added by calculating the sine 
and cosine normalized latitude and longitude. (See Eq.\eqref{equ:coordinate-cos-norm}\ and Eq.\eqref{equ:coordinate-sin-norm})
Time of day is added as a group of 2 dimensional sine and cosine normalized local time.
Time of year is added as a group of 1 dimensional sine and cosine normalized 
current UTC time (in seconds) divided by total seconds per year.

\paragraph{Lateral Boundary Conditions} The choice of Lateral Boundary Conditions, 
is subject several constraints. To be eligible for LBC adaptive training, the 
global DLWP model has to:
\begin{inparaenum}[1)]
    \item share the same variables with the LAM; 
    \item leave sufficient years out of the training set, such that a realistic LBCs can be reproduced;
    \item use a consistent autoregressive rollout strategy;
\end{inparaenum}
As such we have chosen FourCastNet(SFNO) in this paper as it satisfies all 
criteria.

\label{equ:coordinate-cos-norm}
\begin{equation}
C = \frac{\cos\left(C \times \frac{\pi}{180}\right) - \frac{1}{N} \sum_{i=1}^{N} \cos\left(C_i \times \frac{\pi}{180}\right)}{\sqrt{\frac{1}{N} \sum_{i=1}^{N} \left(\cos\left(C_i \times \frac{\pi}{180}\right) - \frac{1}{N} \sum_{j=1}^{N} \cos\left(C_j \times \frac{\pi}{180}\right)\right)^2}}
\end{equation}

\begin{equation}
\label{equ:coordinate-sin-norm}
C = \frac{\sin\left(C \times \frac{\pi}{180}\right) - \frac{1}{N} \sum_{i=1}^{N} \sin\left(C_i \times \frac{\pi}{180}\right)}{\sqrt{\frac{1}{N} \sum_{i=1}^{N} \left(\sin\left(C_i \times \frac{\pi}{180}\right) - \frac{1}{N} \sum_{j=1}^{N} \sin\left(C_j \times \frac{\pi}{180}\right)\right)^2}}
\end{equation}

\section{Model Details}
\label{model-details}
We delineate the design and implementation details of the our \textit{MetMamba} model in this section.
The main difference to two other types of DLWP-LAMs is \textit{MetMamba}'s ability to process spatial-temporal
inputs natively, without the need for an additional 3D convolution encoder. 
Aside from sharing the same structure, auxiliary information encoder and decoder,
all DLWP-LAMs share the same embedding size and similar parameter count.
The SwinV2 based LAM serves as a baseline, the AFNO-Swin based LAM serves as a incremental update with
relatively higher efficiency and performance. See {Table~\ref{tab:model_detail}}. 

For the detailed design of the SwinV2 block, we refer to the Swin-Transformer-V2 paper \cite{liu2022swin}.
For AFNO-Swin Fusion block, we follow the dual branch design from Cross-ViT \cite{chen2021crossvit}, with one AFNO block
for each of the four model stages, the branches are mixed together with FFT truncation operations.

\subsection{Model Design}
\label{model-design}
\begin{figure}
\centering
\includegraphics[width=1.\linewidth]{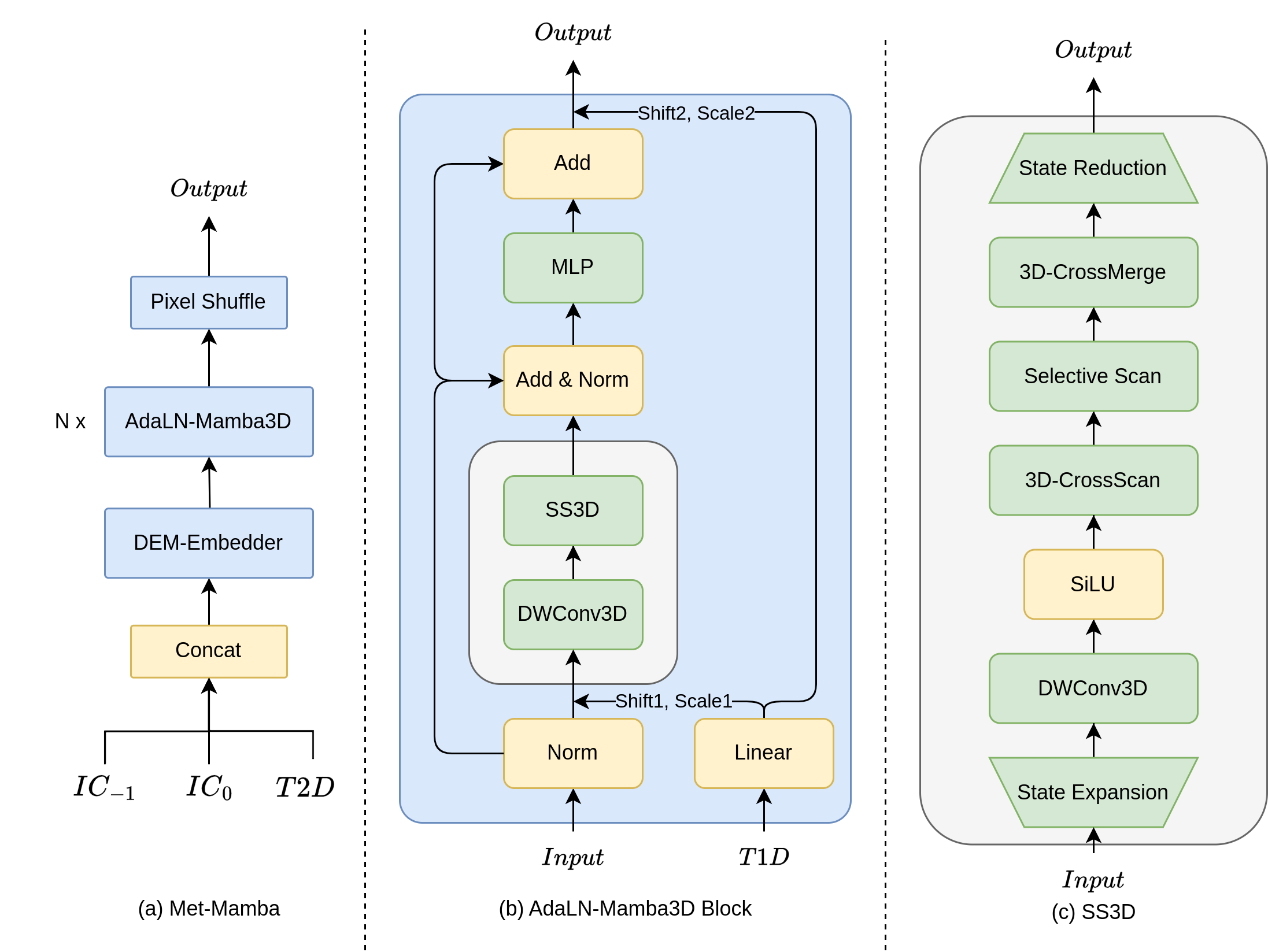}
\caption{\textbf{Met-Mamba Architecture} 
(a) Met-Mamba model architecture, 
the model takes primary inputs: two initial conditions \(IC_{0}\) at initialization time \(t_{0}\) and \(IC_{-1}\) 
at previous (6 hours before) initialization time \(t_{0}-6h\), and auxiliary inputs: DEM masks, elapsed time etc, 
processed by a specific embedder, linear layer, conditioning module or by simple concatenation.
The spatial-temporal tensor is than processed by \(N\) Mamba3D blocks, conditioned by elapsed time (year).
The output is decoded with a simple linear layer and pixel shuffle operation.
(b) AdaLN-Mamba3D Block, a block that utilizes depth-wise 3D convolution and mamba's 
selective scan to achieve token (spatial, temporal) mixing and channel mixing. 
The block integrates seasonal variation via the use of adaLN.
(c) SS3D operator, a module that flattens and rearranges input spatial-temporal 
tensor to achieve different memory layouts (scan  routes) for Mamba's selective 
scan to associate information on different directions.
}
\label{fig:model_architecture}
\vspace{-1em}
\end{figure}

\begin{table}[t]
\centering
\small
\addtolength{\tabcolsep}{-5.pt}
\begin{tabular}{c|c|c|c|c|c}
\toprule 
\textbf{model} & \textbf{embed size} & \textbf{patch size} & \textbf{depth} & \textbf{\# of params} & \textbf{samples/s} \\
\midrule
MetSwin & 192 & (1, 1) & 22 & 19.4M & 6.02 \\
MetAFNOSwinFusion & 192 & (1, 1) & 16+4 & 18.1M & 5.93 \\
MetMamba-2D & 192 & (1, 1) & 22 & 21.8M & 3.10 \\
MetMamba-3D & 192 & (1, 1) & 22 & 20.0M & 1.72 \\

\bottomrule
\end{tabular}
\normalsize
\vspace{1em}
\caption{
  DLWP-LAM Models used in this paper, note that the parameter regression of \textit{MetMamba-3D} 
  is due to its reduced embedding size of $(H, W)$, rather than the 
  $(C, H, W)$ used for other models.
  Static and auxiliary data are used only as input (static and time variant)
}
\label{tab:model_detail}
\end{table}

\paragraph{Scan routes}

To flatten a spatial-temporal block of $(T, H, W)$ for Mamba-3D to scan, there exists 6 permutations of this block and 6 corresponding traversal routes:

\begin{enumerate}[(1)]
    \item \texttt{Horizontal-Vertical-Temporal} ($(T, H, W)$) : $K$ row first 2D scans;
    \item \texttt{Vertical-Horizontal-Temporal} ($(T, W, H)$) : $K$ column first 2D scans;
    \item \texttt{Horizontal-Temporal-Vertical} ($(H, T, W)$) : Row first cross temporal scan;
    \item \texttt{Vertical-Temporal-Horizontal} ($(W, T, H)$) : Column first cross temporal scan;
    \item \texttt{Temporal-Horizontal-Vertical} ($(H, W, T)$) : Temporal first psuedo 2D row scan;
    \item \texttt{Temporal-Vertical-Horizontal} ($(W, H, T)$) : Temporal first psuedo 2D column scan;
\end{enumerate} 

where $K$ is the number of initial conditions, each route has a respective reverse route, making a total of 12 routes.
These routes are designed to enable the model to build a effective receptive field (ERF) that can associate information with nearby patches 
across both spatial and temporal dimensions. We have observed large performance discrepancies when using different groupings of these routes,
with the best combination being routes \texttt{Horizontal-Vertical-Temporal} and \texttt{Temporal-Vertical-Horizontal}. Note that while this grouping
is the best route grouping based on limited experiments, heuristics and speculative convergence behaviour, a better grouping may still exist.

\paragraph{Usage of SS3D}

There are several ways we can apply this SS3D block, we had limited experiments with the following 3:
\begin{enumerate}[(1)]
    \item Use as a drop-in encoder for encoding two input initial conditions;
    \item Use in backbone, to process spatial-temporal relations for inputs with two initial conditions;
    \item Use in backbone, to process spatial relations for inputs with one initial condition;
\end{enumerate} 
And we have found that, the best performance is yielded by model using a full spatial-temporal 3D backbone, we suspect this
scan pattern and implementation is insufficient to exploit the spatial relationship but good enough to spatial temporal 
information.

\subsection{Ablation Study}
\label{subsec:ablation}
We replaced modules in the encoding and backbone stages of the model to observe their effect on forecasting capabilities.

\subsubsection{DEM embedder}
\begin{figure}[htbp]
    \centering
    \includegraphics[width=\textwidth,height=\textheight,keepaspectratio]{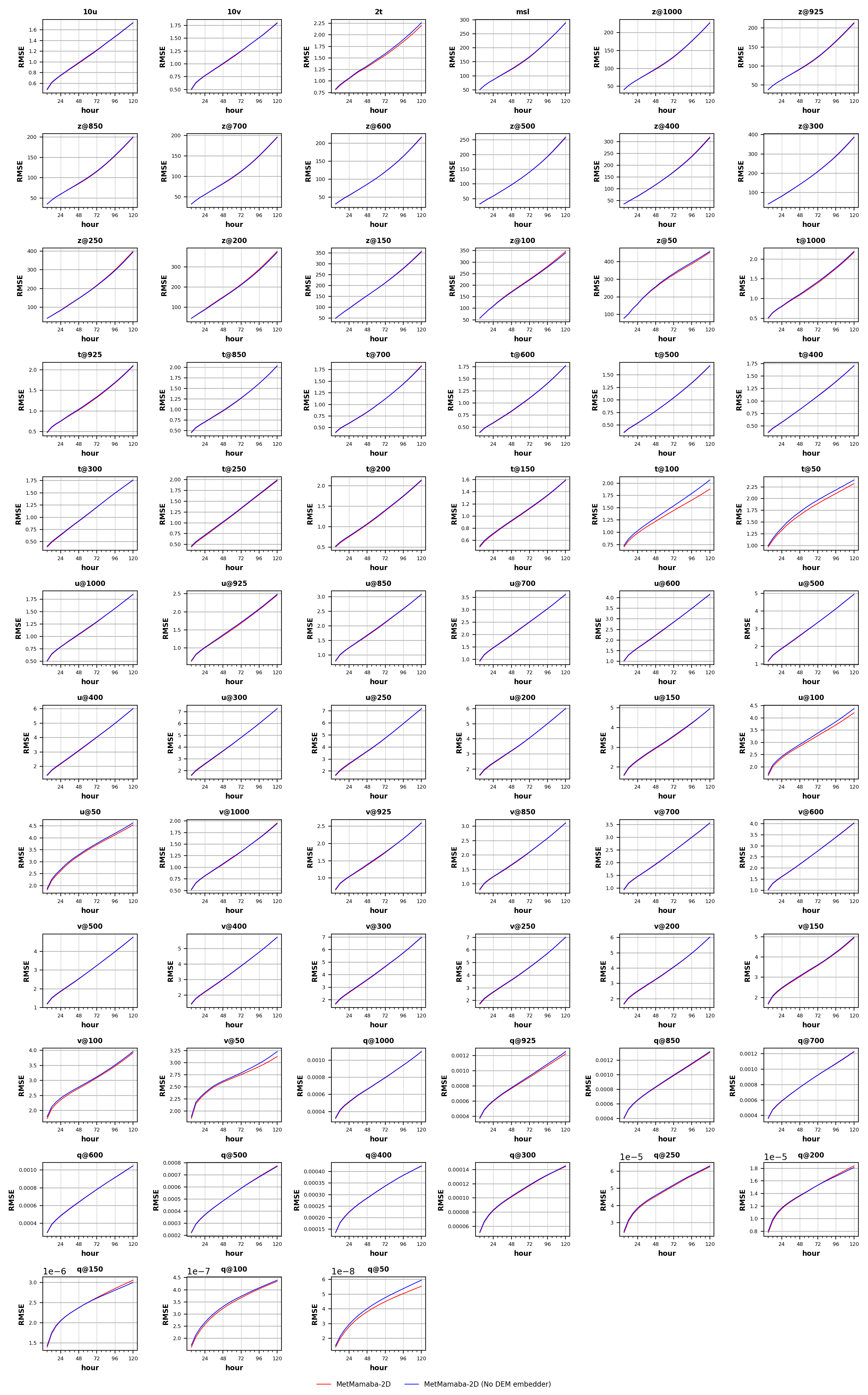}
    \caption{\textbf{Root Mean Squared Error (RMSE)}, MetMamba-2D when comparing 
      against the same model but with the DEM embedder replaced with regular
      linear layers
    \label{fig:dem_abaltion}
}
\end{figure}

We replace the DEM embedder with a all-linear module, slight degradation of performance on \texttt{2t} and variables higher than 
\texttt{150 hPa} is observed, see Figure~\ref{fig:dem_abaltion}

\subsubsection{AdaLN Conditioning}
\begin{figure}[t]
    \centering
    \includegraphics[width=\textwidth,height=\textheight,keepaspectratio]{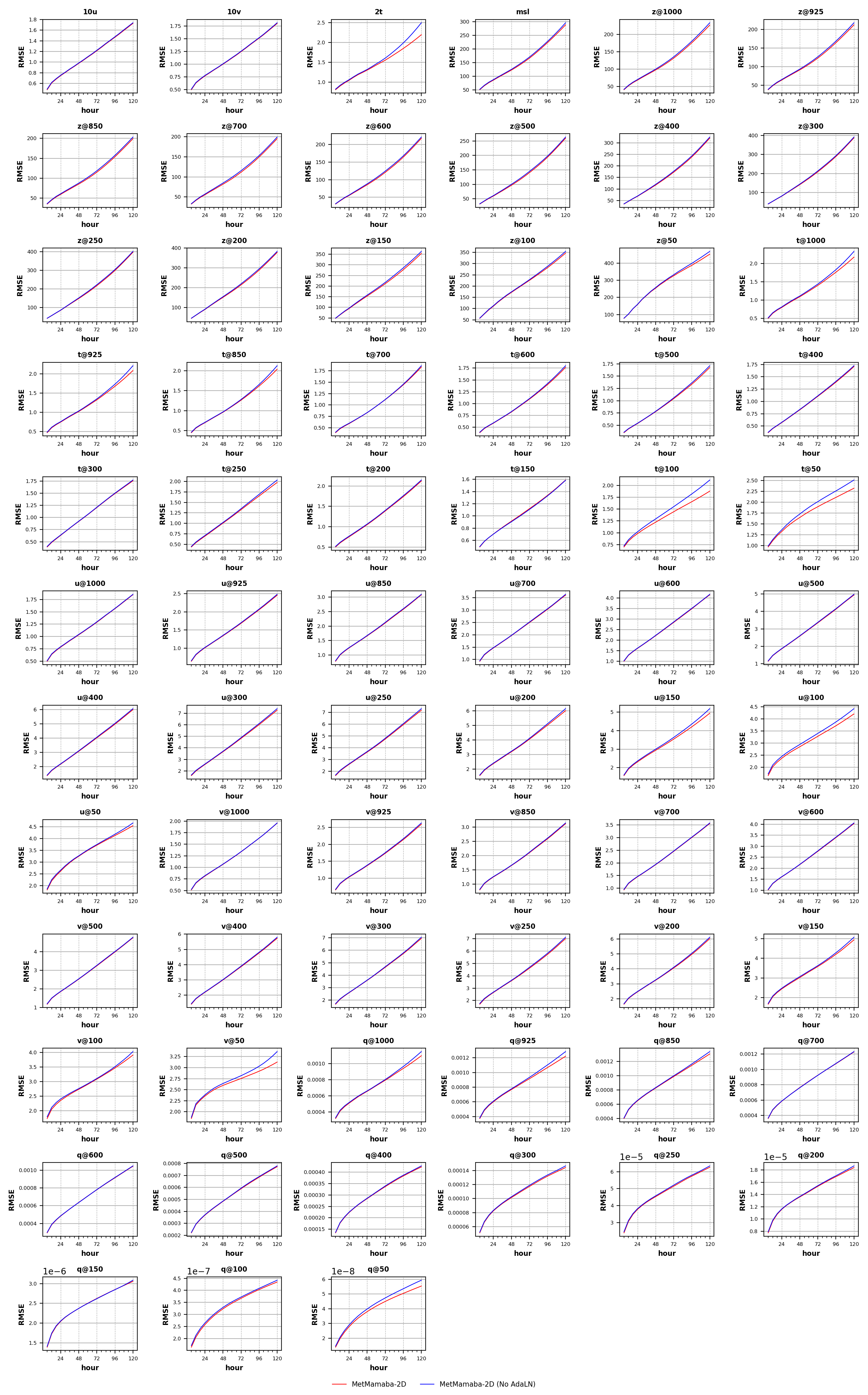}
    \caption{\textbf{Root Mean Squared Error (RMSE)}, MetMamba-2D when comparing 
      against the same model but with the elapsed year time information 
      supplied as a expanded channel, rather than as through adaLN
    }
    \label{fig:adaln_abaltion}
\end{figure}

We removed the adaLN module from the model and supply the elapsed year time related information via expanded 2 dimensional tensors. 
Performance regression can be observed over all variables, see Figure~\ref{fig:adaln_abaltion}

\subsubsection{Mamba3D}
\label{mamba-3d}
\begin{figure}[t]
    \centering
    \includegraphics[width=\textwidth,height=\textheight,keepaspectratio]{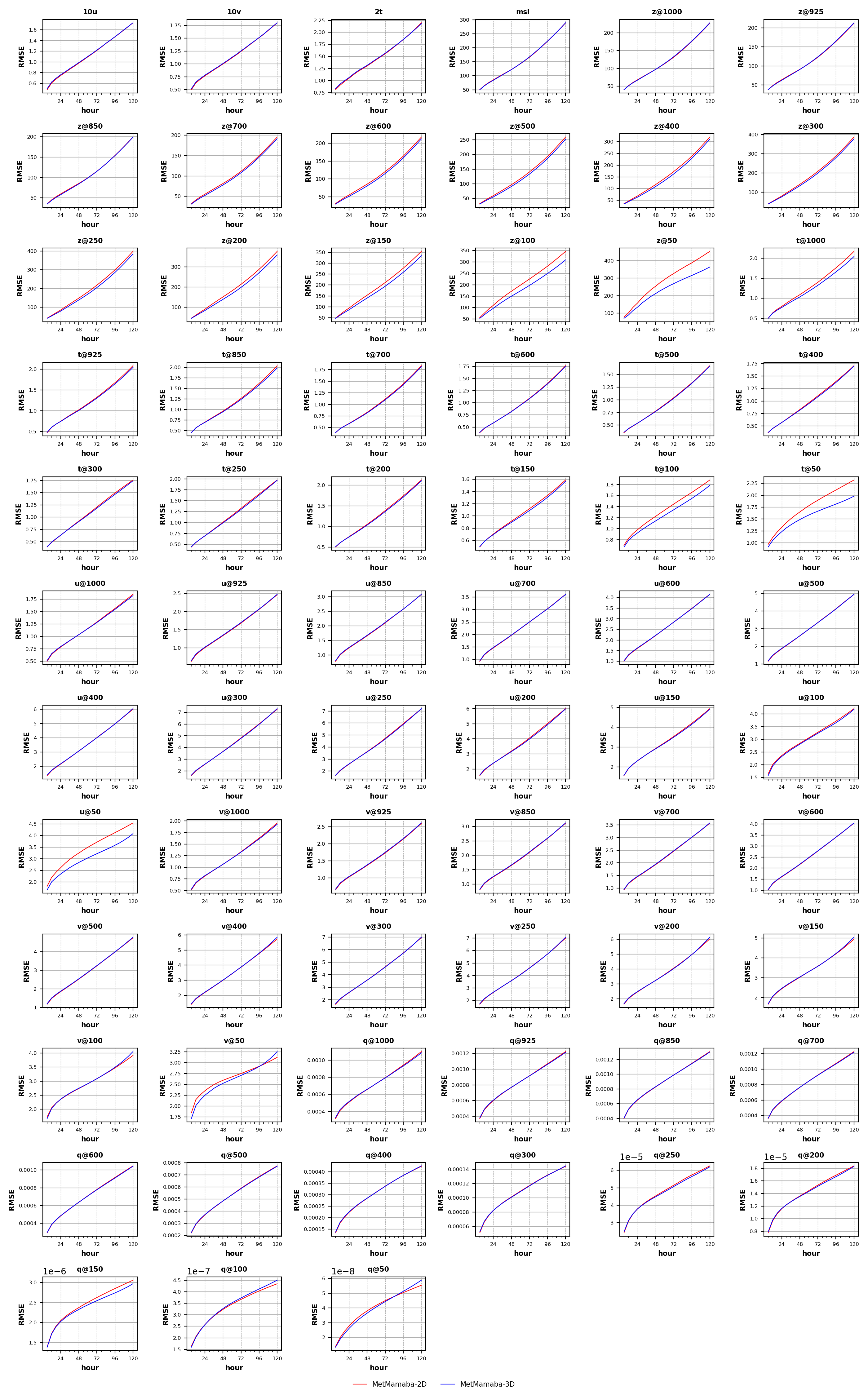}
    \caption{\textbf{Root Mean Squared Error (RMSE)}, MetMamba (2D) when comparing 
      against the MetMamba (3D) with the same overarching architecture 
    }
    \label{fig:3d_ablation}
\end{figure}

We replace spatial-temporal Mamba3D blocks with vanilla Mamba2D, which comes natively with VMamba, 
and observed a slight performance degradation across all variables, with geopotential \texttt{z}
and variables above \texttt{150 hPa} seeing the most regression, see Figure~\ref{fig:3d_ablation}

\section{Detailed Results}
\label{sec:detailed-results}
In this appendix showcase additional results from the \textit{MetMamba} model, 
when compared against FourCastNet(SFNO) and ERA5.

\subsection{RMSE}
\label{subsec:rmse}
\begin{figure}[htbp]
    \centering
    \includegraphics[width=\textwidth,height=\textheight,keepaspectratio]{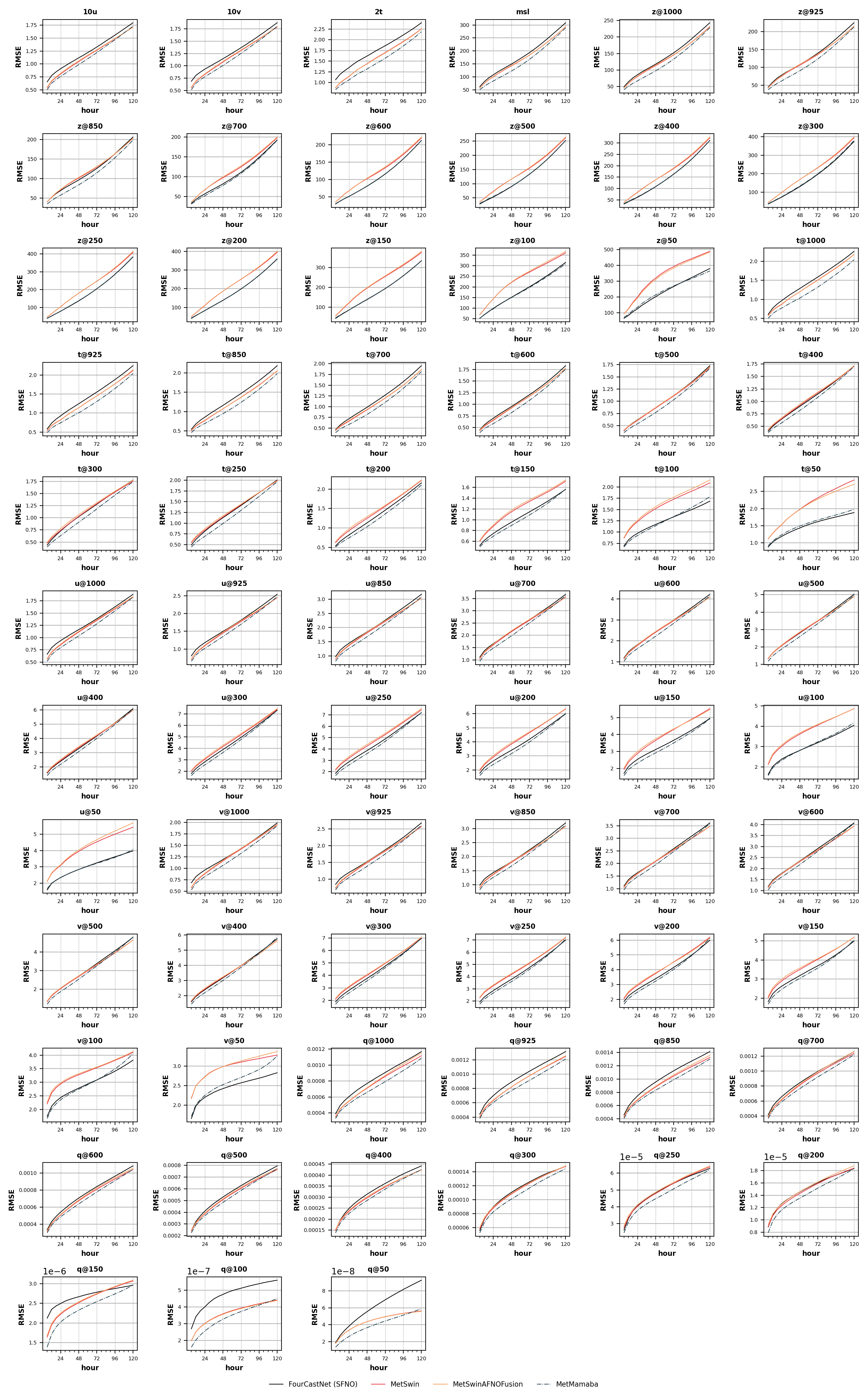}
    \caption{\textbf{Root Mean Squared Error (RMSE)}, 3 types of LAM comparing against their global host model FourCastNet (SFNO)
      \label{fig:all_model_line}
}
\end{figure}

The DLWP-LAMs are able to surpass or match the performance of host 
FourCastNet (SFNO) with significantly less training data, with \textit{MetMamba}
being the only one  beats SFNO with notable margins on almost all variables with
a few exceptions of \texttt{v@50} and \texttt{v@100}, see Figure~\ref{fig:all_model_line}

\subsection{Example Forecasts}
\begin{figure}[htbp]
    \centering
    \includegraphics[width=\textwidth,height=\textheight,keepaspectratio]{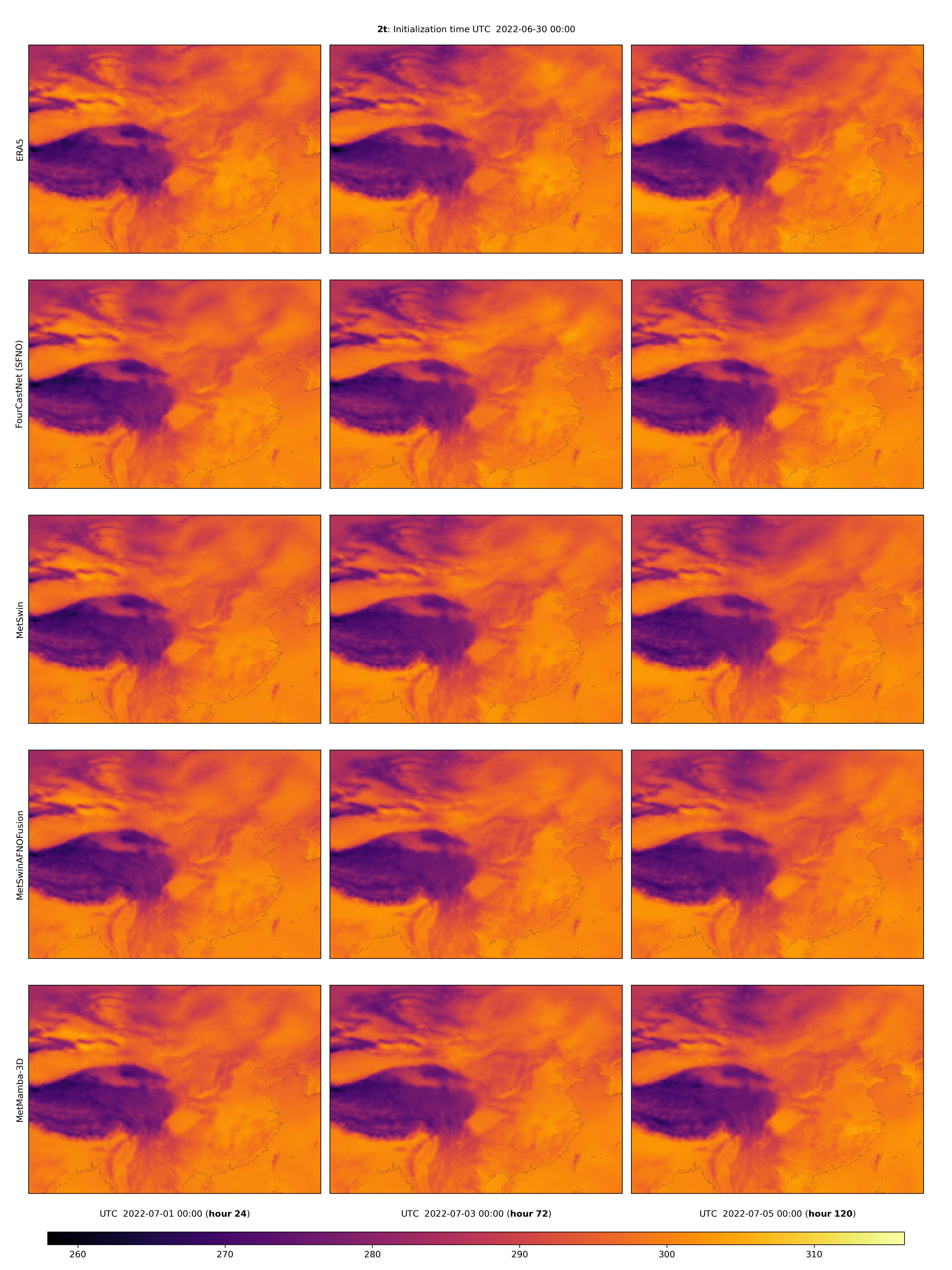}
    \caption{\textbf{Example Forecasts: 2t}, 3 types of LAM comparing against their global host model FourCastNet (SFNO)
      initialized at \texttt{2022-6-30 00:00:00 UTC}, showing forecast at 24, 72 and 120 hour lead times
    \label{fig:fcst_sample_2t}
  }
\end{figure}

\begin{figure}[htbp]
    \centering
    \includegraphics[width=\textwidth,height=\textheight,keepaspectratio]{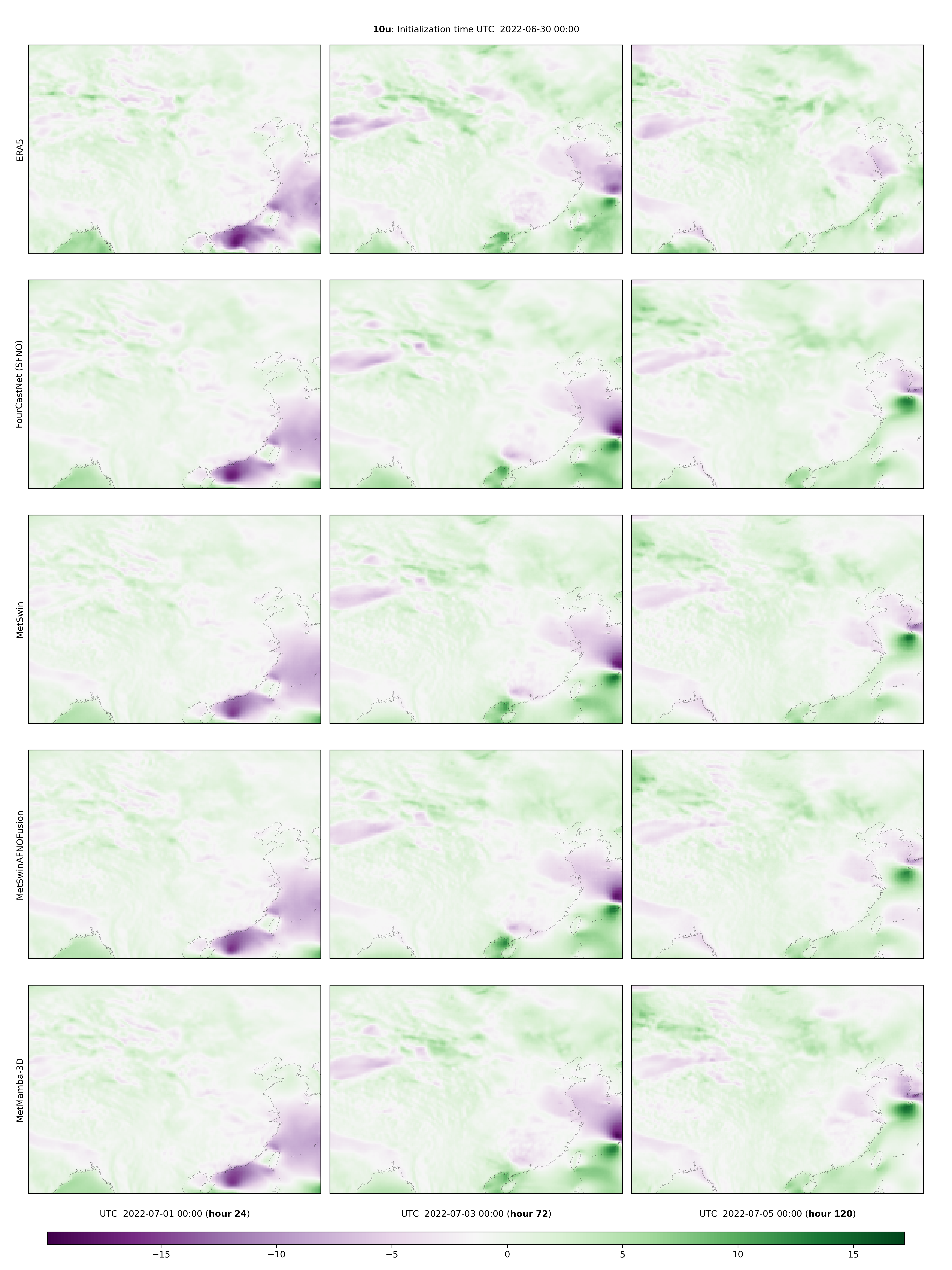}
    \caption{\textbf{Example Forecasts: 10u}, 3 types of LAM comparing against their global host model FourCastNet (SFNO)
      initialized at \texttt{2022-6-30 00:00:00 UTC}, showing forecast at 24, 72 and 120 hour lead times
  }
  \label{fig:fcst_sample_10u}
\end{figure}

\begin{figure}[htbp]
    \centering
    \includegraphics[width=\textwidth,height=\textheight,keepaspectratio]{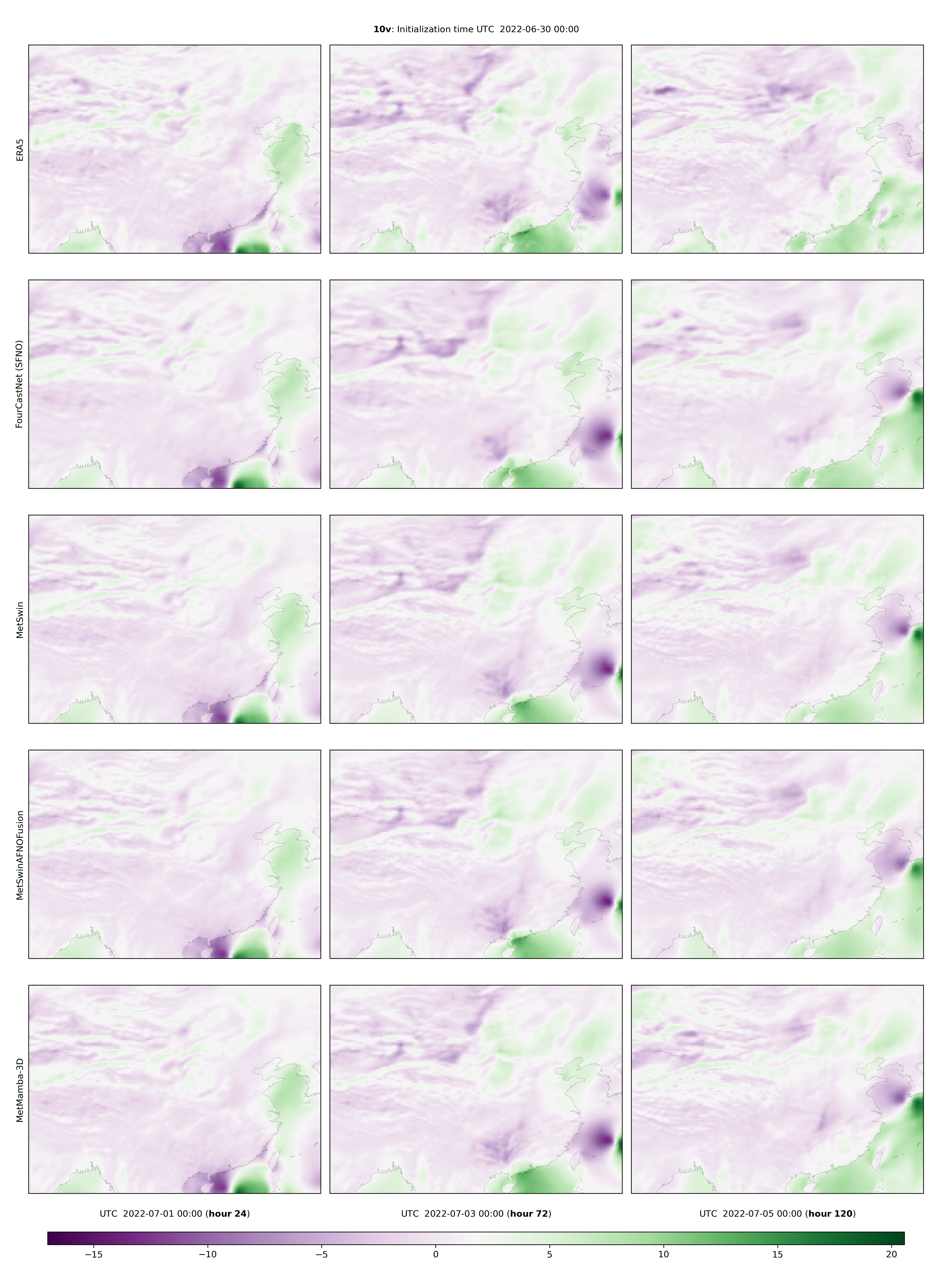}
    \caption{\textbf{Example Forecasts: 10v}, 3 types of LAM comparing against their global host model FourCastNet (SFNO)
      initialized at \texttt{2022-6-30 00:00:00 UTC}, showing forecast at 24, 72 and 120 hour lead times
  }
    \label{fig:fcst_sample_10v}
\end{figure}

\begin{figure}[htbp]
    \centering
    \includegraphics[width=\textwidth,height=\textheight,keepaspectratio]{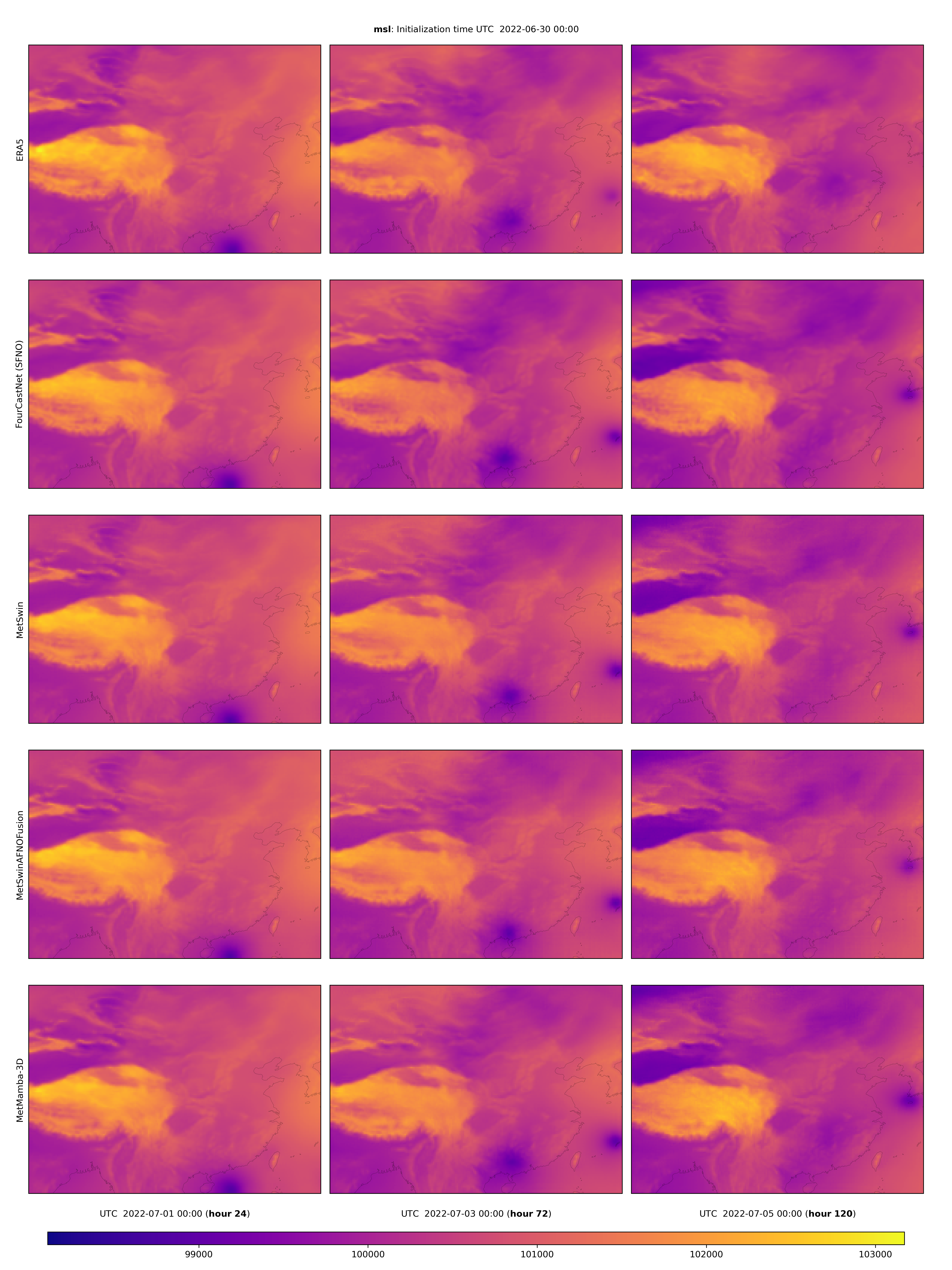}
    \caption{\textbf{Example Forecasts: msl}, 3 types of LAM comparing against their global host model FourCastNet (SFNO)
      initialized at \texttt{2022-6-30 00:00:00 UTC}, showing forecast at 24, 72 and 120 hour lead times
  }
    \label{fig:fcst_sample_msl}
\end{figure}

\begin{figure}[htbp]
    \centering
    \includegraphics[width=\textwidth,height=\textheight,keepaspectratio]{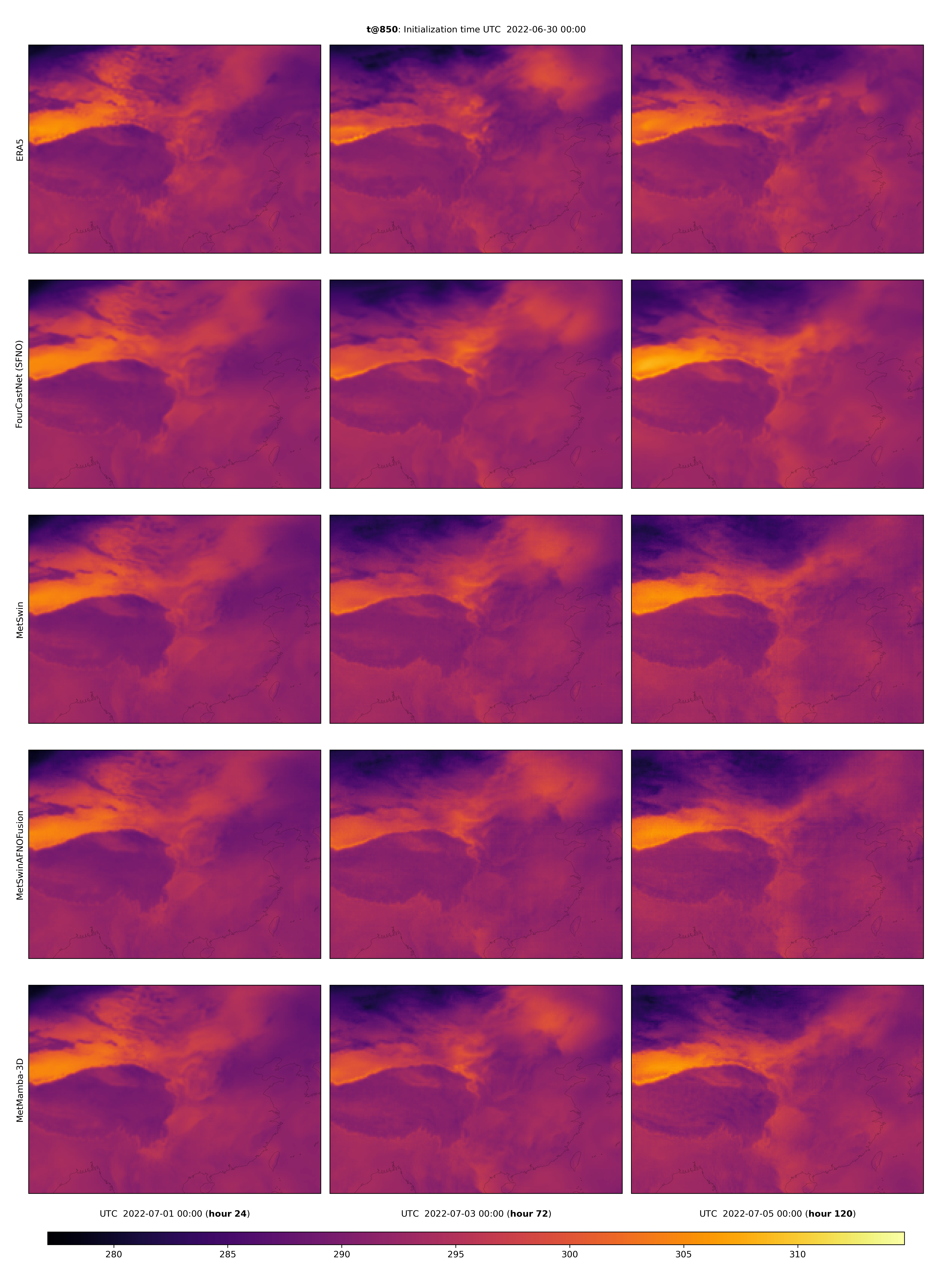}
    \caption{\textbf{Example Forecasts: t@850}, 3 types of LAM comparing against their global host model FourCastNet (SFNO)
      initialized at \texttt{2022-6-30 00:00:00 UTC}, showing forecast at 24, 72 and 120 hour lead times
  }
  \label{fig:fcst_sample_t850}
\end{figure}

\begin{figure}[htbp]
    \centering
    \includegraphics[width=\textwidth,height=\textheight,keepaspectratio]{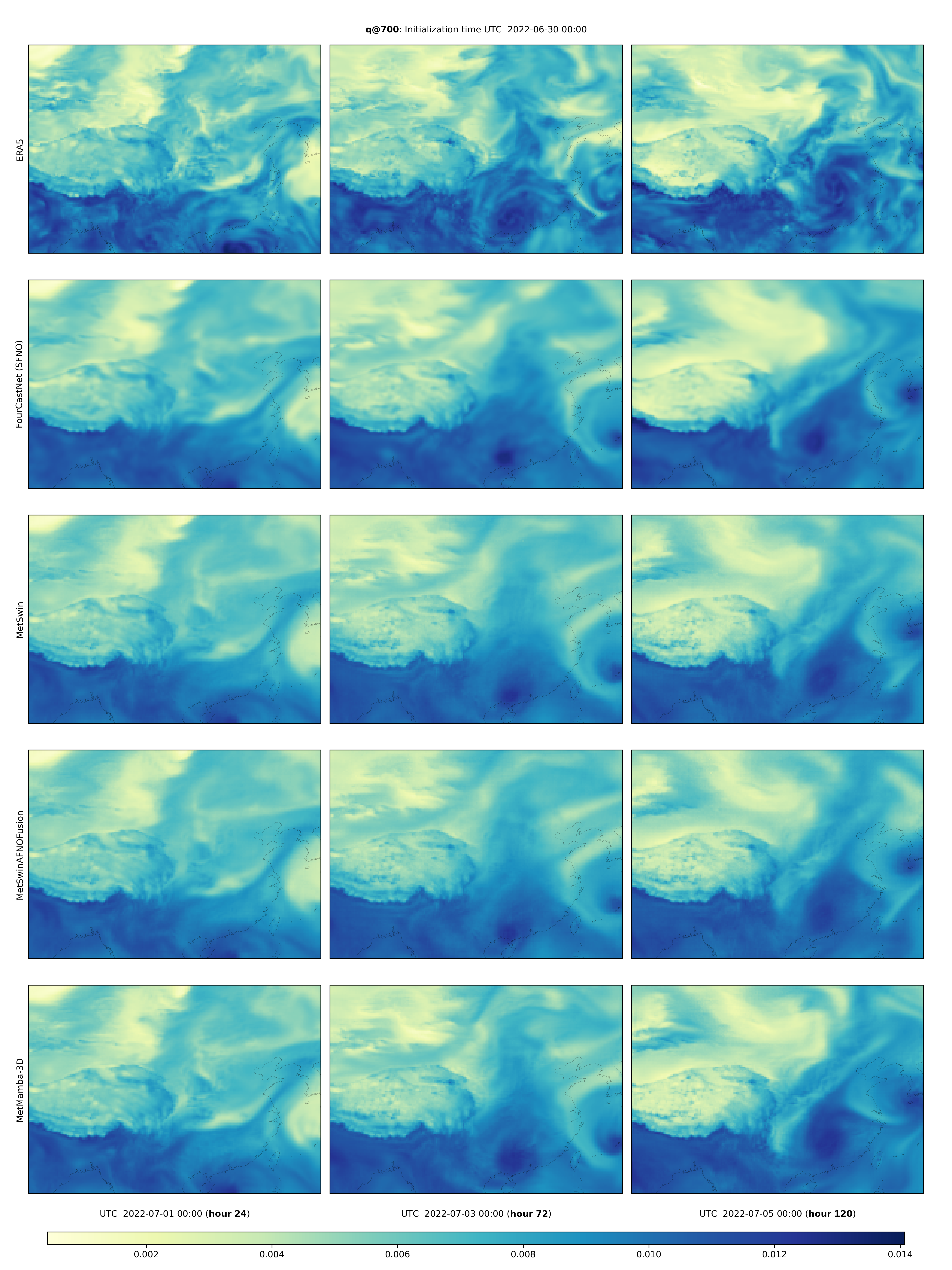}
    \caption{\textbf{Example Forecasts: q@700}, 3 types of LAM comparing against their global host model FourCastNet (SFNO)
      initialized at \texttt{2022-6-30 00:00:00 UTC}, showing forecast at 24, 72 and 120 hour lead times
  }
    \label{fig:fcst_sample_q700}
\end{figure}

\begin{figure}[htbp]
    \centering
    \includegraphics[width=\textwidth,height=\textheight,keepaspectratio]{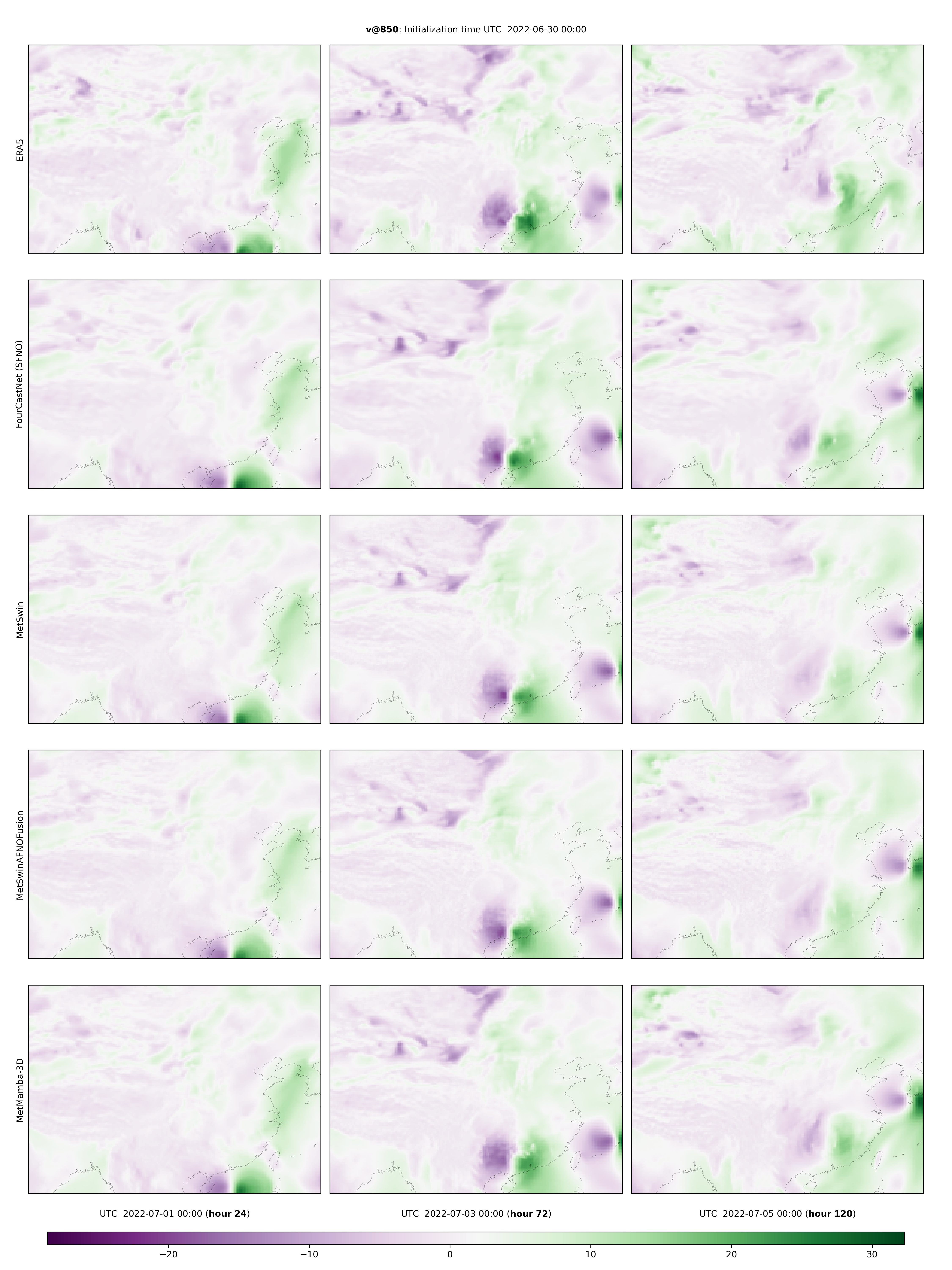}
    \caption{\textbf{Example Forecasts: v@850}, 3 types of LAM comparing against their global host model FourCastNet (SFNO)
      initialized at \texttt{2022-6-30 00:00:00 UTC}, showing forecast at 24, 72 and 120 hour lead times
  }
    \label{fig:fcst_sample_v850}
\end{figure}

\begin{figure}[htbp]
    \centering
    \includegraphics[width=\textwidth,height=\textheight,keepaspectratio]{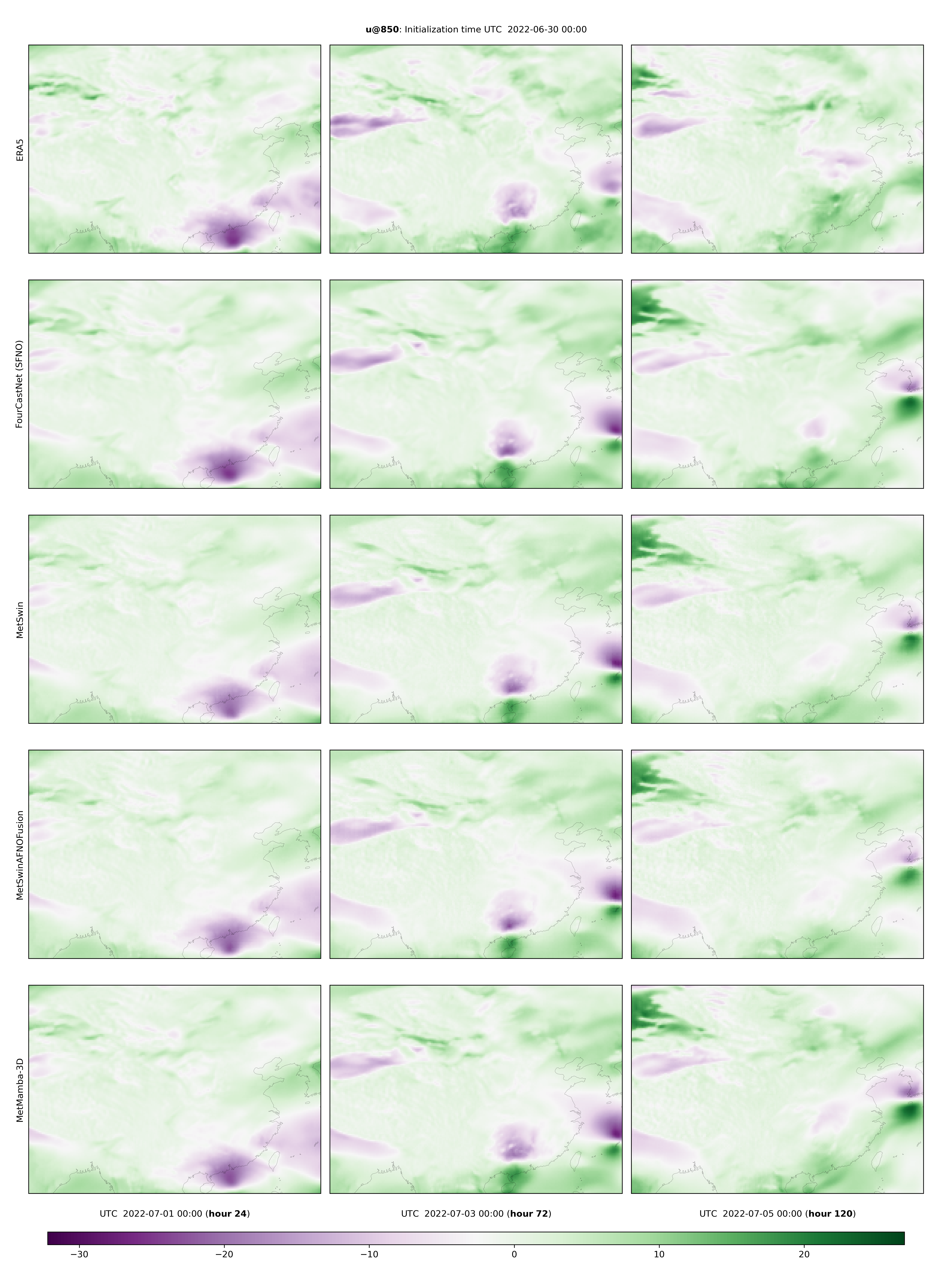}
    \caption{\textbf{Example Forecasts: u@850}, 3 types of LAM comparing against their global host model FourCastNet (SFNO)
      initialized at \texttt{2022-6-30 00:00:00 UTC}, showing forecast at 24, 72 and 120 hour lead times
  }
    \label{fig:fcst_sample_u850}
\end{figure}

Here we showcase forecast results at lead time 24, 72 and 120 hours, 
initialized with ERA5 data at \texttt{2022-6-30 00:00:00 UTC}. See Figure~\ref{fig:fcst_sample_2t}--\ref{fig:fcst_sample_u850}
\label{subsec:example_forecasts}

\subsection{Spatial Distribution of Error}
\label{subsec:spatial-distribution-of-error}
Here we showcase the (averaged on test year 2022) spatial distribution of error (RMSE) of MetMamba versus 
FourCastNet(SFNO) across all headline variables. Both DLWP models are evaluated against ERA5. 

\begin{figure}[tbp]
     \centering
     \normalspatialsubfig{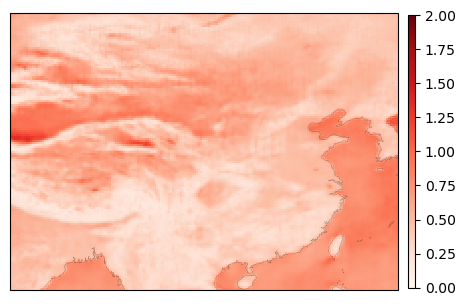}{MetMamba}{24}
     \normalspatialsubfig{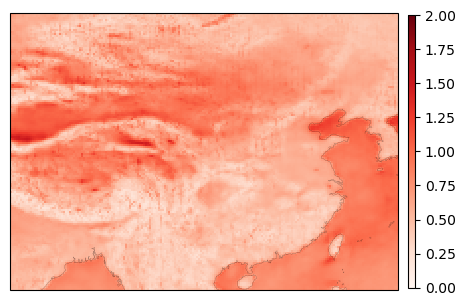}{SFNO}{24}
     \normalspatialsubfig{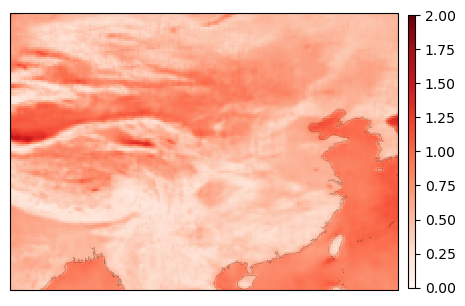}{MetMamba}{36}
     \normalspatialsubfig{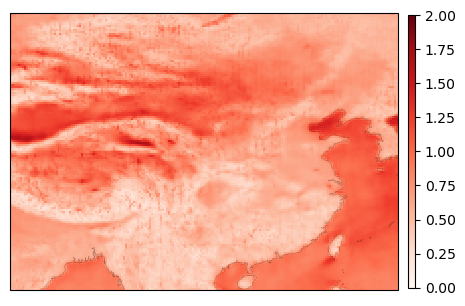}{SFNO}{36}
     \normalspatialsubfig{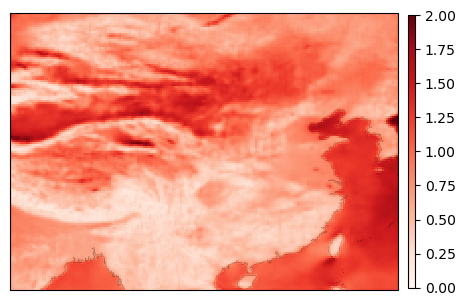}{MetMamba}{72}
     \normalspatialsubfig{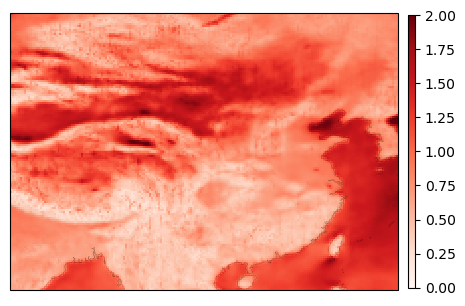}{SFNO}{72}
     \normalspatialsubfig{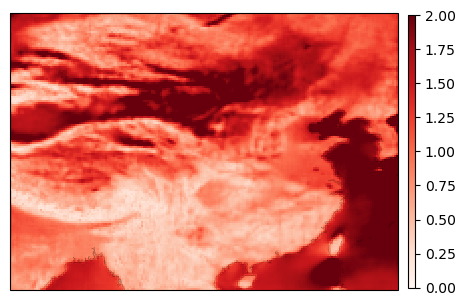}{MetMamba}{120}
     \normalspatialsubfig{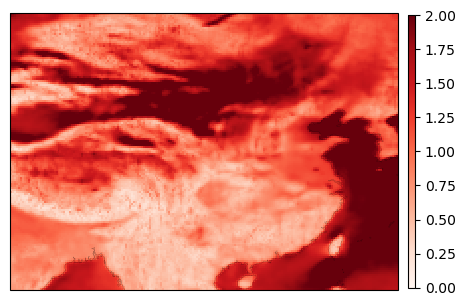}{SFNO}{120}
    \caption{
       \textbf{Root Mean Squared Error (RMSE) of FourCastNet (SFNO) and MetMamba}
       of variable \texttt{10u},
       when both compared against ERA5.
       Darker red indicates worse performance observed at this grid point.
     } 
    \label{fig:spatial_errors_comp_10u}
\end{figure}

\begin{figure}[tbp]
     \centering
     \normalspatialsubfig{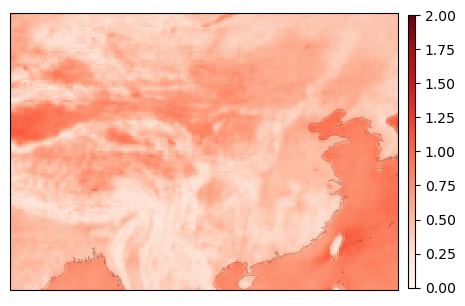}{MetMamba}{24}
     \normalspatialsubfig{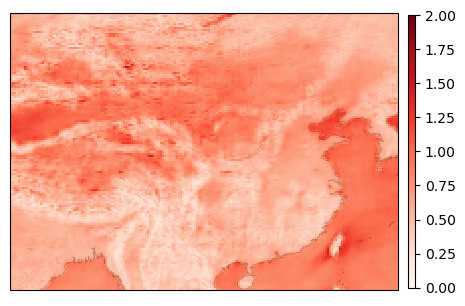}{SFNO}{24}
     \normalspatialsubfig{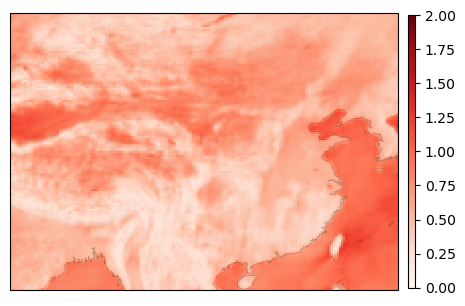}{MetMamba}{36}
     \normalspatialsubfig{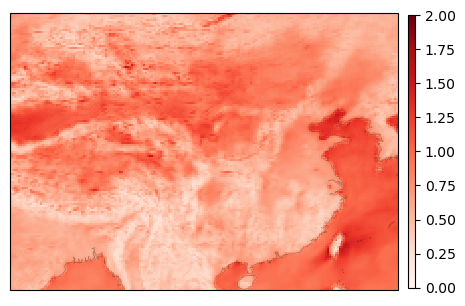}{SFNO}{36}
     \normalspatialsubfig{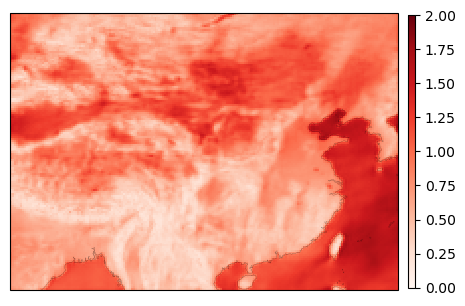}{MetMamba}{72}
     \normalspatialsubfig{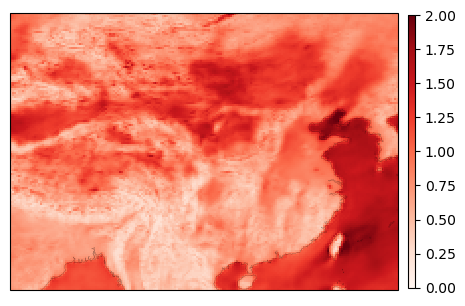}{SFNO}{72}
     \normalspatialsubfig{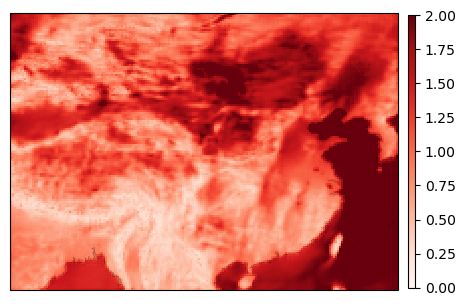}{MetMamba}{120}
     \normalspatialsubfig{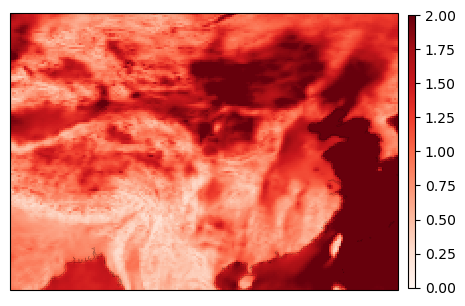}{SFNO}{120}
    \caption{
       \textbf{Root Mean Squared Error (RMSE) of FourCastNet (SFNO) and MetMamba}
       of variable \texttt{10v},
       when both compared against ERA5.
       Darker red indicates worse performance observed at this grid point.
     } 
    \label{fig:spatial_errors_comp_10v}
\end{figure}

\begin{figure}[tbp]
     \centering
     \normalspatialsubfig{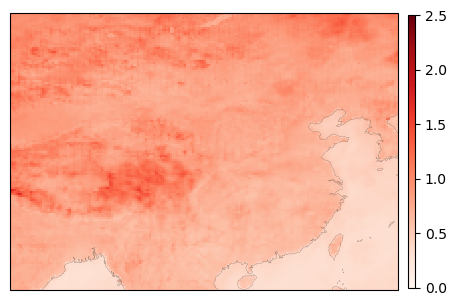}{MetMamba}{24}
     \normalspatialsubfig{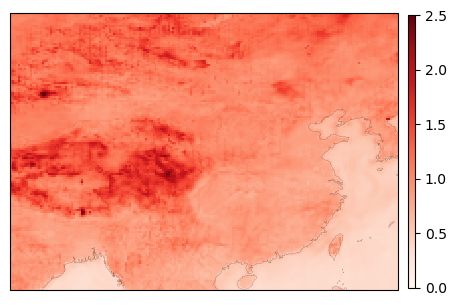}{SFNO}{24}
     \normalspatialsubfig{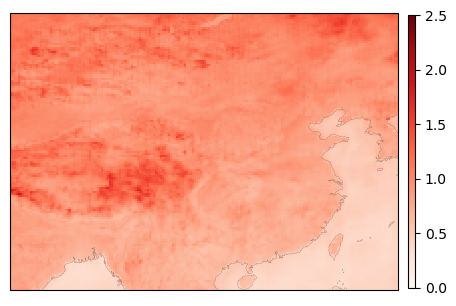}{MetMamba}{36}
     \normalspatialsubfig{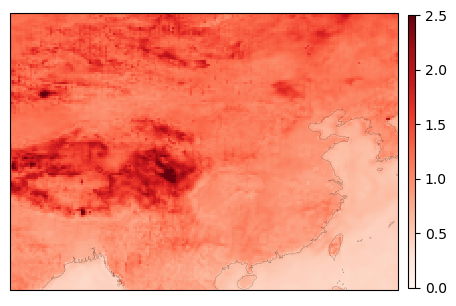}{SFNO}{36}
     \normalspatialsubfig{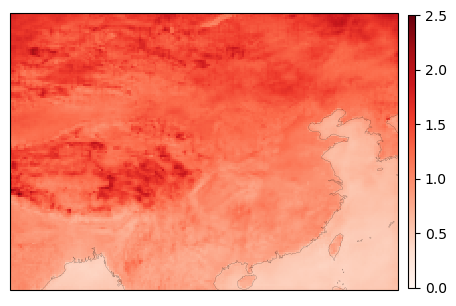}{MetMamba}{72}
     \normalspatialsubfig{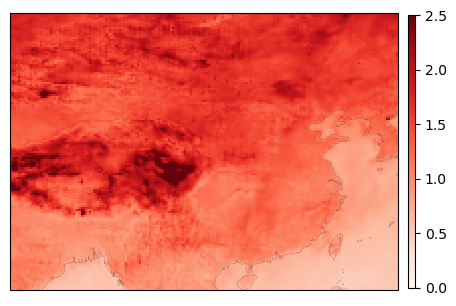}{SFNO}{72}
     \normalspatialsubfig{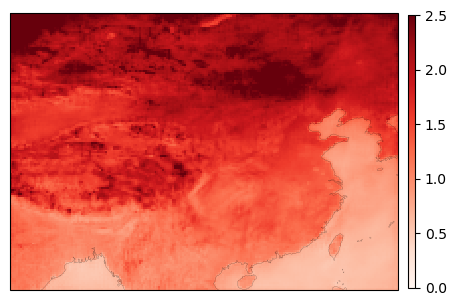}{MetMamba}{120}
     \normalspatialsubfig{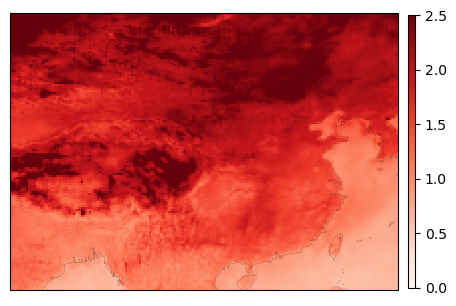}{SFNO}{120}
    \caption{
       \textbf{Root Mean Squared Error (RMSE) of FourCastNet (SFNO) and MetMamba}
       of variable \texttt{2t},
       when both compared against ERA5.
       Darker red indicates worse performance observed at this grid point.
     } 
    \label{fig:spatial_errors_comp_2t}
\end{figure}

\begin{figure}[tbp]
     \centering
     \normalspatialsubfig{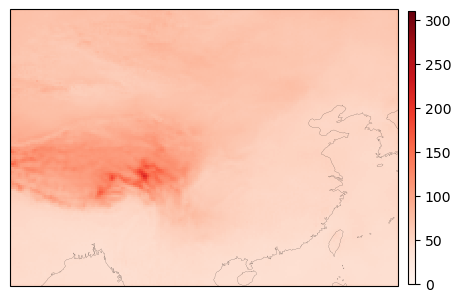}{MetMamba}{24}
     \normalspatialsubfig{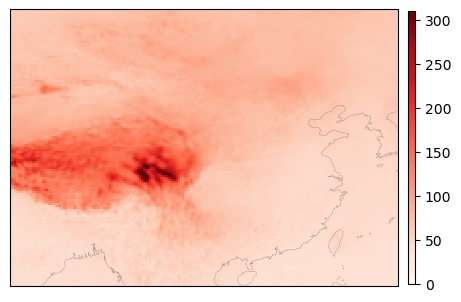}{SFNO}{24}
     \normalspatialsubfig{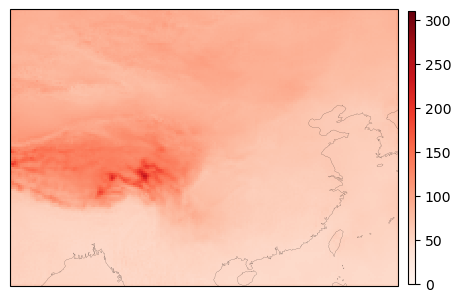}{MetMamba}{36}
     \normalspatialsubfig{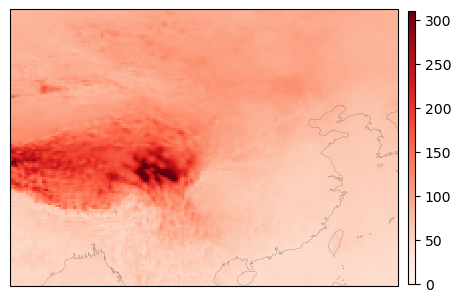}{SFNO}{36}
     \normalspatialsubfig{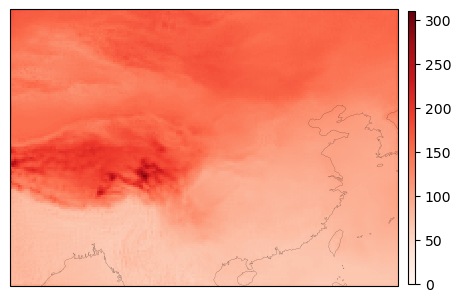}{MetMamba}{72}
     \normalspatialsubfig{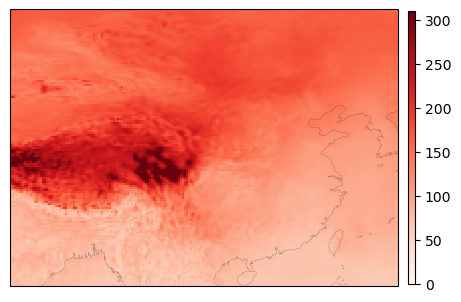}{SFNO}{72}
     \normalspatialsubfig{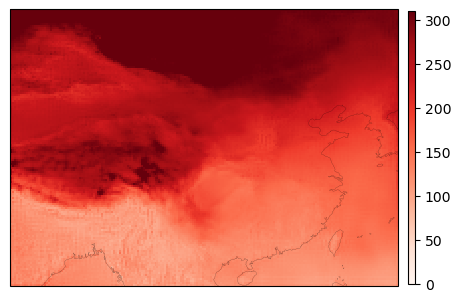}{MetMamba}{120}
     \normalspatialsubfig{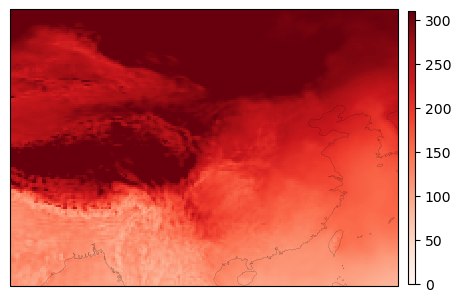}{SFNO}{120}
    \caption{
       \textbf{Root Mean Squared Error (RMSE) of FourCastNet (SFNO) and MetMamba}
       of variable \texttt{msl},
       when both compared against ERA5.
       Darker red indicates worse performance observed at this grid point.
     } 
    \label{fig:spatial_errors_comp_msl}
\end{figure}

\begin{figure}[tbp]
     \centering
     \normalspatialsubfig{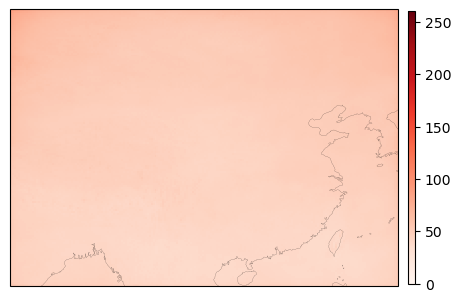}{MetMamba}{24}
     \normalspatialsubfig{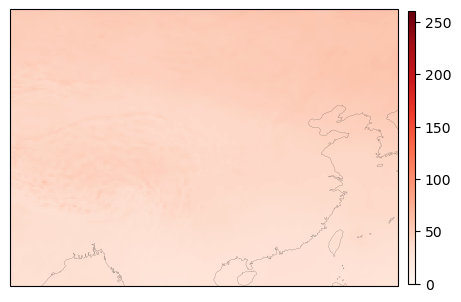}{SFNO}{24}
     \normalspatialsubfig{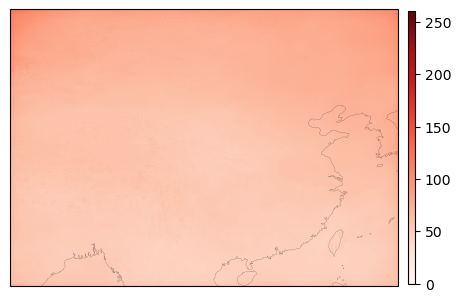}{MetMamba}{36}
     \normalspatialsubfig{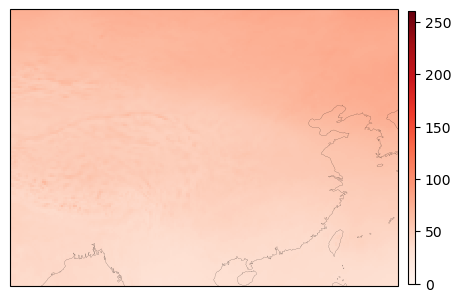}{SFNO}{36}
     \normalspatialsubfig{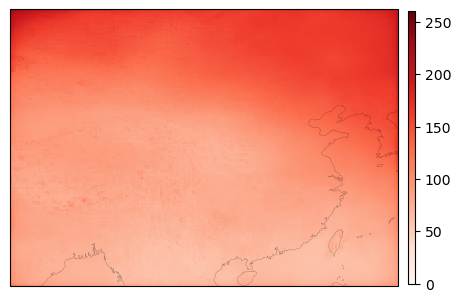}{MetMamba}{72}
     \normalspatialsubfig{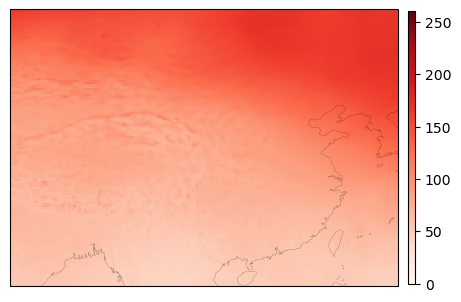}{SFNO}{72}
     \normalspatialsubfig{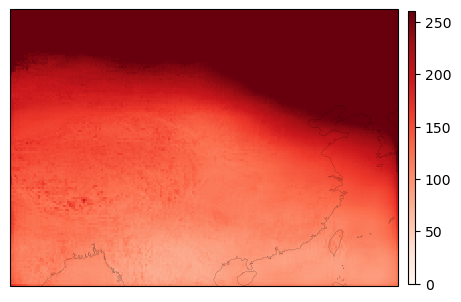}{MetMamba}{120}
     \normalspatialsubfig{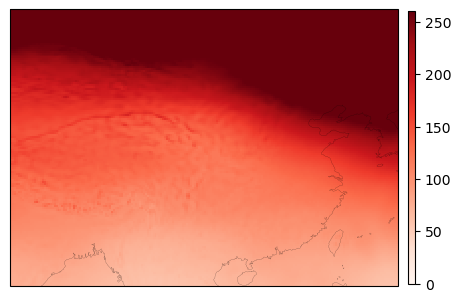}{SFNO}{120}
    \caption{
       \textbf{Root Mean Squared Error (RMSE) of FourCastNet (SFNO) and MetMamba}
       of variable \texttt{z@500},
       when both compared against ERA5.
       Darker red indicates worse performance observed at this grid point.
     } 
    \label{fig:spatial_errors_comp_z500}
\end{figure}

\begin{figure}[tbp]
     \centering
     \normalspatialsubfig{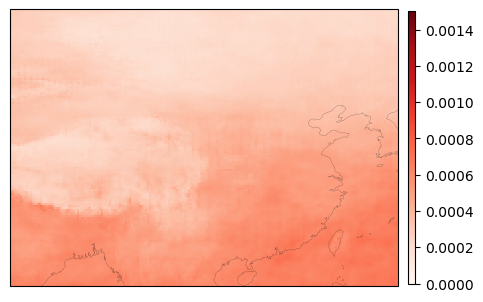}{MetMamba}{24}
     \normalspatialsubfig{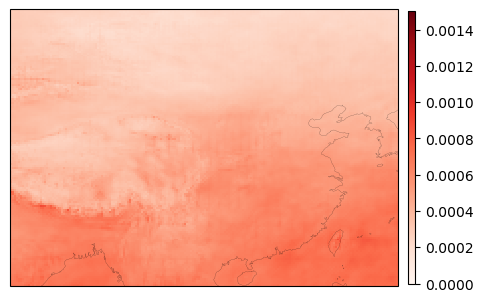}{SFNO}{24}
     \normalspatialsubfig{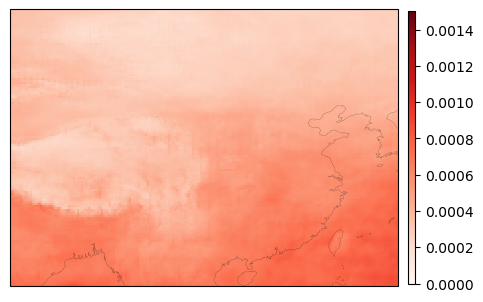}{MetMamba}{36}
     \normalspatialsubfig{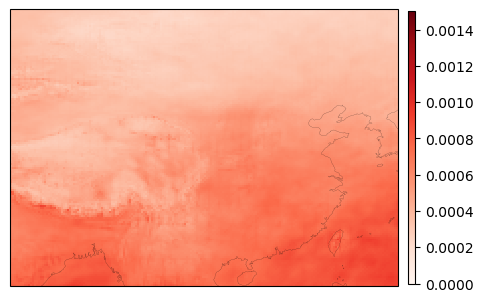}{SFNO}{36}
     \normalspatialsubfig{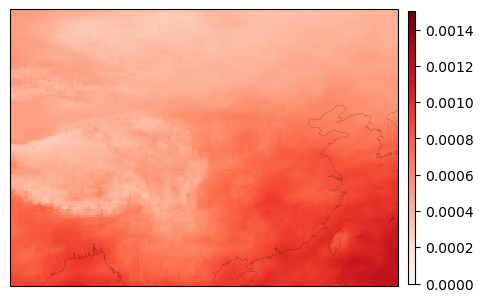}{MetMamba}{72}
     \normalspatialsubfig{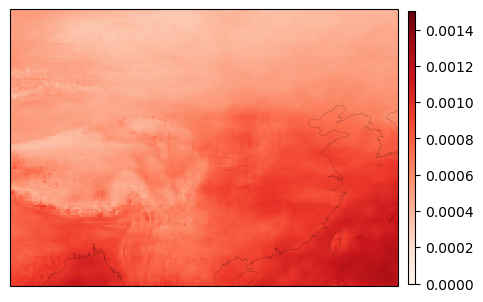}{SFNO}{72}
     \normalspatialsubfig{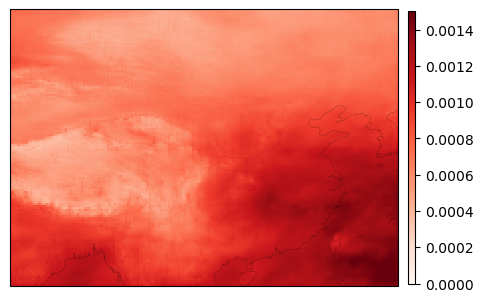}{MetMamba}{120}
     \normalspatialsubfig{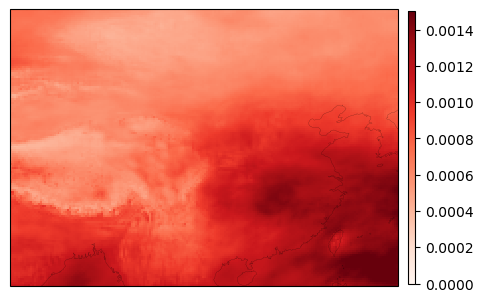}{SFNO}{120}
    \caption{
       \textbf{Root Mean Squared Error (RMSE) of FourCastNet (SFNO) and MetMamba}
       of variable \texttt{q@700},
       when both compared against ERA5.
       Darker red indicates worse performance observed at this grid point.
     } 
    \label{fig:spatial_errors_comp_q700}
\end{figure}

\begin{figure}[tbp]
     \centering
     \normalspatialsubfig{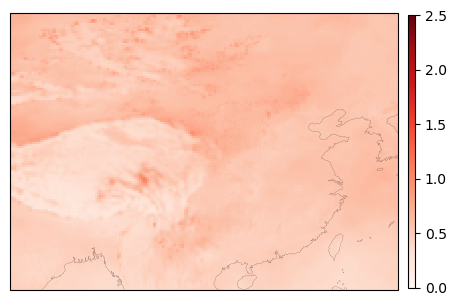}{MetMamba}{24}
     \normalspatialsubfig{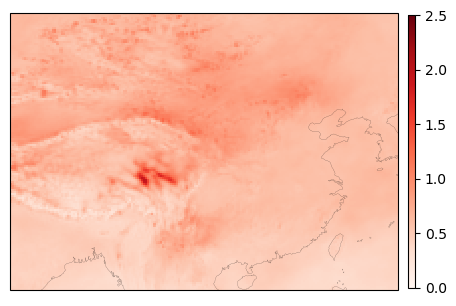}{SFNO}{24}
     \normalspatialsubfig{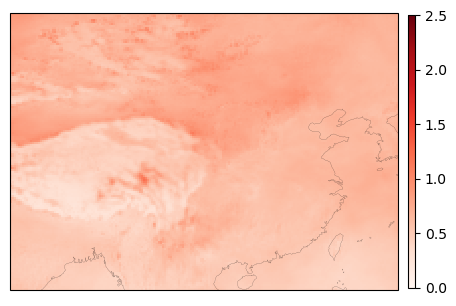}{MetMamba}{36}
     \normalspatialsubfig{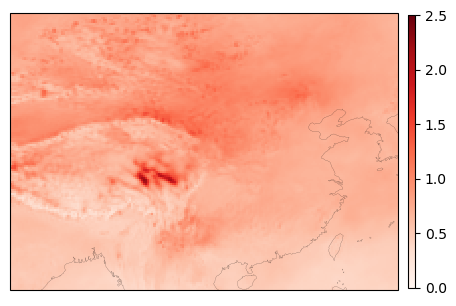}{SFNO}{36}
     \normalspatialsubfig{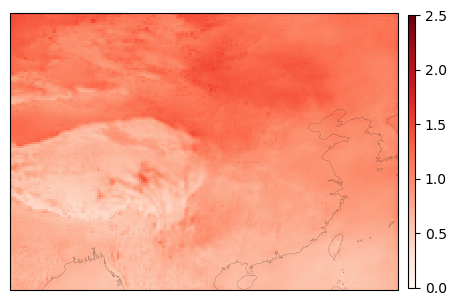}{MetMamba}{72}
     \normalspatialsubfig{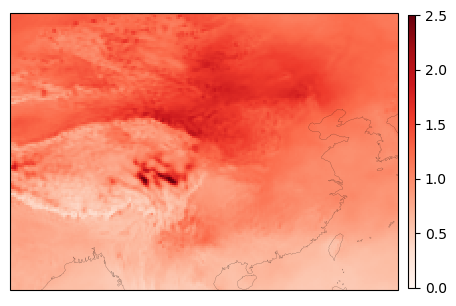}{SFNO}{72}
     \normalspatialsubfig{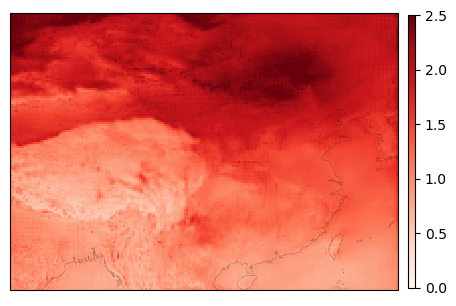}{MetMamba}{120}
     \normalspatialsubfig{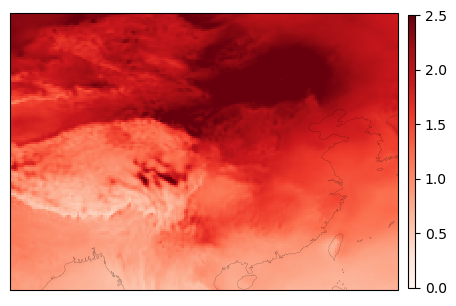}{SFNO}{120}
    \caption{
       \textbf{Root Mean Squared Error (RMSE) of FourCastNet (SFNO) and MetMamba}
       of variable \texttt{t@850},
       when both compared against ERA5.
       Darker red indicates worse performance observed at this grid point.
     } 
    \label{fig:spatial_errors_comp_t850}
\end{figure}

\begin{figure}[tbp]
     \centering
     \normalspatialsubfig{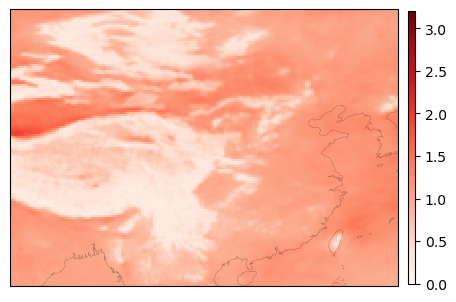}{MetMamba}{24}
     \normalspatialsubfig{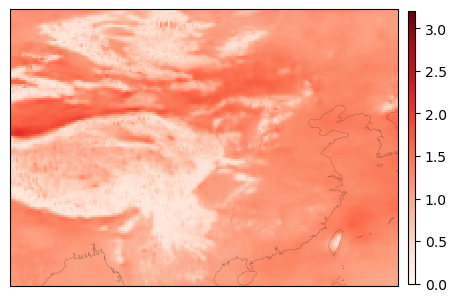}{SFNO}{24}
     \normalspatialsubfig{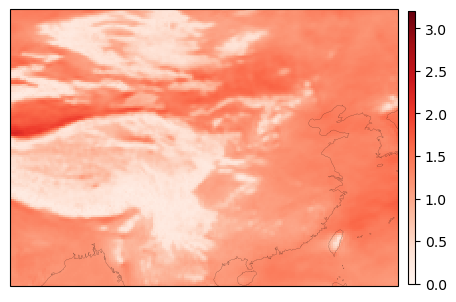}{MetMamba}{36}
     \normalspatialsubfig{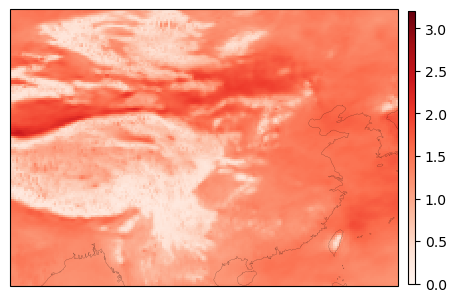}{SFNO}{36}
     \normalspatialsubfig{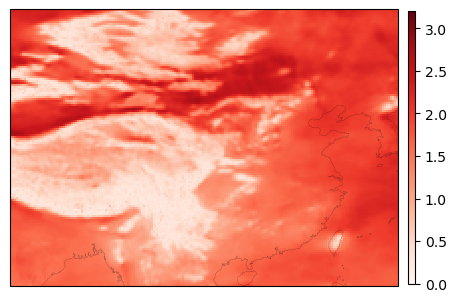}{MetMamba}{72}
     \normalspatialsubfig{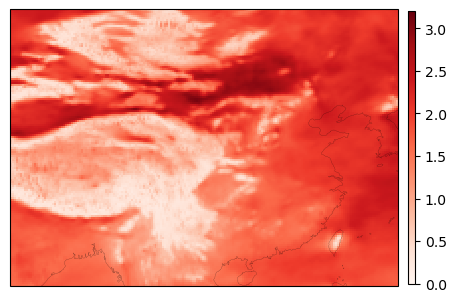}{SFNO}{72}
     \normalspatialsubfig{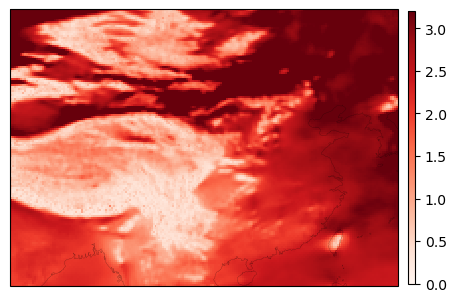}{MetMamba}{120}
     \normalspatialsubfig{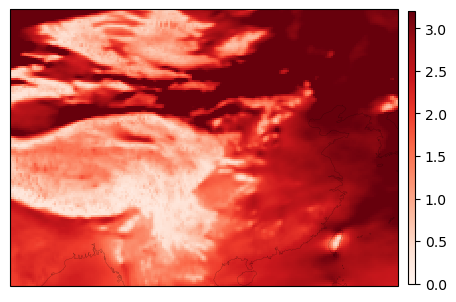}{SFNO}{120}
    \caption{
       \textbf{Root Mean Squared Error (RMSE) of FourCastNet (SFNO) and MetMamba}
       of variable \texttt{u@850},
       when both compared against ERA5.
       Darker red indicates worse performance observed at this grid point.
     } 
    \label{fig:spatial_errors_comp_u850}
\end{figure}

\begin{figure}[tbp]
     \centering
     \normalspatialsubfig{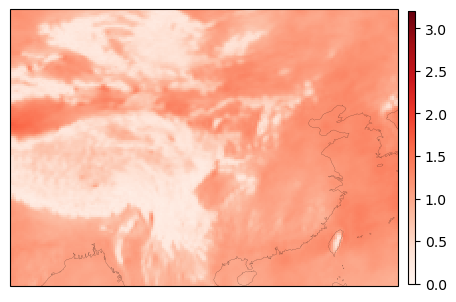}{MetMamba}{24}
     \normalspatialsubfig{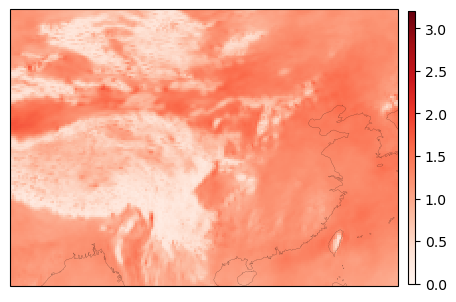}{SFNO}{24}
     \normalspatialsubfig{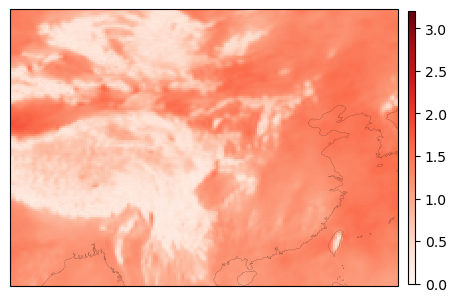}{MetMamba}{36}
     \normalspatialsubfig{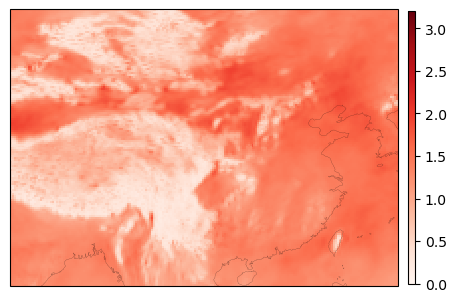}{SFNO}{36}
     \normalspatialsubfig{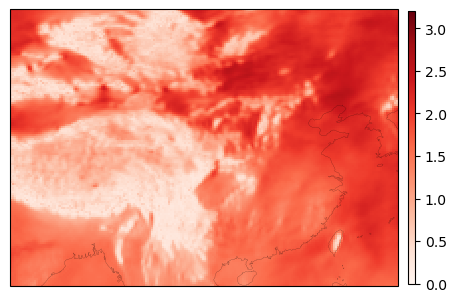}{MetMamba}{72}
     \normalspatialsubfig{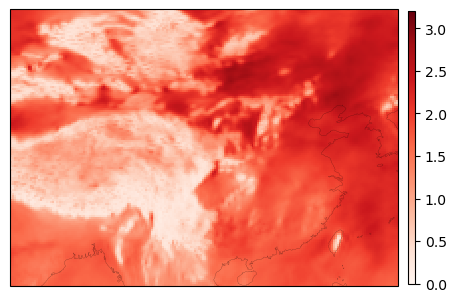}{SFNO}{72}
     \normalspatialsubfig{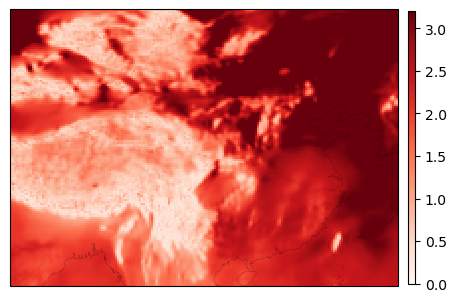}{MetMamba}{120}
     \normalspatialsubfig{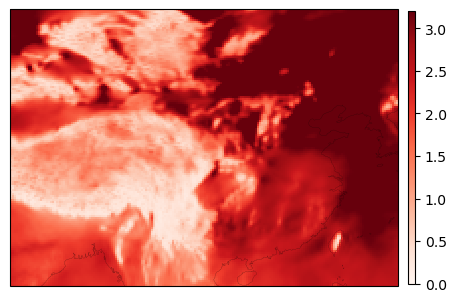}{SFNO}{120}
    \caption{
       \textbf{Root Mean Squared Error (RMSE) of FourCastNet (SFNO) and MetMamba}
       of variable \texttt{v@850},
       when both compared against ERA5.
       Darker red indicates worse performance observed at this grid point.
     } 
    \label{fig:spatial_errors_comp_v850}
\end{figure}

For side by side comparisons, see Figure~\ref{fig:spatial_errors_comp_10u}--\ref{fig:spatial_errors_comp_v850}.

\begin{figure}[tbp]
     \centering
     \spatialsubfig{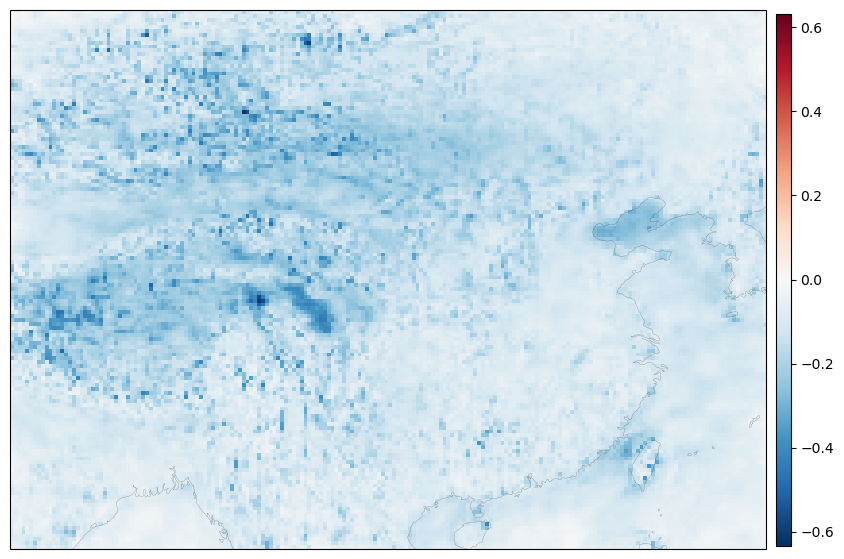}{24}
     \spatialsubfig{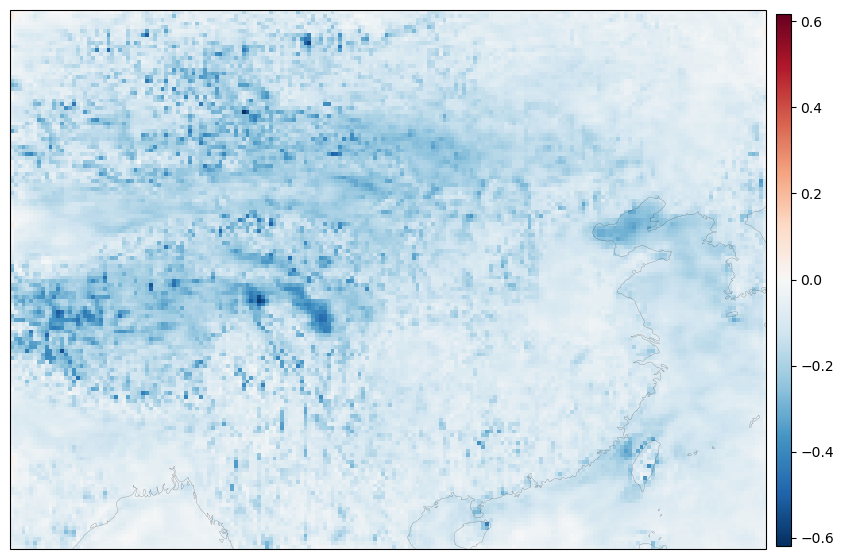}{36}
     \spatialsubfig{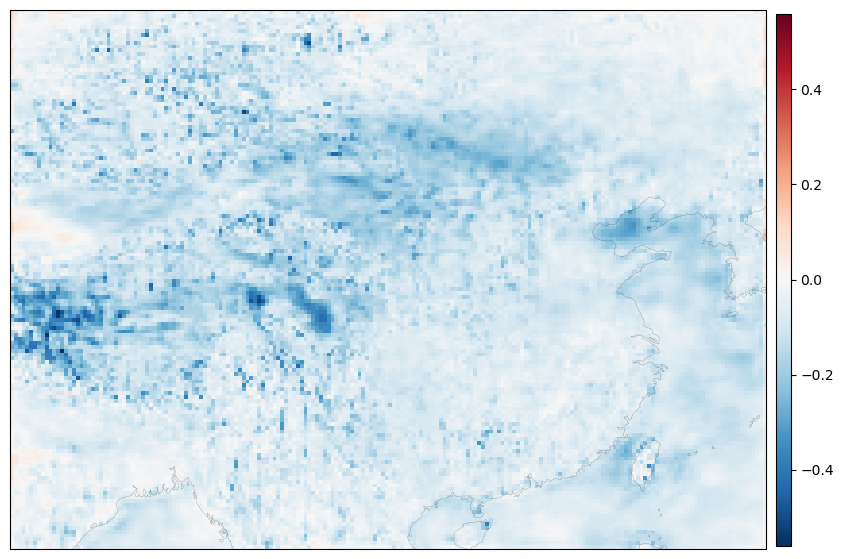}{72}
     \spatialsubfig{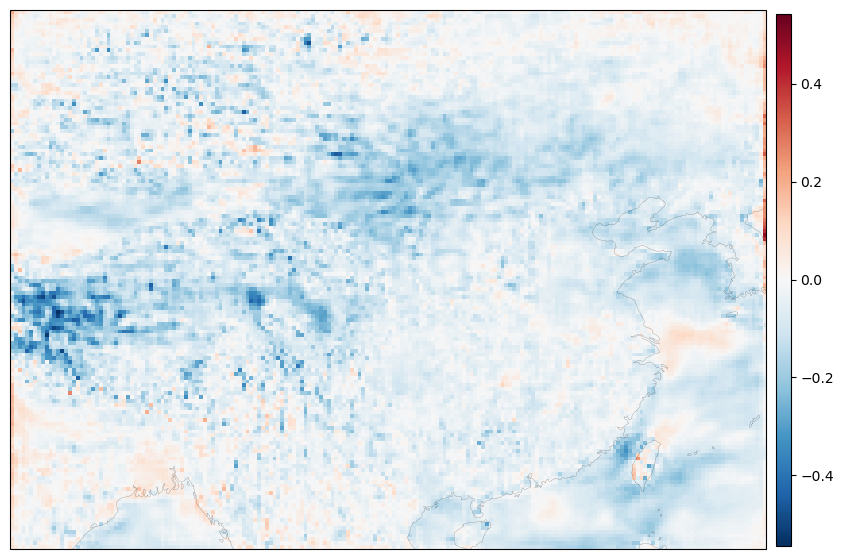}{120}
    \caption{
       \textbf{Root Mean Squared Error (RMSE) difference between FourCastNet (SFNO) and MetMamba}
       of variable \texttt{10u},
       when both compared against ERA5.
       Blue indicates better performance observed at this grid point, red indicates worse performance observed at this grid point.
     } 
    \label{fig:spatial_errors_10u}
\end{figure}

\begin{figure}[tbp]
     \centering
     \spatialsubfig{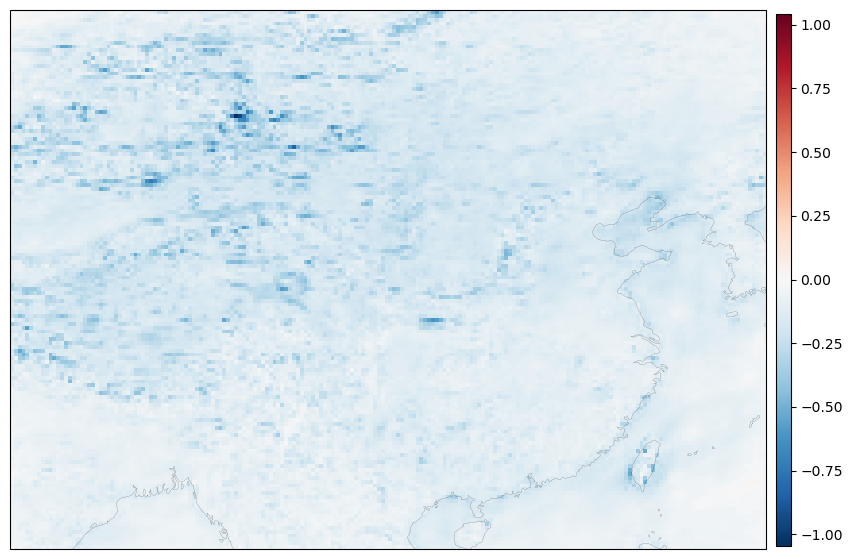}{24}
     \spatialsubfig{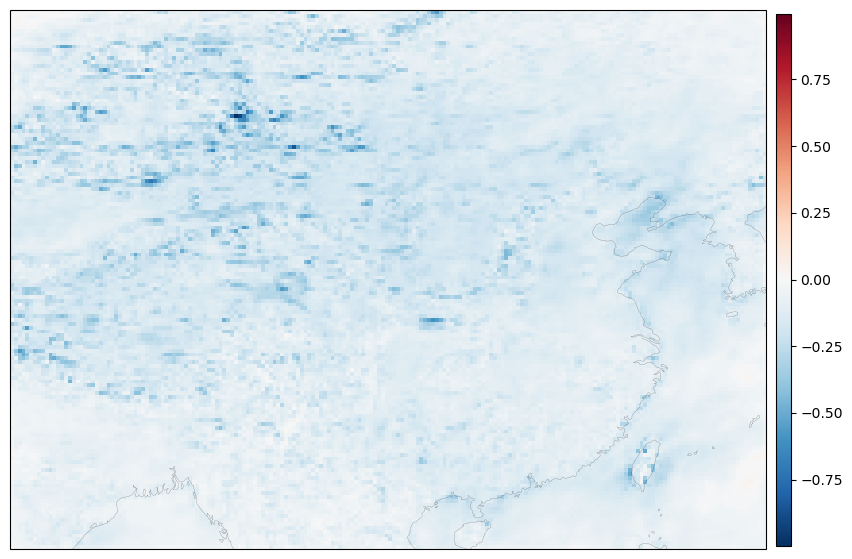}{36}
     \spatialsubfig{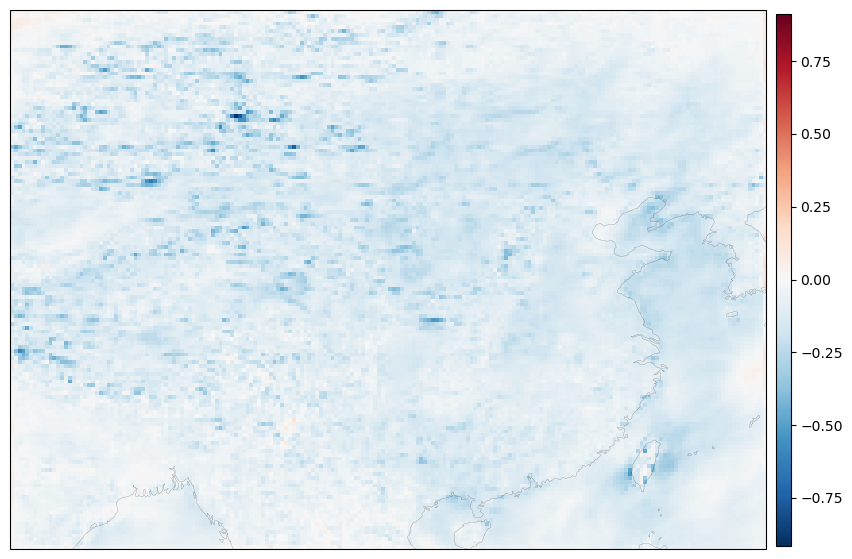}{72}
     \spatialsubfig{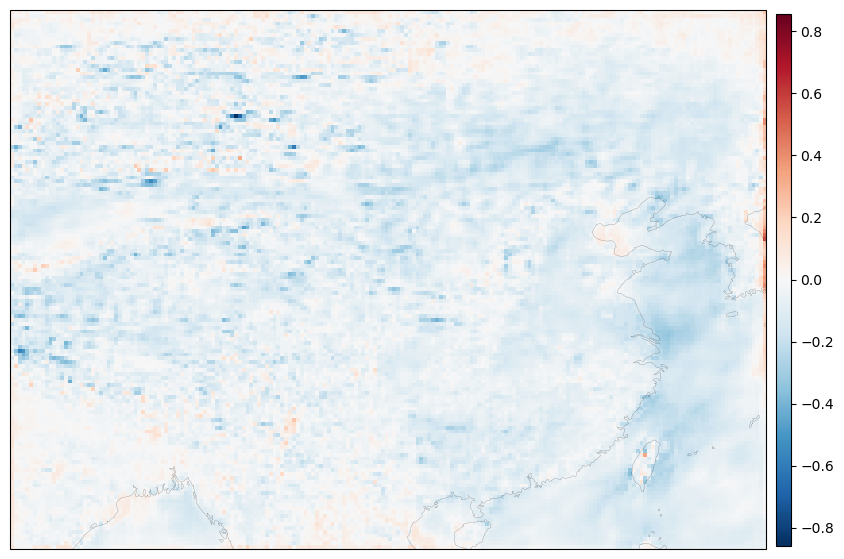}{120}
     \caption{
       \textbf{Root Mean Squared Error (RMSE) difference between FourCastNet (SFNO) and MetMamba}
       of variable \texttt{10v},
       when both compared against ERA5.
       Blue indicates better performance observed at this grid point, red indicates worse performance observed at this grid point.
     } 
    \label{fig:spatial_errors_10v}
\end{figure}

\begin{figure}[tbp]
     \centering
     \spatialsubfig{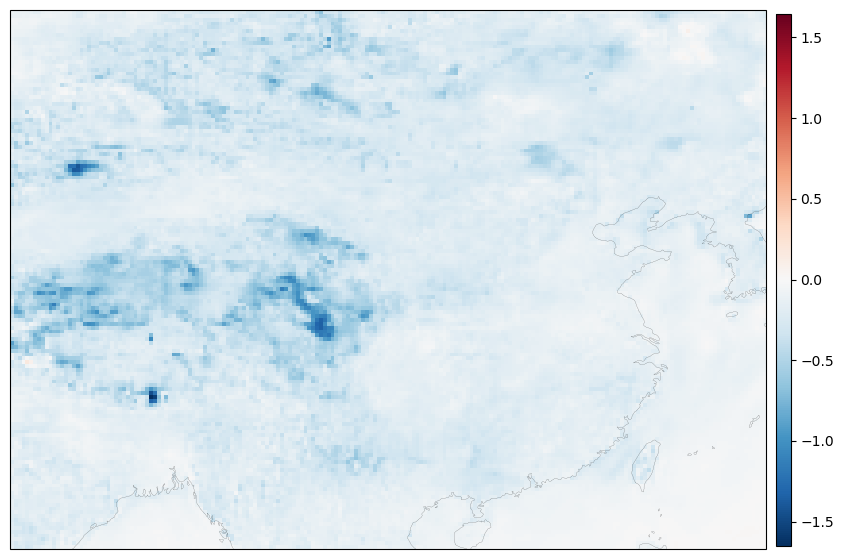}{24}
     \spatialsubfig{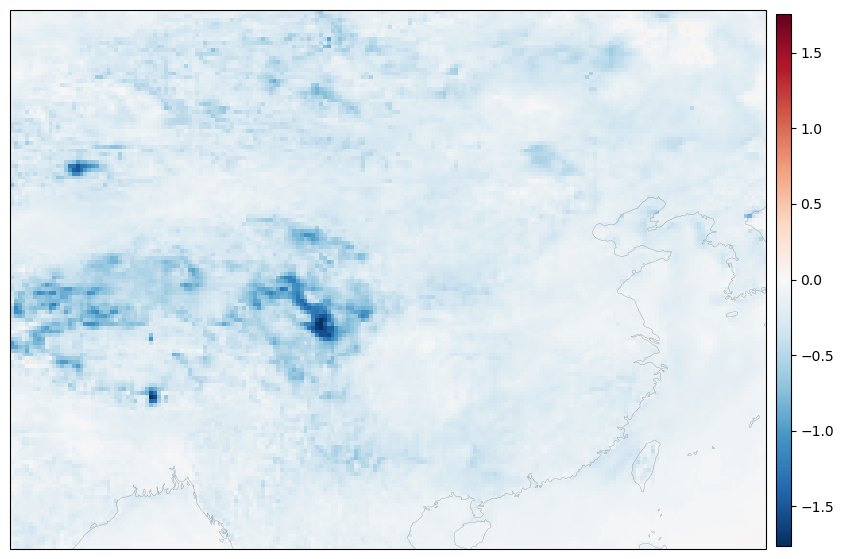}{36}
     \spatialsubfig{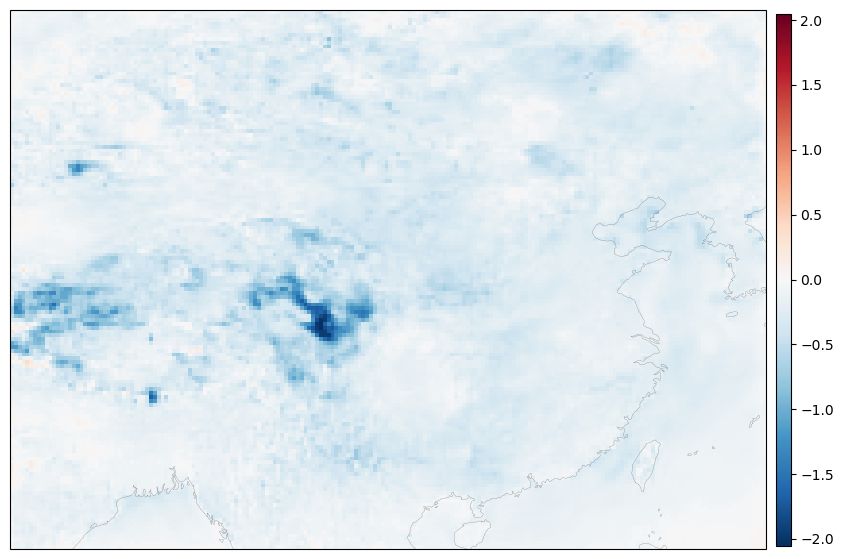}{72}
     \spatialsubfig{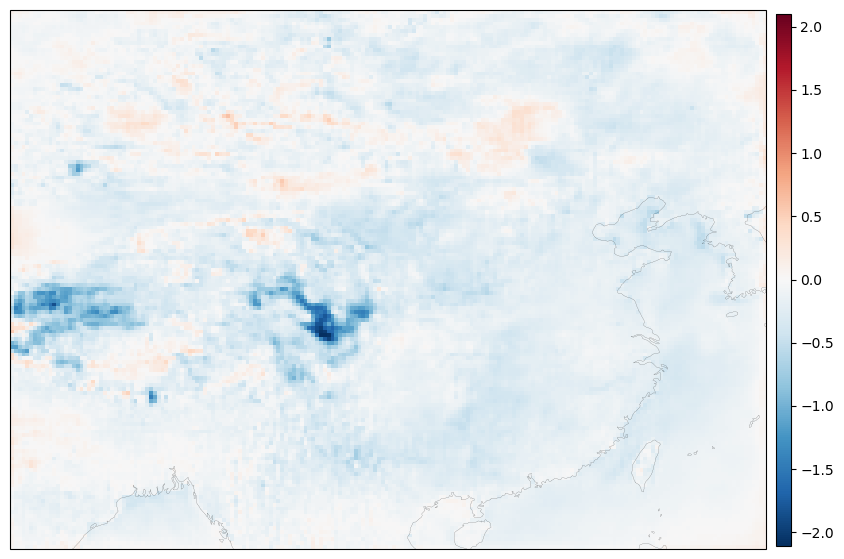}{120}
    \caption{
       \textbf{Root Mean Squared Error (RMSE) difference between FourCastNet (SFNO) and MetMamba}
       of variable \texttt{2t},
       when both compared against ERA5.
       Blue indicates better performance observed at this grid point, red indicates worse performance observed at this grid point.
     } 
    \label{fig:spatial_errors_2t}
\end{figure}

\begin{figure}[tbp]
     \centering
     \spatialsubfig{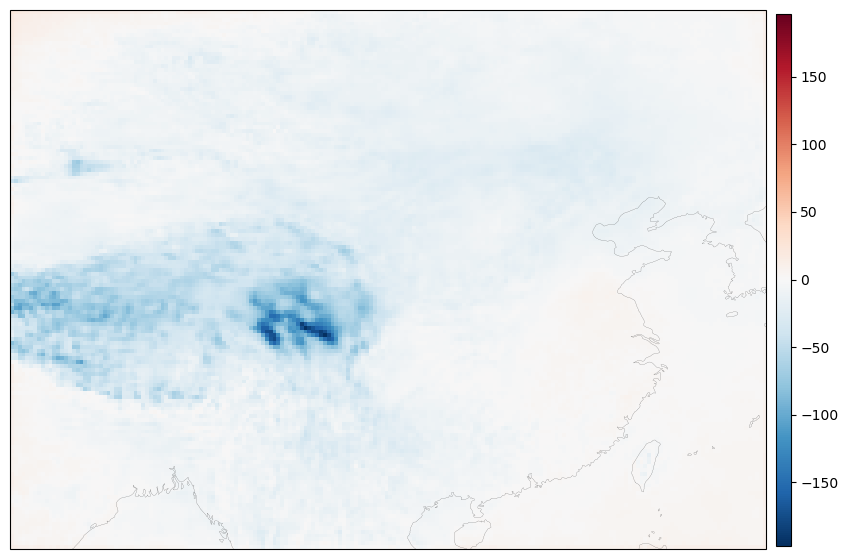}{24}
     \spatialsubfig{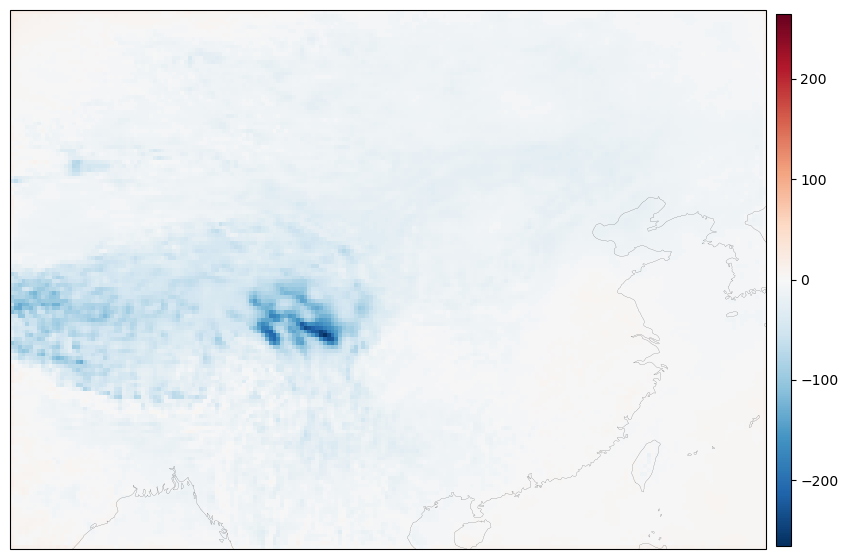}{36}
     \spatialsubfig{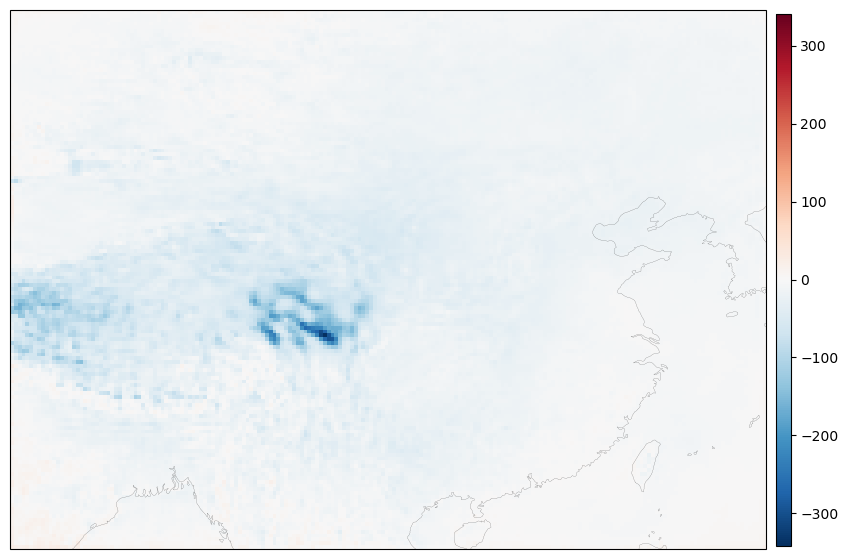}{72}
     \spatialsubfig{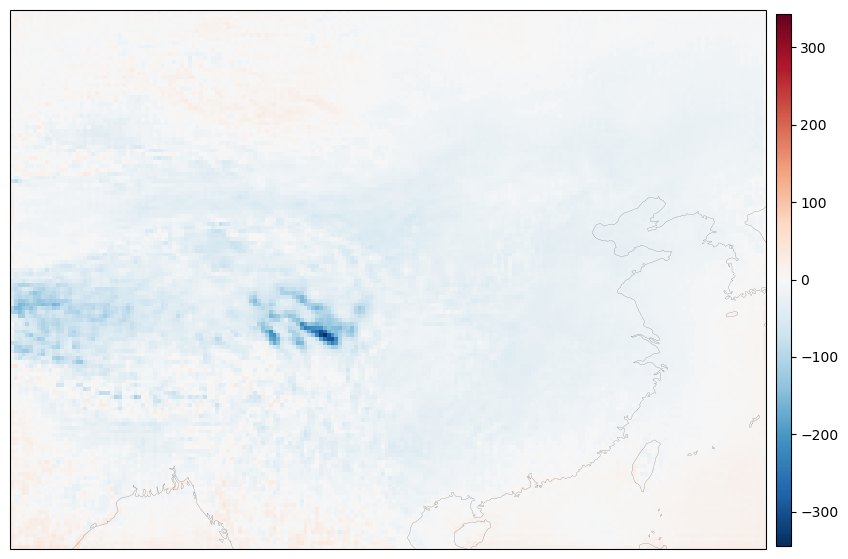}{120}
     \caption{
       \textbf{Root Mean Squared Error (RMSE) difference between FourCastNet (SFNO) and MetMamba}
       of variable \texttt{msl},
       when both compared against ERA5.
       Blue indicates better performance observed at this grid point, red indicates worse performance observed at this grid point.
     }
    \label{fig:spatial_errors_msl}
\end{figure}

\begin{figure}[tbp]
     \centering
     \spatialsubfig{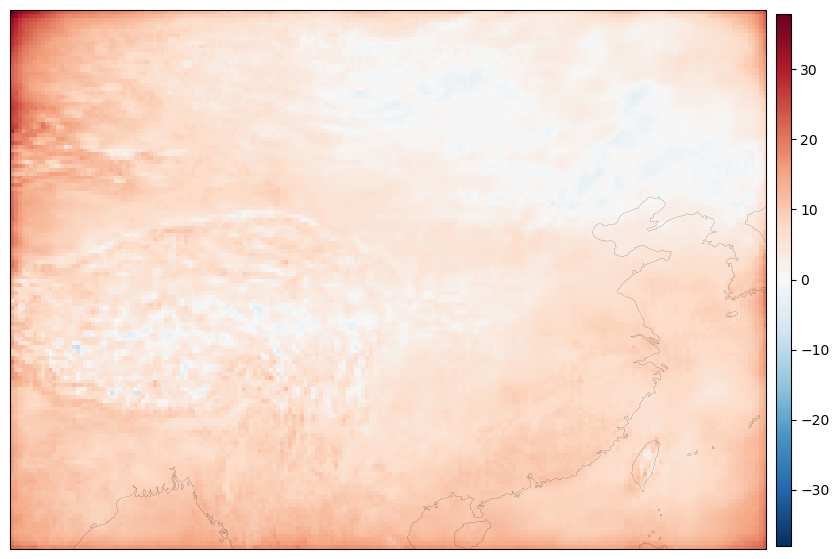}{24}
     \spatialsubfig{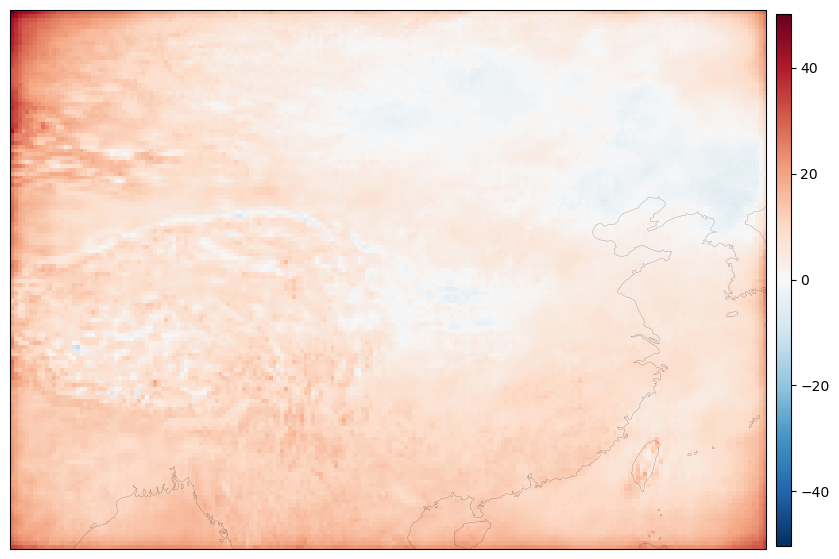}{36}
     \spatialsubfig{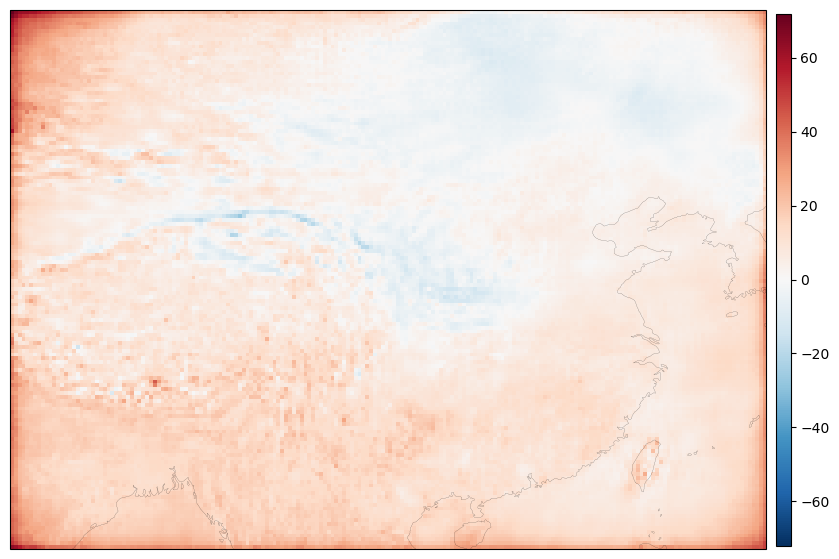}{72}
     \spatialsubfig{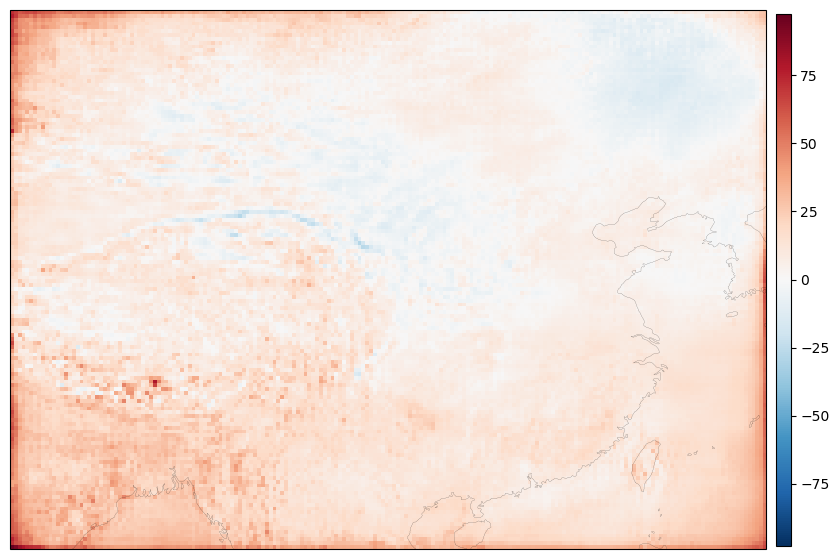}{120}
     \caption{
       \textbf{Root Mean Squared Error (RMSE) difference between FourCastNet (SFNO) and MetMamba}
       of variable \texttt{z@500},
       when both compared against ERA5.
       Blue indicates better performance observed at this grid point, red indicates worse performance observed at this grid point.
     }
    \label{fig:spatial_errors_z500}
\end{figure}

\begin{figure}[tbp]
     \centering
     \spatialsubfig{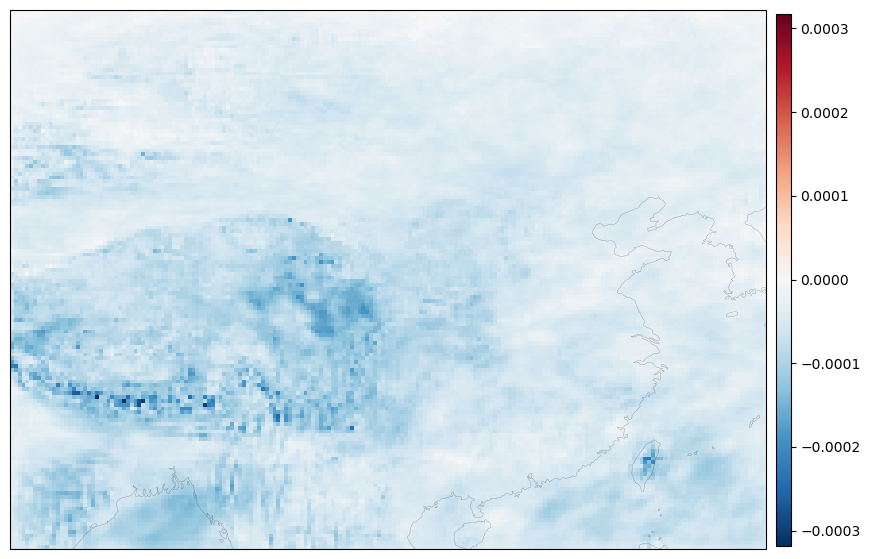}{24}
     \spatialsubfig{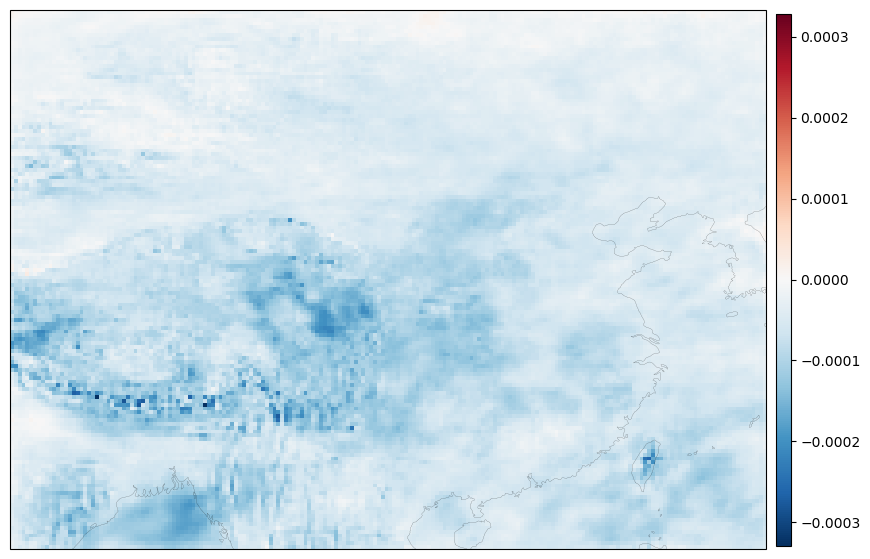}{36}
     \spatialsubfig{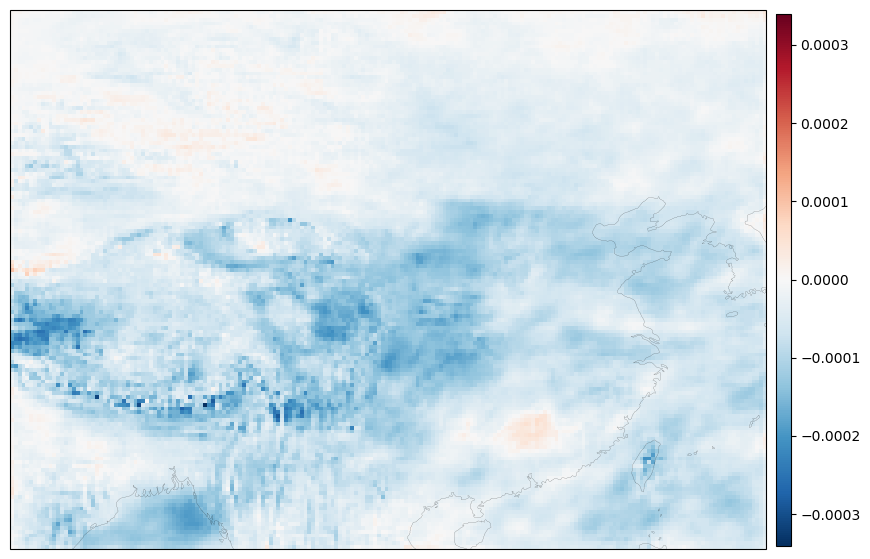}{72}
     \spatialsubfig{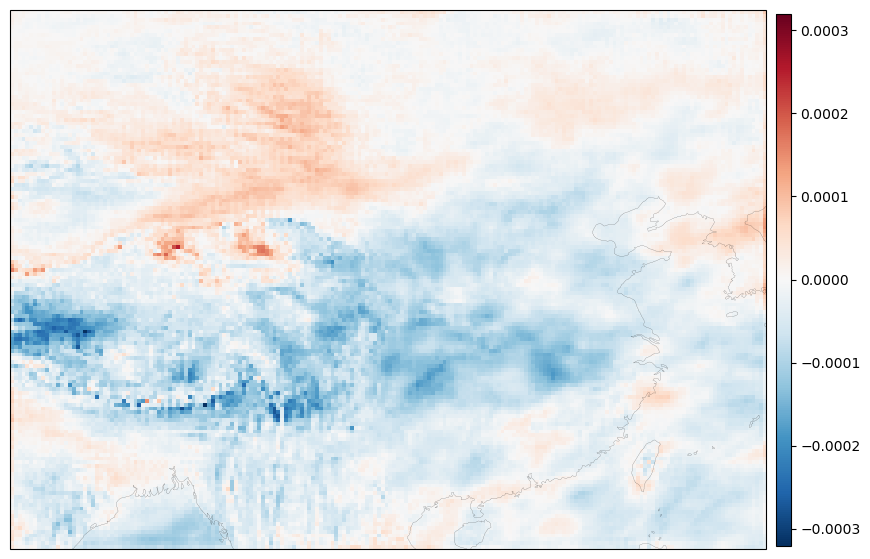}{120}
     \caption{
       \textbf{Root Mean Squared Error (RMSE) difference between FourCastNet (SFNO) and MetMamba}
       of variable \texttt{q@700},
       when both compared against ERA5.
       Blue indicates better performance observed at this grid point, red indicates worse performance observed at this grid point.
     }
    \label{fig:spatial_errors_q700}
\end{figure}

\begin{figure}[tbp]
     \centering
      \spatialsubfig{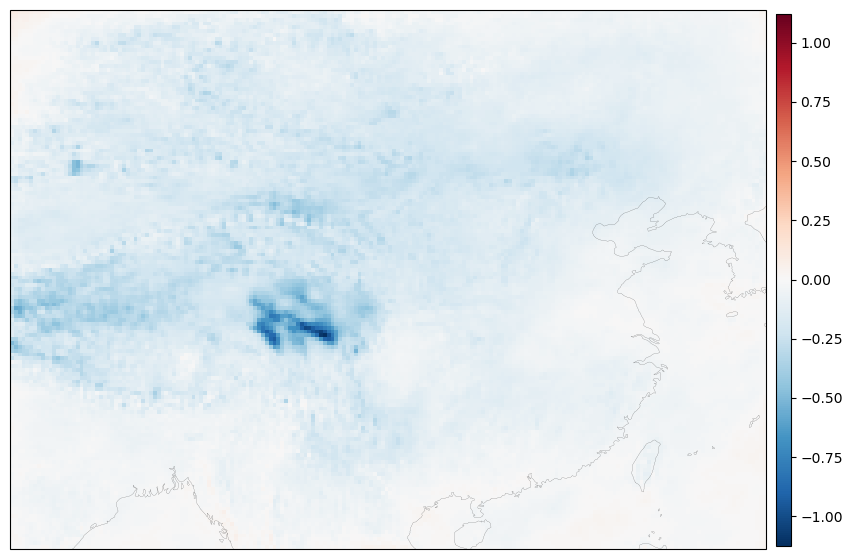}{24}
      \spatialsubfig{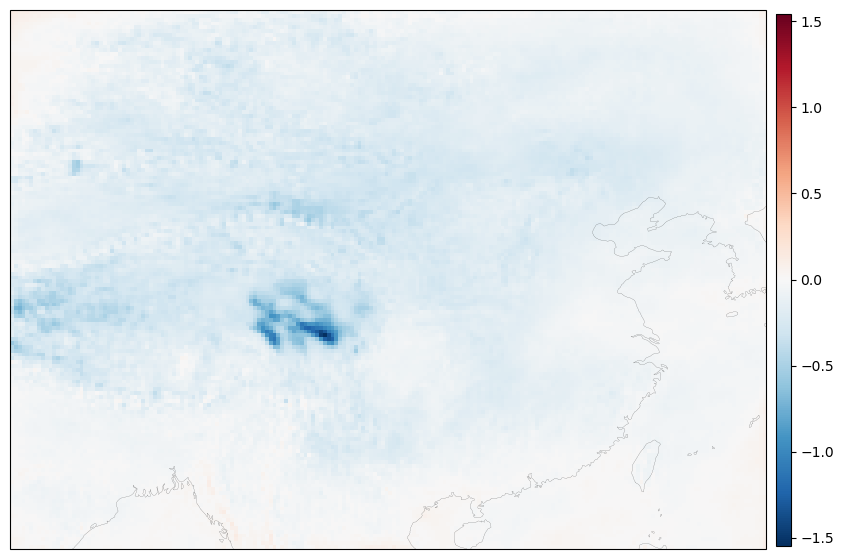}{36}
     \spatialsubfig{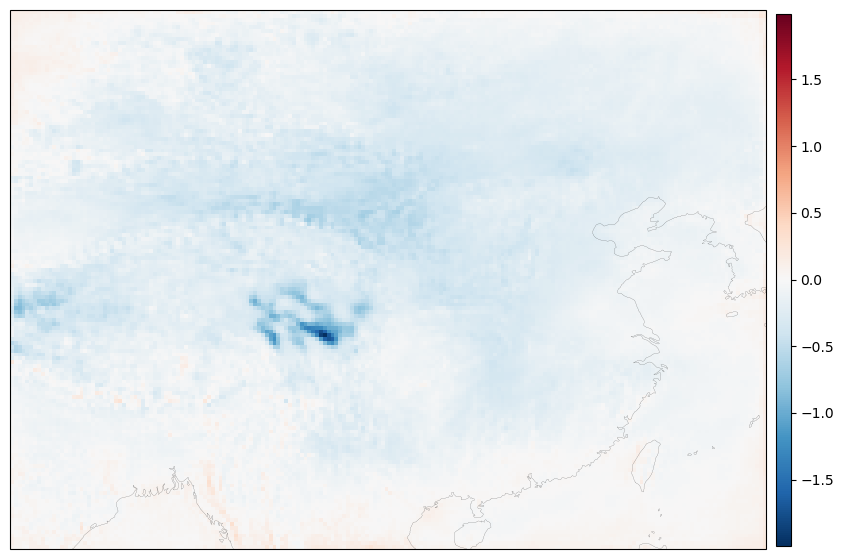}{72}
     \spatialsubfig{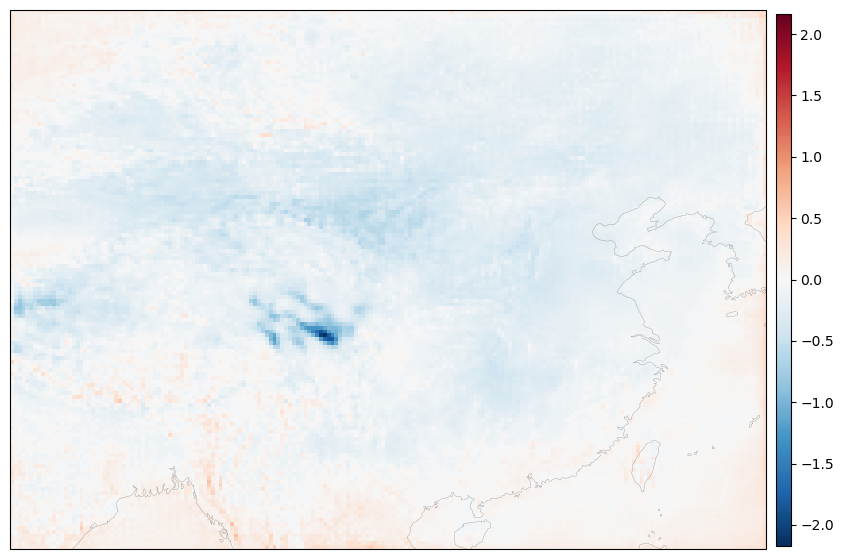}{120}
     \caption{
       \textbf{Root Mean Squared Error (RMSE) difference between FourCastNet (SFNO) and MetMamba}
       of variable \texttt{t@850},
       when both compared against ERA5.
       Blue indicates better performance observed at this grid point, red indicates worse performance observed at this grid point.
     }
    \label{fig:spatial_errors_t850}
\end{figure}

\begin{figure}[tbp]
     \centering
      \spatialsubfig{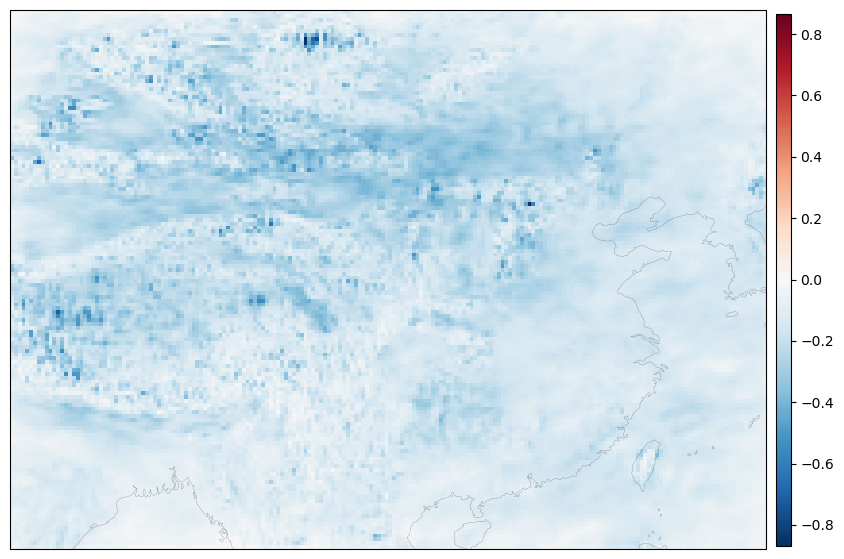}{24}
      \spatialsubfig{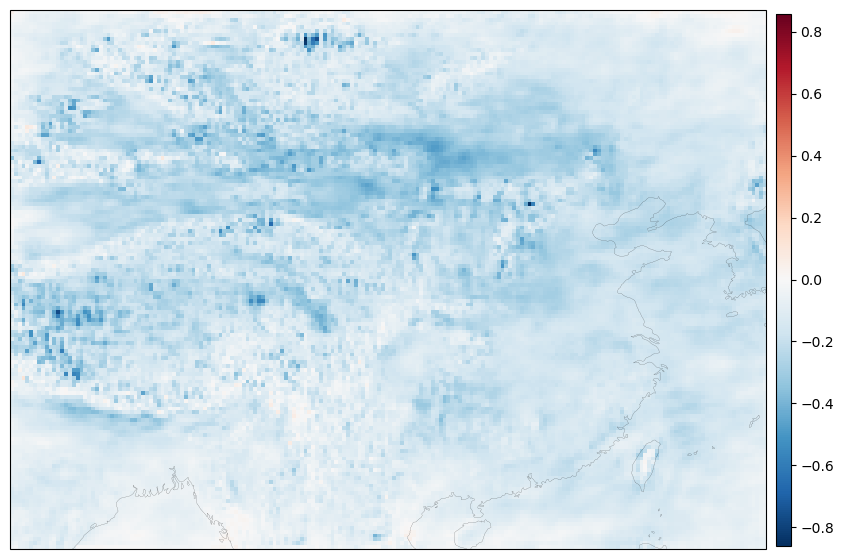}{36}
     \spatialsubfig{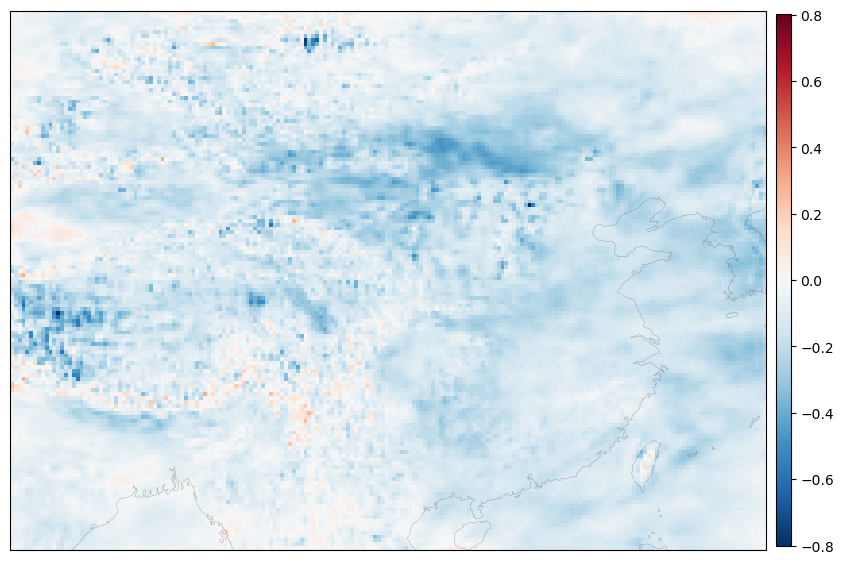}{72}
     \spatialsubfig{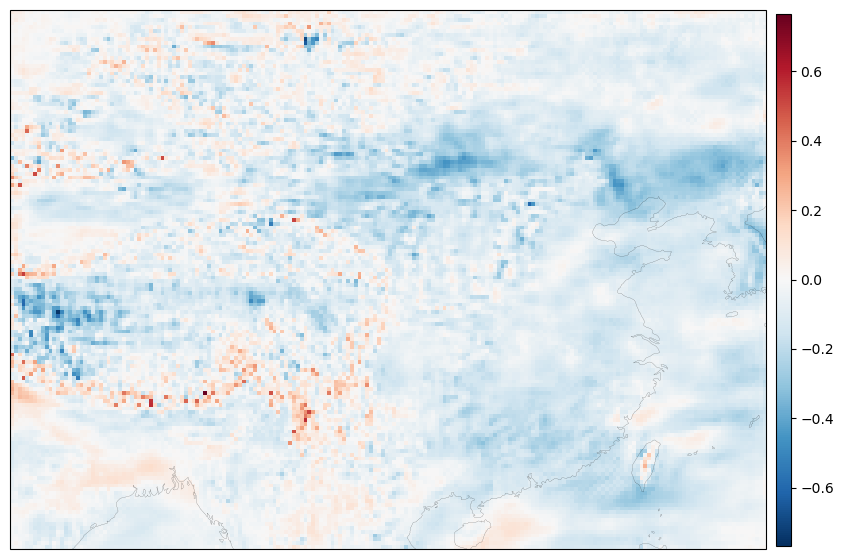}{120}
     \caption{
       \textbf{Root Mean Squared Error (RMSE) difference between FourCastNet (SFNO) and MetMamba}
       of variable \texttt{u@850},
       when both compared against ERA5.
       Blue indicates better performance observed at this grid point, red indicates worse performance observed at this grid point.
     }
    \label{fig:spatial_errors_u850}
\end{figure}

\begin{figure}[tbp]
     \centering
      \spatialsubfig{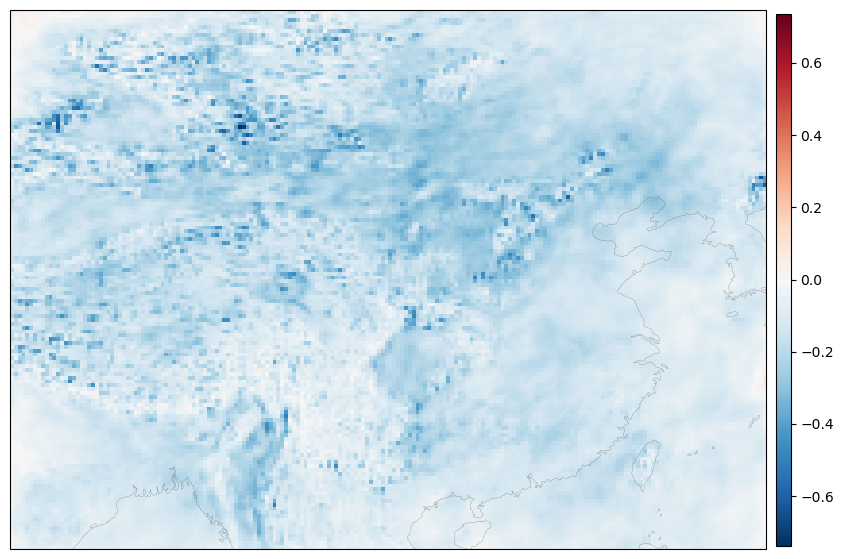}{24}
      \spatialsubfig{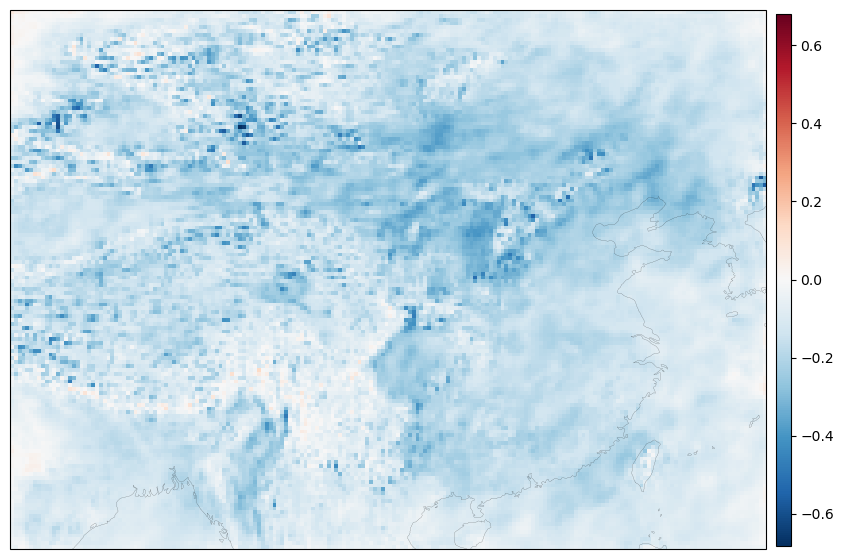}{36}
     \spatialsubfig{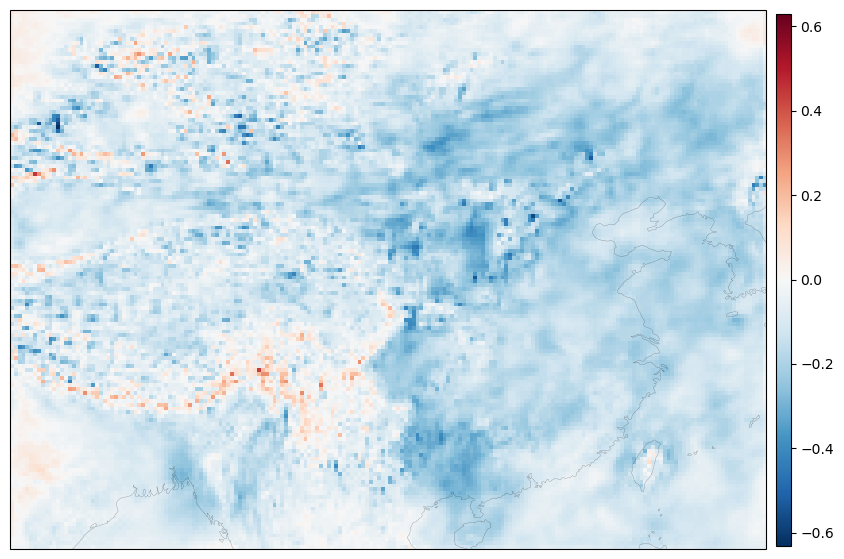}{72}
     \spatialsubfig{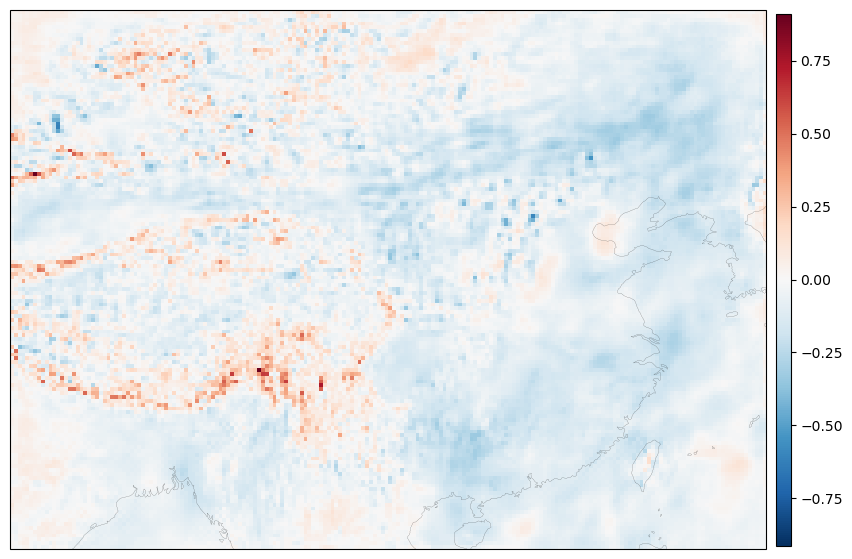}{120}
     \caption{
       \textbf{Root Mean Squared Error (RMSE) difference between FourCastNet (SFNO) and MetMamba}
       of variable \texttt{v@850},
       when both compared against ERA5.
       Blue indicates better performance observed at this grid point, red indicates worse performance observed at this grid point.
     }
    \label{fig:spatial_errors_v850}
\end{figure}

For comparisons of error differential, see Figure~\ref{fig:spatial_errors_10u}--\ref{fig:spatial_errors_v850}

In the illustrated examples, we observed degraded performance along
the edges and in long lead times, suggesting the LBCs still played a role in limiting the performance of LAM.
We also note the performance improvements on mountainous areas in a majority of variables, with the exception of 
\texttt{u@850} and \texttt{v@850}. Large error differential spikes can also be 
found in north-western China, where the climate tends to be drier, this could be 
a manifestation of the overestimation of water contents in the Mamba based model. 

\subsection{Effects of Lateral Boundary Conditions}
\label{subsec:effects-of-lateral-boundary-conditions}
\begin{figure}[htbp]
    \centering
    \includegraphics[width=\textwidth,height=\textheight,keepaspectratio]{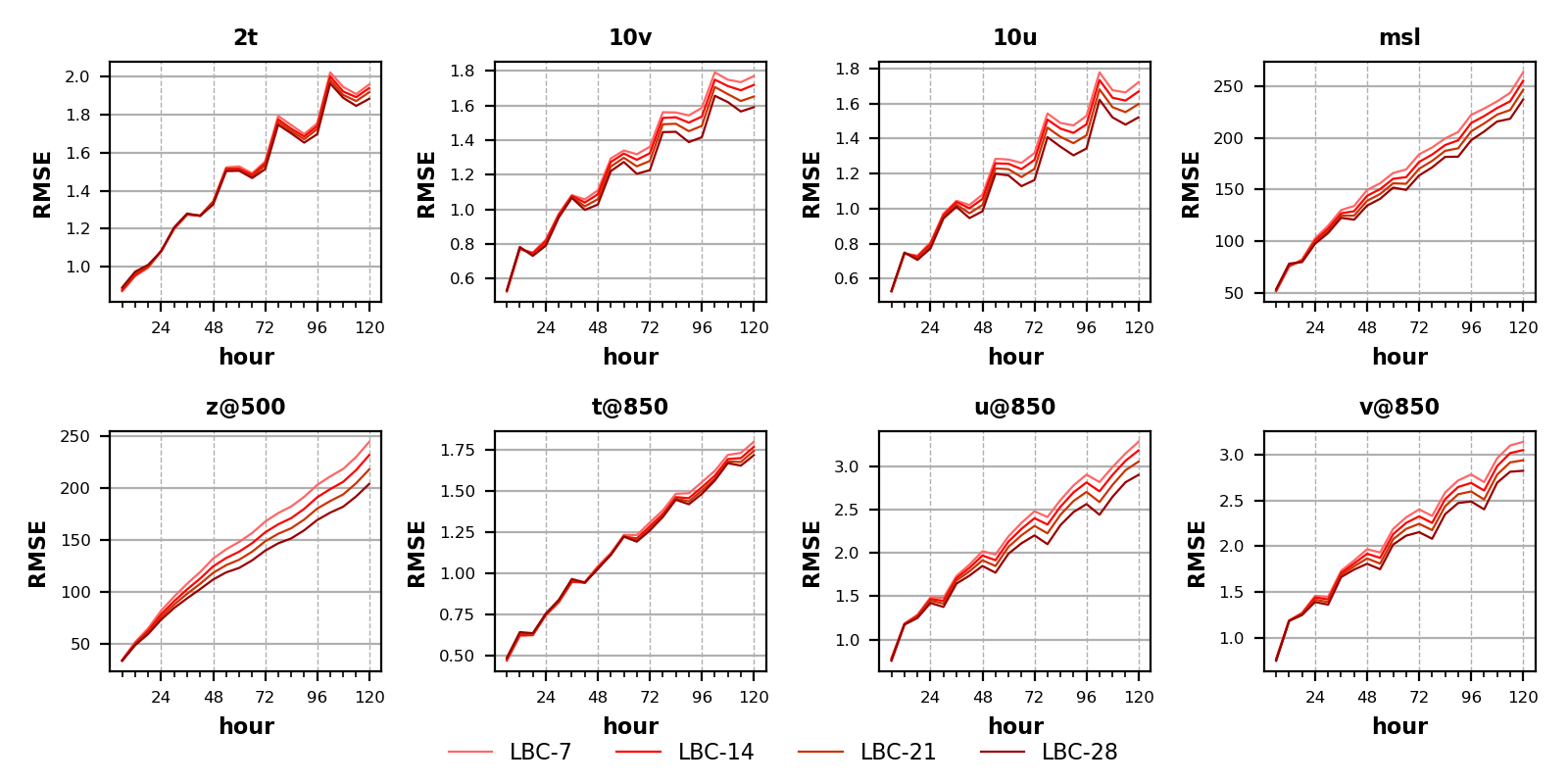}
    \caption{\textbf{Root Mean Squared Error difference between different LBC sizes}.
    } 
    \label{fig:lbc_size}
  
\end{figure}

We run a scaled down toy model on different $W$ sizes of the Lateral Boundary Conditions.
Larger $W$ yields better performance. See Figure~\ref{fig:lbc_size}

\subsection{Spectra}
\label{spectra}
\begin{figure}[tbp]
     \centering
     \normalsubfig{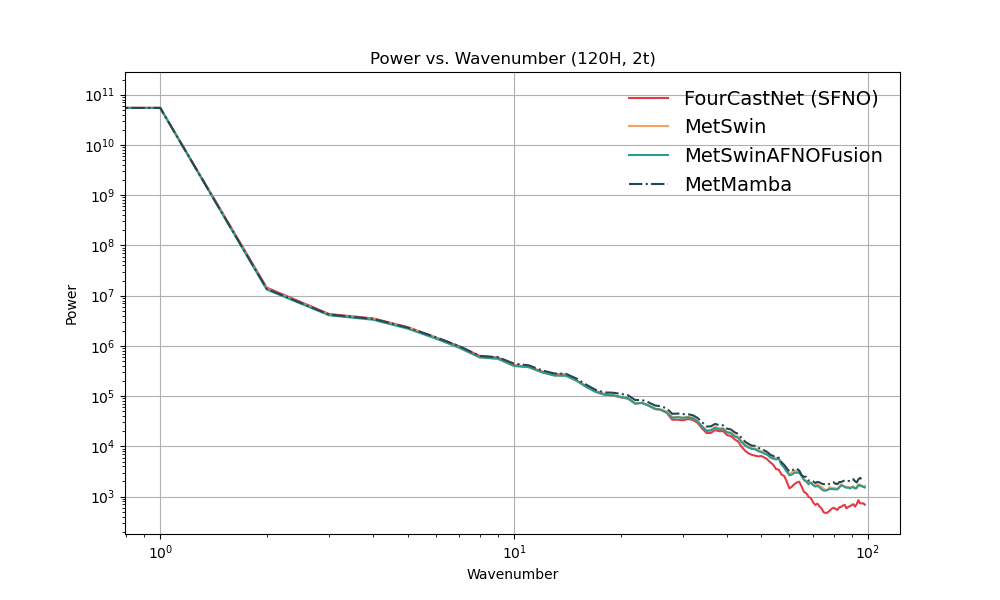}{2t}
     \normalsubfig{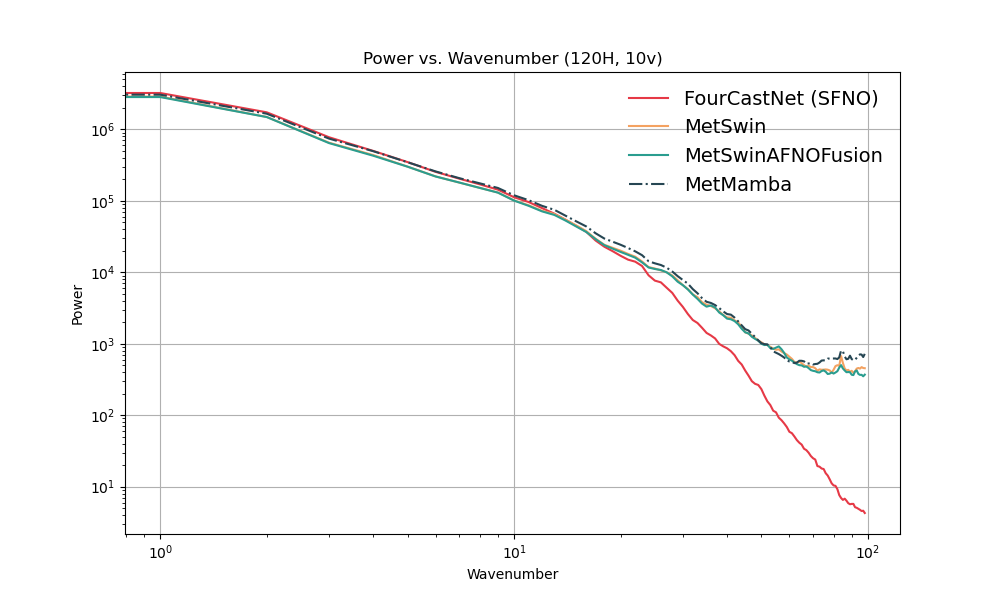}{10v}
     \normalsubfig{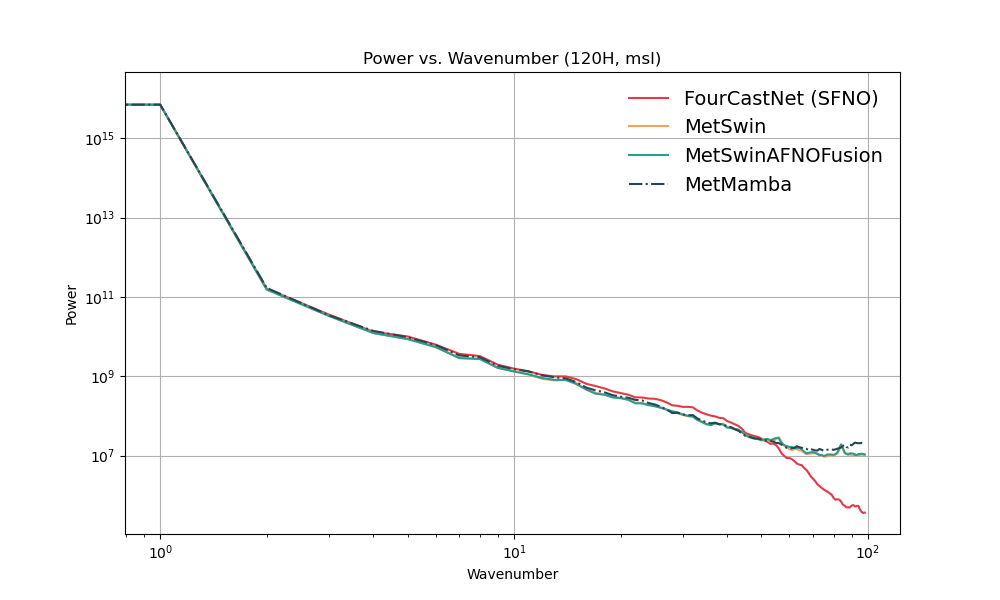}{msl}
     \normalsubfig{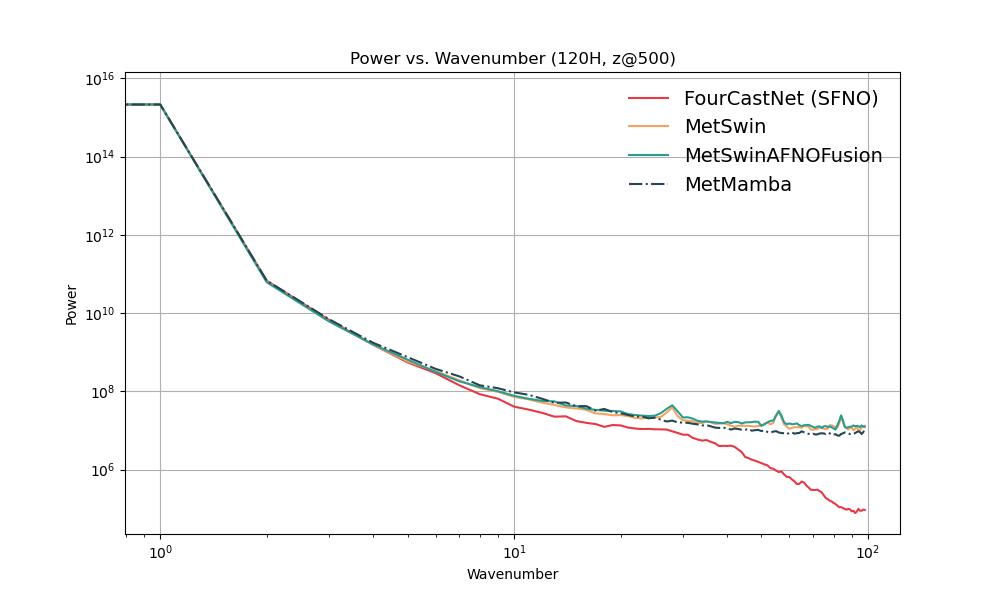}{z@500}
     \normalsubfig{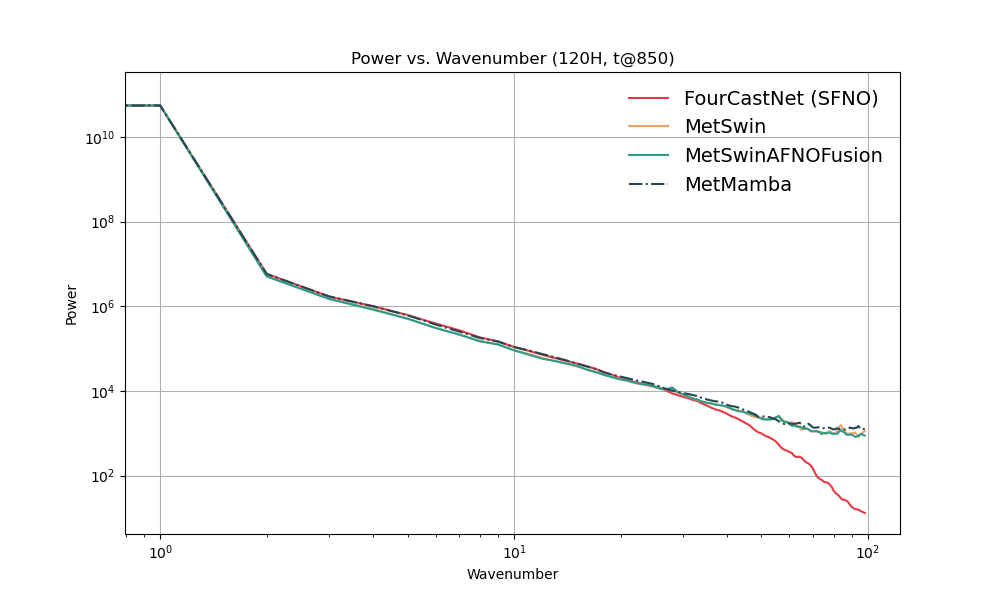}{t@850}
     \normalsubfig{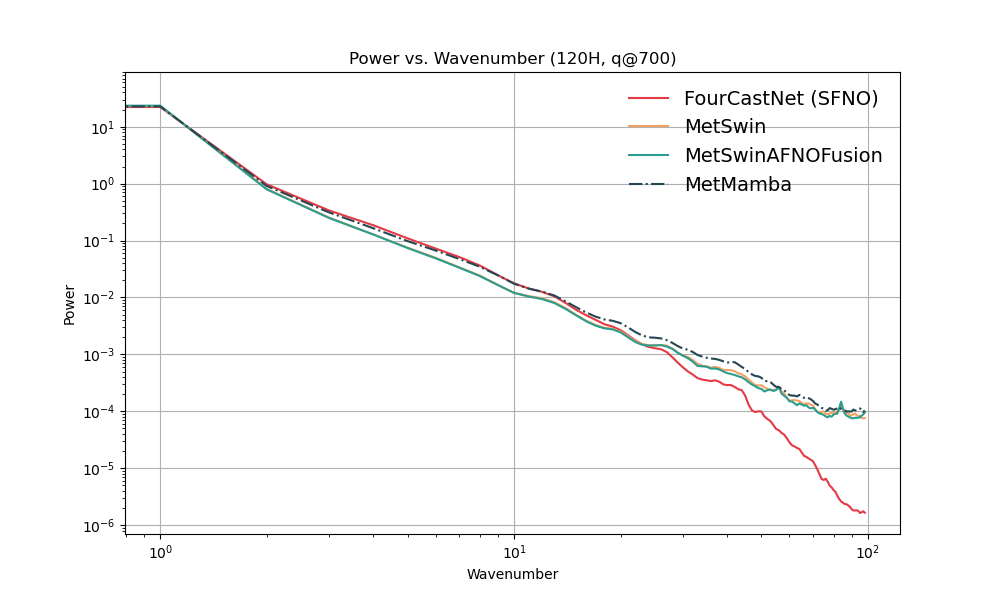}{q@700}
     \normalsubfig{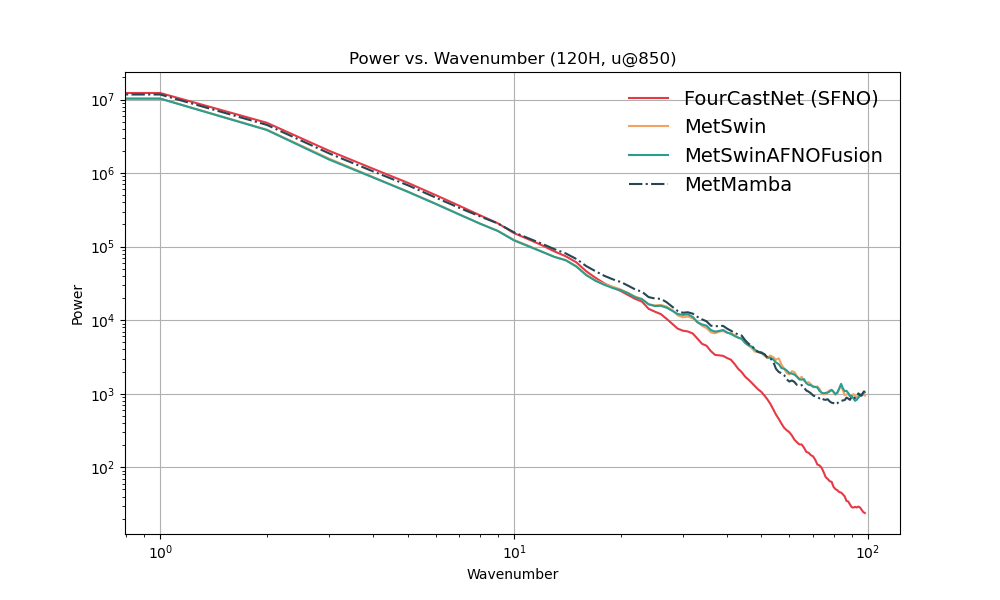}{u@850}
     \normalsubfig{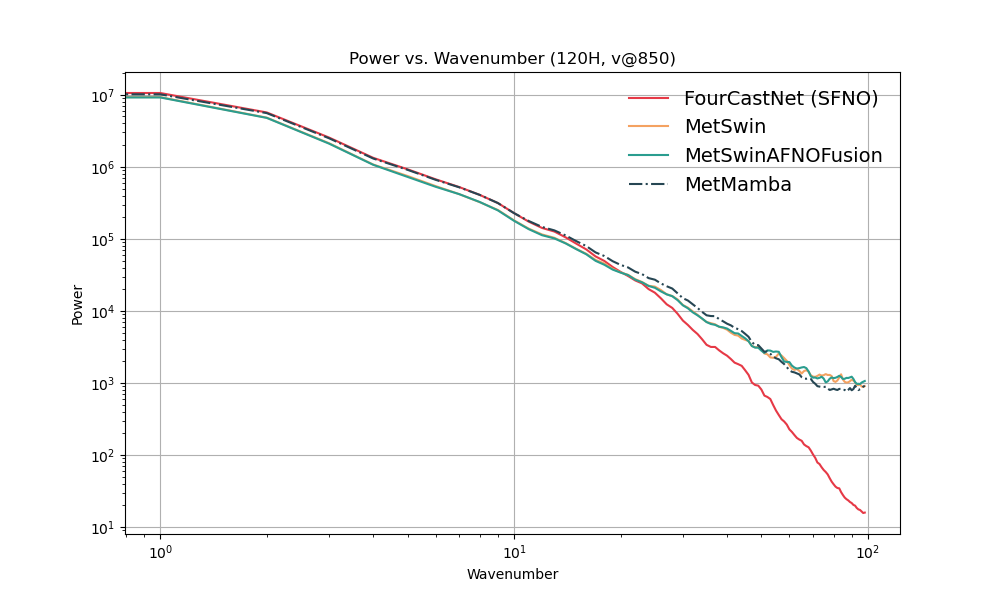}{v@850}
     \caption{Power spectrum of different DLWP models at forecast lead time 120 hour}
    \label{fig:power-hour120}
\end{figure}

We follow WeatherBench2\cite{rasp2024weatherbench2benchmarkgeneration} and adopt
its zonal energy spectrum calculation in a limited area setting. See power spectrum
analysis of 8 headline variables in \reffig{power-hour120}. The DLWP-LAMs 
have markedly better finer detail retention than its host model FourCastNet(SFNO),
with \textit{MetMamba} taking first place among majority of evaluated variables.

\end{document}